\documentclass[final,10pt,a4paper]{article}
\usepackage{times}
\usepackage{a4wide} 
\usepackage{amsmath}
\usepackage{amssymb}
\usepackage{amsfonts}
\usepackage{stmaryrd}
\usepackage{amsthm}
\usepackage{mathrsfs}
\usepackage[english]{babel}
\usepackage{color,graphicx}
\usepackage[all]{xy}

 \theoremstyle{plain}
 \newtheorem{theorem}{Theorem}[section]
 \newtheorem{lemma}[theorem]{Lemma}
 \newtheorem{corollary}[theorem]{Corollary}
 \newtheorem{proposition}[theorem]{Proposition}
 \theoremstyle{definition}
 
 \newtheorem{definition2}[theorem]{Definition}
 
 \newtheorem{example2}[theorem]{Example}
 
 \theoremstyle{remark}

 \numberwithin{equation}{section}
 
\theoremstyle{remark}

\newcommand{\comment} [1]{}

\def\ok#1{\mbox{\raisebox{0ex}[1ex][1ex]{$#1$}}}
\def \tuple#1{\langle #1 \rangle}

\newcommand{\sra}{{\shortrightarrow}}

\newcommand{\ud}{\mbox{\raisebox{0ex}[1ex][1ex]{$\:\stackrel{{\scriptscriptstyle
\mathrm{def}}}{=}\:$}}}

\newcommand{\ra}{\rightarrow}
\newcommand{\Lra}{\Leftrightarrow}
\newcommand{\Ra}{\Rightarrow}

\newcommand{\AP}{{\mathit{AP}}}
\newcommand{\APb}{{\boldsymbol{AP}}}

\newcommand{\Op}{{\mathit{Op}}}
\newcommand{\Opb}{{\boldsymbol{Op}}}
\newcommand{\States}{{\mathit{States}}}

\newcommand{\AStates}{{\mathit{A}\hspace{-1pt}\mathit{States}}}
\newcommand{\AbsDom}{{\mathrm{AbsDom}}}

\newcommand{\cL}{{\mathcal{L}}}
\newcommand{\cM}{{\mathcal{M}}}
\newcommand{\cS}{{\mathcal{S}}}
\newcommand{\cX}{{\mathcal{X}}}
\newcommand{\sS}{{\mathscr{S}}}

\newcommand{\cP}{{\mathcal{P}}}
\newcommand{\cK}{{\mathcal{K}}}
\newcommand{\cA}{{\mathcal{A}}}
\newcommand{\cT}{{\mathcal{T}}}

\newcommand{\bP}{{\mathbb{P}}}
\newcommand{\bN}{{\mathbb{N}}}
\newcommand{\bZ}{{\mathbb{Z}}}

\newcommand{\fL}{\ensuremath{\mathscr{L}}}

\newcommand{\beu}{{{\rm\bf EU}_\sra}}
\newcommand{\SP}{\ensuremath{\mathrm{SP}}}
\newcommand{\ACTL}{\ensuremath{\mathrm{ACTL}}}
\newcommand{\CTL}{\ensuremath{\mathrm{CTL}}}
\newcommand{\LTL}{\ensuremath{\mathrm{LTL}}}
\newcommand{\CTLS}{\ensuremath{\mathrm{CTL}^*}}
\newcommand{\ACTLS}{\ensuremath{\mathrm{ACTL}^*}}
\newcommand{\CTLX}{\ensuremath{\mathrm{CTL}\mbox{-}\mathrm{X}}}
\newcommand{\CTLSX}{\ensuremath{\mathrm{CTL}^*\mbox{-}\mathrm{X}}}
\newcommand{\EX}{\mathrm{EX}}
\newcommand{\Sign}{\mathit{Sign}}

\newcommand{\pret}{\ensuremath{\widetilde{\pre}}}
\newcommand{\postt}{\ensuremath{\widetilde{\post}}}

\newcommand{\grasse}[1]{\llbracket {#1} \rrbracket}

\newcommand*{\gdca}{(\alpha ,C,A,\gamma)}
\newcommand*{\gdcas}{(\alpha ,\wp(\Sigma),A,\gamma)}

\DeclareMathOperator{\Algbis}{{Alg}_{\mathrm{bis}}}
\DeclareMathOperator{\Algsimeq}{{Alg}_{\mathrm{simeq}}}
\DeclareMathOperator{\AD}{{AD}}
\DeclareMathOperator{\sq}{{sq}}
\DeclareMathOperator{\id}{id}
\DeclareMathOperator{\pre}{pre}
\DeclareMathOperator{\post}{{post}}
\DeclareMathOperator{\uco}{uco}
\DeclareMathOperator{\ucop}{uco^{\mathrm{par}}}
\DeclareMathOperator{\Absp}{Abs^{\mathrm{par}}}

\DeclareMathOperator{\lfp}{lfp}
\DeclareMathOperator{\gfp}{gfp}

\DeclareMathOperator{\img}{img}
\DeclareMathOperator{\Part}{Part}
\DeclareMathOperator{\PreOrd}{PreOrd}

\DeclareMathOperator{\preord}{preord}

\DeclareMathOperator{\Fun}{Fun}
\DeclareMathOperator{\pr}{par}

\DeclareMathOperator{\ad}{ad}
\DeclareMathOperator{\adp}{ad^p}
\DeclareMathOperator{\add}{ad^d}

\DeclareMathOperator{\GV}{\mathrm{GV}}

\DeclareMathOperator{\Ord}{\mathrm{Ord}}
\DeclareMathOperator{\Abs}{\mathrm{Abs}}

\begin{document}

\title{\Large \bf Generalized Strong Preservation\\ by Abstract Interpretation}

\author{\normalsize {\sc Francesco Ranzato} ~~~ {\sc Francesco Tapparo}\\
\normalsize Dipartimento di Matematica Pura ed Applicata, Universit\`a
di Padova\\
\normalsize Via Belzoni 7, 35131 Padova, Italy\\
\normalsize \texttt{franz$@$math.unipd.it} ~~~~ \texttt{tapparo$@$math.unipd.it}
}

\date{}
\pagestyle{plain}

\maketitle

\begin{abstract}
Standard abstract model checking relies on abstract Kripke structures
which approximate concrete models by gluing together indistinguishable
states, namely by a partition of the concrete state space.  Strong
preservation for a specification language $\fL$ encodes the
equivalence of concrete and abstract model checking of formulas in
$\fL$. We show how abstract interpretation can be used to design
abstract models that are more general than abstract Kripke
structures. Accordingly, strong preservation is generalized to
abstract interpretation-based models and precisely related to the
concept of completeness in abstract interpretation.  The problem of
minimally refining an abstract model in order to make it strongly
preserving for some language $\fL$ can be formulated as a minimal
domain refinement in abstract interpretation in order to get
completeness w.r.t.\ the logical/temporal operators of $\fL$. It turns
out that this refined strongly preserving abstract model always exists
and can be characterized as a greatest fixed point.  As a consequence,
some well-known behavioural equivalences, like bisimulation,
simulation and stuttering, and their corresponding partition
refinement algorithms can be elegantly characterized in abstract
interpretation as completeness properties and refinements.\\[10pt]
\end{abstract}

\section{Introduction}\label{intro}
The design of an abstract model checking framework always includes 
a preservation result, roughly stating that for any
formula $\varphi$ specified in some temporal language $\fL$, if
$\varphi$ holds on an abstract model then $\varphi$ also 
holds on the concrete model. On the other hand, \emph{strong
preservation} means that a formula of $\fL$ holds on an abstract
model if and only if it holds on the concrete model.  Strong
preservation is highly desirable since it allows to draw consequences
from negative answers on the abstract side~\cite{cgp99}.
%

\medskip
\noindent
\textbf{Generalized Strong Preservation.}
The relationship between abstract interpretation and abstract model
checking has been the subject of a number of works (see e.g.\ 
\cite{cgl94,cpy95,CC99,CC00,dams96,dgg97,GQ01,loi95,mas02,mas04,RT02,sch04}). 
This paper follows the standard abstract interpretation approach
\cite{CC77,CC79} where
abstract domains are specified by Galois connections, namely pairs of
abstraction and concretization maps $\alpha$/$\gamma$. We deal with generic (temporal)
languages $\fL$ of state formulae 
that are inductively generated by some given sets of atomic
propositions and operators.  
The interpretation $\boldsymbol{p}$
of atomic propositions $p\in \AP$ as subsets of $\States$ and of 
operators $f\in \Op$ as mappings $\boldsymbol{f}$ on $\wp(\States)$ is determined by   
a suitable semantic structure $\cS$, e.g.\ a Kripke structure, so that
the concrete semantics 
$\grasse{\varphi}_\cS \in \wp(\States)$ of a formula
$\varphi\in \fL$ is the
set of states making $\varphi$ true w.r.t.\ $\cS$. 
\emph{Abstract semantics} 
can be systematically defined by standard abstract interpretation.
The powerset $\wp(\States)$ plays the role of concrete semantic
domain so that abstract domains range in $\AbsDom(\wp(\States))$. 
Any abstract domain $A\in \AbsDom(\wp(States))$ 
induces an abstract semantic structure $\cS^A$ where atoms $p$ are
abstracted to $\alpha(\boldsymbol{p})$ and operators $f$ are interpreted
as best correct approximations on $A$, that is $\alpha \circ
\boldsymbol{f} \circ \gamma$. Thus, $A$ 
determines an abstract semantics 
$\grasse{\varphi}_\cS^A \in A$ that evaluates formulae
$\varphi\in \fL$ in the abstract domain $A$. \\
\indent
It turns out that this approach generalizes standard abstract model
checking \cite{cgl94,cgp99}. Given a Kripke structure $\cK
= (\States, \ra)$ (for simplicity we omit here a labeling function
for atomic propositions), a standard abstract model is specified as an
abstract Kripke structure $\cA = (\AStates,
\ok{\ra^\sharp})$ where the set $\AStates$ of abstract states is defined by a
surjective map $h:\States \ra \AStates$. Thus, $\AStates$ determines a partition of
$\States$ and vice versa. It turns out 
that state partitions are particular
abstract domains. In fact, the lattice of
partitions of $\States$ is an abstract interpretation of the lattice
of abstract domains $\AbsDom(\wp(\States))$ so that the abstract state space
$\AStates$ corresponds to a particular abstract domain $\ad(\AStates) \in
\AbsDom(\wp(\States))$. Abstract domains 
that can be derived from a state
partition are called \emph{partitioning}. 
The interpretation of the language
$\fL$ w.r.t.\ the abstract Kripke structure $\cA$ determines an abstract semantic
function $\grasse{\varphi}_\cA \in \AStates$.  
The abstract Kripke structure $\cA$ strongly preserves $\fL$ when for any
$\varphi \in \fL$ and $s\in \States$,
it turns out that $h(s)\in \grasse{\varphi}_\cA \: \Lra \: s\in \grasse{\varphi}_\cK$. \\
\indent
Strong preservation can then be generalized from standard abstract
models to abstract interpretation-based models.  Given a generalized abstract
model $A\in \AbsDom(\wp(\States))$, the induced abstract semantics 
$\grasse{\cdot}_\cS^A$ is strongly
preserving for $\fL$ when for any $\varphi\in \fL$ and
$S\in \wp(\States)$, $\alpha(S)\leq_A \grasse{\varphi}_\cS^A\:\Lra S\subseteq
\grasse{\varphi}_\cS$. It turns out that this is an abstract domain
property, because any abstract semantics $\grasse{\cdot}^\sharp:\fL
\ra A$ that evaluates formulae in the abstract domain $A$ is strongly
preserving for $\fL$ if and only if $\grasse{\cdot}_\cS^A$ is.
Standard strong preservation becomes a particular instance, namely 
an abstract Kripke structure
strongly preserves $\fL$ if and only if the corresponding
partitioning abstract model 
strongly preserves $\fL$.  On the other hand, generalized strong
preservation may work where standard strong preservation may fail,
namely it may happen that
although 
a strongly preserving abstract semantics on a partitioning abstract
model $\ad(\AStates)$ exists this cannot be derived from a strongly preserving
abstract Kripke structure on $\AStates$.

\medskip
\noindent
\textbf{Generalized Strong Preservation and Complete Abstract Interpretations.} 
Given a language $\fL$ and a Kripke structure $\cK=(\States,\ra)$, a
well-known key problem is to compute the smallest abstract state space
$\AStates_\fL$, when this exists, such that one can define an abstract
Kripke structure $\cA_{\fL}=(\AStates_\fL,\ok{\ra^\sharp})$
that strongly preserves $\fL$.  This problem admits solution for a
number of well-known temporal languages like $\CTL$ (or, equivalently,
the $\mu$-calculus), $\ACTL$ and $\CTLX$ (i.e.\ $\CTL$ without the
next-time operator $\mathrm{X}$). A number of algorithms for solving
this problem exist, like those by Paige and Tarjan \cite{pt87} for
$\CTL$, by Henzinger et al.~\cite{hhk95}, Bustan and
Grumberg~\cite{bg03} and Tan and Cleaveland \cite{TC01} for $\ACTL$,
and Groote and Vaandrager~\cite{gv90} for $\CTLX$. These are coarsest
partition refinement algorithms: given a language $\fL$ and a
partition $P$ of $\States$, which is determined by a state labeling,
these algorithms can be viewed as computing the coarsest partition
$P_\fL$ that refines $P$ and strongly preserves $\fL$.  It is worth
remarking that most of these algorithms have been designed for
computing well-known behavioural equivalences used in process algebra
like bisimulation (for $\CTL$), simulation (for $\ACTL$) and
divergence-blind stuttering (for $\CTLX$) equivalence.
%
Our abstract
interpretation-based framework allows to give a generalized view of the
above partition refinement algorithms.  
We show that the most abstract
domain $\AD_\fL \in \AbsDom(\wp(\States))$ that strongly preserves a
given language $\fL$ always exists. 
It turns out that $\AD_\fL$ is a
partitioning abstract domain if and only if
$\fL$ includes full propositional logic, that is when $\fL$ is closed under
logical conjunction and negation. Otherwise, a proper loss of
information occurs when abstracting $\AD_\fL$ to the corresponding
partition $P_\fL$.
Moreover, for some languages $\fL$, 
it may happen that one cannot define an
abstract Kripke structure on the abstract state space $P_{\fL}$
that strongly
preserves $\fL$ 
whereas the most abstract strongly preserving
semantics in $\AbsDom(\wp(\States))$ instead exists.   
\\
\indent
The concept of 
\emph{complete} abstract interpretation is well known \cite{CC79,jacm}. 
This encodes an ideal situation where the abstract semantics
coincides with the abstraction of the concrete semantics.  We
establish a precise correspondence between generalized strong
preservation of abstract models and completeness in abstract
interpretation. Our results are based on the notion of
\emph{forward complete} abstract domain. An abstract domain $A$ 
is forward complete for a concrete semantic
function $\boldsymbol{f}$ when
for any $a\in A$,
$\boldsymbol{f}(\gamma(a)) =
\gamma(\alpha(\boldsymbol{f}(\gamma(a))))$, namely when no loss of
precision occurs by approximating in $A$ a computation
$\boldsymbol{f}(\gamma(a))$. This notion of forward completeness is
dual and orthogonal to the standard definition of
completeness in abstract interpretation.    
Giacobazzi et al.~\cite{jacm} showed how complete abstract domains can
be systematically and
constructively derived from noncomplete abstract domains 
by minimal refinements. This can be done for forward
completeness as well. Given any domain
$A$, the most abstract domain that refines $A$ and is forward
complete for $\boldsymbol{f}$ does exist and can be characterized
as a greatest fixpoint. Such a domain is called the \emph{forward}
\emph{complete shell} of $A$ for $\boldsymbol{f}$. 
It turns out that strong preservation is related to forward
completeness as follows. As
described above, the most abstract domain $\AD_\fL$ that strongly
preserves $\fL$ always exists. It turns out that $\AD_\fL$ coincides
with the forward complete shell for the operators of $\fL$ 
of a basic abstract domain determined by the state labeling.
This characterization provides an elegant
generalization of  partition refinement algorithms used in standard
abstract model checking.  As a consequence of these results, we derive 
a novel characterization of the corresponding behavioural equivalences
in terms of forward completeness of abstract domains. For example, it
turns out that a partition $P$ is a bisimulation on some Kripke
structure $\cK$
if and only if the corresponding partitioning
abstract domain $\ad(P)$ is forward complete
for the standard predecessor transformer $\pre_{\ra}$ in $\cK$. 

\section{Basic Notions}

\subsection{Notation and Preliminaries} \label{not}
Let $X$ be any set.
$\Fun(X)$ denotes the set of 
functions $f:X^n\ra X$, for some $n\geq 0$, called arity of $f$.  
Following a standard
convention, when $n=0$, $f$ is meant to be a specific object of $X$.
The arity of $f$ is also
denoted by $\sharp (f)\geq 0$.  $\id$ denotes the identity map.  
If $F\subseteq \Fun(X)$ and $Y\subseteq X$ then $F(Y)\ud \{
f(\vec{y})~|~ f\in F,\, \vec{y}\in \ok{Y^{\sharp(f)}}\}$, namely $F(Y)$
is the set of images of $Y$ for each function in $F$. If $f:X\ra Y$
then the image of $f$ is also denoted by $\img(f)=\{f(x)\in Y~|~ x\in X\}$. 
If $f:X \ra Y$ and $g: Y \ra Z$ then $g \circ f : X\ra Z$ denotes the
composition of $f$ and $g$, i.e.\ $g \circ f = \lambda x. g(f(x))$.  
The complement operator for the universe set $X$ is 
$\complement :\wp(X)\ra \wp(X)$, where $\complement
(S)=X\smallsetminus S$.
When writing a set $S$ of subsets of a given set, like a partition,
we often write $S$ in a compact form like 
$\{1, 12, 13\}$ or $\{[1], [12], [13]\}$ that stand for $\{\{1\},
\{1,2\}, \{1,3\}\}$. $\Ord$ denotes the proper
class
of ordinals and $\omega\in\Ord$ denotes the first infinite ordinal. 

Let $\tuple{P,\leq}$ be a poset. Posets are often denoted also by $P_\leq$.
We use the symbol $\sqsubseteq$ to denote pointwise ordering between
functions: If $X$ is any set and $f,g:X \ra P$ then
$f\sqsubseteq g$ if for all $x\in X$, $f(x)\leq g(x)$.  
A mapping $f: P\ra Q$ on posets is continuous when $f$
preserves least upper bounds (lub's) 
of countable chains in $P$, while, dually, it is co-continuous when $f$ 
preserves greatest lower bounds (glb's) 
of countable chains in $P$. A complete lattice $C_\leq$ 
is also denoted by $\tuple{C,\leq,\vee,\wedge,\top,\bot}$ where $\vee$,
$\wedge$, $\top$ and $\bot$ denote, respectively, lub, glb, greatest
element and least element in $C$.  
A mapping $f:C\ra D$ between complete lattices
is additive (co-additive) when for any $Y\subseteq C$, $f(\vee_C
Y)=\vee_D f(Y)$ ($f(\wedge_C
Y)=\wedge_D f(Y)$). 
We denote by $\lfp(f)$ and $\gfp (f)$,
respectively, the least and greatest fixpoint, when they exist, of an
operator $f$ on a poset. 
The well-known Knaster-Tarski's theorem
states that any monotone operator $f:C\ra C$ on a complete
lattice $C$ admits a least fixpoint and the
following characterization holds:
$$\lfp(f) = \wedge \{x\in C~|~ f(x) \leq
x\}=\vee_{\alpha\in \Ord} f^{\alpha,\uparrow} (\bot)$$
where the upper iteration sequence $\{f^{\alpha,\uparrow} (x)\}_{\alpha \in
\Ord}$ of $f$ in $x\in C$ is defined by transfinite induction on $\alpha$ as
usual: 
\begin{itemize}
\item[--] $\alpha=0$:~ $\ok{f^{0,\uparrow} (x)= x}$; 
\item[--] successor ordinal $\alpha=\beta+1$:~ $\ok{f^{\beta+1,\uparrow}
(x)=f(f^{\beta,\uparrow} (x))}$;
\item[--] limit ordinal $\alpha$:~ $\ok{f^{\alpha,\uparrow} (x)}
=\vee_{\beta<\alpha} 
f^{\beta,\uparrow} (x)$. 
\end{itemize}
It is well known that if $f$ is continuous then $\lfp(f)=\vee_{\alpha\in \omega} 
f^{\alpha,\uparrow}(\bot)$. 
Dually, $f$ also admits a greatest fixpoint and the
following characterization holds:
$$\gfp(f) = \vee \{x\in C~|~ x\leq f(x)\} = 
\wedge_{\alpha\in \Ord} f^{\alpha,\downarrow} (\top),$$ 
where the lower iteration sequence $\{f^{\alpha,\downarrow} (x)\}_{\alpha \in
\Ord}$ of $f$ in $x\in C$ is defined as the upper iteration sequence
but for the case of limit ordinals: $\ok{f^{\alpha,\downarrow} (x)}
=\wedge_{\beta<\alpha} 
f^{\beta,\downarrow} (x)$.

Let $\Sigma$ be any set. $\PreOrd(\Sigma)$ denotes the set of preorder
relations on $\Sigma$, that is $R\subseteq \Sigma\times \Sigma$ is a
preorder on $\Sigma$ if $R$ is reflexive and transitive.  
$\Part(\Sigma)$ denotes the
set of partitions of $\Sigma$. Sets in a partition $P$
are called blocks of $P$. 
If $\equiv\;\subseteq \Sigma \times \Sigma $ is an equivalence relation
then we denote by $P_\equiv\in \Part(\Sigma)$ the corresponding partition of
$\Sigma$. Vice versa, if $P\in \Part(\Sigma)$ then 
$\equiv_P\,\subseteq \Sigma \times \Sigma $ denotes 
the corresponding equivalence relation on
$\Sigma$. $\Part(\Sigma)$ is endowed with the
following standard partial order $\preccurlyeq$: 
$P_1
\preccurlyeq P_2$, i.e.\ $P_2$ is coarser than $P_1$ (or $P_1$ refines
$P_2$) iff $\forall B\in P_1. \exists B' \in P_2 .\: B \subseteq
B'$. It is well known that 
$\tuple{\Part(\Sigma), \preccurlyeq}$ is a complete lattice.

A transition system $\cT=(\Sigma ,\sra)$ consists of a (possibly
infinite) set $\Sigma$ of
states  and a transition relation $\sra \subseteq \Sigma \times
\Sigma$. 
As usual \cite{cgp99}, we assume
that
the relation $\sra$ is total, i.e., for any $s\in \Sigma $ there
exists some $t\in \Sigma $ such that $s\sra t$, so that any maximal path in
$\cT$ is necessarily
infinite.  
$\cT$ is  finitely branching when  
for any $s\in \Sigma $, $\{t\in \Sigma ~|~ s \sra t \}$ is a finite
set. 
The
pre/post transformers on $\wp(\Sigma)$ are defined as usual:
\begin{itemize}
\item[--] $\pre_\sra \ud \lambda Y. \{ a\in \Sigma ~|~\exists b\in Y.\; a \sra
b\}$;
\item[--] $\pret_\sra \ud \complement \circ \pre_\sra \circ \complement = 
\lambda Y. \{ a\in \Sigma  ~|~\forall b\in \Sigma . (a \sra
b \Rightarrow  b\in Y)\}$; 
\item[--] $\post_\sra \ud \lambda Y. \{ b\in \Sigma ~|~\exists a\in Y.\; a \sra
b\}$;
\item[--] $\postt_\sra \ud \complement \circ \post_\sra \circ \complement = 
\lambda Y. \{ b\in \Sigma ~|~\forall a\in \Sigma . (a \sra
b \Rightarrow a\in Y)\}$.
\end{itemize}
Let us observe that $\pre_\sra$ and $\post_\sra$ are additive operators on
$\wp(\Sigma)_\subseteq$ while $\pret_\sra$ and
$\postt_\sra$ are co-additive.

If $R\subseteq \Sigma_1 \times \Sigma_2$ is any relation then the
relations $\ok{R^{\exists\exists}},\ok{R^{\forall\exists}}\subseteq
\wp(\Sigma_1)\times \wp(\Sigma_2)$ are defined as follows:
\begin{itemize}
\item[--] $(S_1,S_2)\in \ok{R^{\exists\exists}}~$ iff $~\exists s_1 \in
S_1.\exists s_2 \in S_2.\, (s_1,s_2)\in R$;

\item[--] $(S_1,S_2)\in \ok{R^{\forall\exists}}~$ iff $~\forall s_1 \in
S_1.\exists s_2 \in S_2.\, (s_1,s_2)\in R$.
\end{itemize}

\subsection{Abstract Interpretation and Completeness}\label{aic}

\subsubsection{Abstract Domains}\label{absdom}

In standard Cousot and Cousot's abstract interpretation,
abstract domains can be equivalently specified either by Galois
connections, i.e.\ adjunctions, or by upper closure operators
(uco's)~\cite{CC77,CC79}.  
Let us recall these standard notions.

\paragraph{Galois Connections and Insertions.}
If $A$ and $C$ are posets and $\alpha :C\ra A$ and $\gamma :A\ra C$
are monotone functions such that $\forall c\in C.\: c \leq_C \gamma
(\alpha (c))$ and $\alpha(\gamma (a)) \leq_A a$ then the quadruple
$\gdca$ is called a Galois connection (GC for short) between $C$ and
$A$.  If in addition $\alpha\circ \gamma =\lambda x.x$ then $\gdca$ is
a Galois insertion (GI for short) of $A$ in $C$.  In a GI, $\gamma$ is
1-1 and $\alpha$ is onto.  Let us also recall that the notion of GC is
equivalent to that of adjunction: if $\alpha :C\ra A$ and $\gamma
:A\ra C$ then $\gdca$ is a GC iff $\forall c\in C.\forall a\in A.\;
\alpha (c)\leq_{A} a \Lra c\leq_C \gamma (a)$.  The map $\alpha$
($\gamma$) is called the left- (right-) adjoint to $\gamma$
($\alpha$).  It turns out that one adjoint map $\alpha$/$\gamma$
uniquely determines the other adjoint map $\gamma$/$\alpha$ as
follows. On the one hand, a map $\alpha:C\ra A$ admits a necessarily
unique right-adjoint map $\gamma:A\ra C$ iff $\alpha$ preserves
arbitrary lub's; in this case, we have that $\gamma \ud \lambda a.
\vee_C \{ c\in C~|~ \alpha(c) \leq_A a\}$. 
On the other hand, a map $\gamma:A\ra C$ admits a
necessarily unique left-adjoint map $\alpha:C\ra A$ iff $\gamma$
preserves arbitrary glb's; in this case, 
$\alpha \ud \lambda c. \wedge_A \{ a\in A~|~ c \leq_C \gamma(a)\}$. 
  In particular, we have that in any GC
$\gdca$ between complete lattices it turns out that $\alpha$ is
additive and $\gamma$ is co-additive.  Also, if $\gdca$ is a GI and
$C$ is a complete lattice then $A$ is a complete lattice as well and
$\tuple{A,\leq_A}$ is order-isomorphic to
$\tuple{\img(\gamma),\leq_C}$.

We  assume the standard abstract interpretation framework, where concrete and
abstract domains, $C$ and $A$, are complete lattices 
related by abstraction and concretization maps $\alpha$ and $\gamma$ forming a GC
$\gdca$. $A$ is called an abstraction of $C$
and $C$ a concretization of $A$.  The ordering relations on concrete and
abstract domains describe the relative precision of domain values:
$x\leq y$ means that $y$ is an approximation of $x$ or, equivalently,
$x$ is more precise than $y$.
Galois connections allow to relate the concrete and
abstract notions of relative precision: an abstract value $a\in A$ approximates a
concrete value $c\in C$ when $\alpha(c) \leq_A a$, or, equivalently
(by adjunction), $c\leq_C \gamma(a)$. As a key consequence of
requiring a Galois connection, it turns out that $\alpha(c)$ is the
best possible approximation in $A$ of $c$, that is $\alpha(c) = \wedge
\{a\in A~|~ c  \leq_C \gamma(a)\}$ holds. 
If $\gdca$ is a GI then each value
of the abstract domain $A$ is useful in representing $C$, because all
the values in $A$ represent distinct members of $C$, being $\gamma$
1-1.  Any GC can be lifted to a GI by identifying in an equivalence class
those values of the abstract domain with the same concretization.
$\Abs(C)$ denotes the set of abstract domains of $C$ 
and  we write $A\in \Abs(C)$ to mean that the abstract
domain $A$ is related to $C$
through a GI $(\alpha,C,A,\gamma)$. An abstract domain $A$ is
disjunctive when the corresponding concretization map $\gamma$ is additive.

\paragraph{Closure Operators.}
An (upper) closure operator, or simply a
closure, on a poset $P_\leq$ is an operator $\mu: P\rightarrow P$ that
is
monotone, idempotent and extensive, i.e., $\forall x\in P.\; x\leq
\mu (x)$.  Dually, lower closure operators 
are monotone, idempotent, and restrictive, i.e., $\forall x\in P.\; 
\mu (x) \leq x$.  
$\uco(P)$ denotes the set of closure operators on $P$.
Let $\tuple{C, \leq ,\vee ,\wedge ,\top ,\bot}$ be a complete lattice.
A closure $\mu
\in\uco(C)$ is uniquely determined by its image $\img(\mu)$, which 
coincides with its set of fixpoints, as follows: $\mu=
\lambda y.\wedge \{ x\in \img(\mu)~|~ y\leq x\}$. Also, 
$X\subseteq C$ is the image of some closure operator $\mu_X$ on $C$
iff $X$ is a Moore-family of $C$, i.e., $X=\cM (X)\ud \{\wedge S~|~
S\subseteq X\}$~---~where $\wedge \varnothing=\top \in \cM (X)$. In
other terms, $X$ is a Moore-family of $C$ when $X$ is meet-closed.  In
this case, $\mu_X=\lambda y.\wedge \{ x\in X~|~ y\leq x\}$ is the
corresponding closure operator on $C$.  For any $X\subseteq C$, $\cM
(X)$ is called the Moore-closure of $X$ in $C$, i.e., $\cM (X)$ is the
least (w.r.t.\ set inclusion) subset of $C$ which contains $X$ and is
a Moore-family of $C$. Moreover, it turns out that for any $\mu \in
\uco(C)$ and any Moore-family $X\subseteq C$, $\mu_{\img(\mu)} = \mu$
and $\img(\mu_X)=X$. Thus, closure operators on $C$ are in bijection
with Moore-families of $C$. This allows us to consider a closure
operator $\mu\in \uco(C)$ both as a function $\mu:C\ra C$ and as a
Moore-family $\img(\mu)\subseteq C$.  This is particularly useful and
does not give rise to ambiguity since one can distinguish the use of a
closure $\mu$ as function or set according to the context.

It turns out that $\tuple{\mu,\leq}$ is a complete meet
subsemilattice of $C$, i.e.\ $\wedge$ is its glb, 
but, in general, it is not a complete sublattice of $C$, since
the lub in $\mu$~---~defined by $\lambda Y\subseteq \mu.\, \mu
(\vee Y)$~---~might be different from that in $C$.  In fact, it turns
out that $\mu$ is a complete sublattice of $C$ (namely, $\img(\mu)$
is also join-closed) iff $\mu$ is additive.  

If $C$ is a complete lattice then $\uco(C)$ endowed with the pointwise ordering
$\sqsubseteq$ is a
complete lattice denoted by
$\tuple{\uco(C),\sqsubseteq,\sqcup,\sqcap,\lambda x.\top,\lambda x.x}$,
where for every $\mu,\eta \in\uco(C)$, $\{
\mu_i \}_{i\in I} \subseteq\uco(C)$ and $x\in C$:
\begin{itemize}
\item[--] $\mu \sqsubseteq \eta$ iff $\forall y\in C.\; \mu (y) \leq
 \eta(y)$ iff  $\img(\eta)
\subseteq \img(\mu)$;
\item[--] $(\sqcap_{i\in I} \mu_i)(x) = \wedge_{i\in I} \mu_i (x)$;
\item[--] $x \in \sqcup_{i\in I} \mu_i \:\Lra\:
\forall i\in I.\; x\in \mu_i$;
\item[--] $\lambda x.\top$ is the greatest element, whereas $\lambda x.x$ is
the least element.
\end{itemize}
Thus, the glb in $\uco(C)$ is defined pointwise, while the
lub of a set of closures $\{
\mu_i \}_{i\in I} \subseteq\uco(C)$ is the closure whose image 
is given by the set-intersection $\cap_{i \in I}
\mu_i$.

\paragraph{The Lattice of Abstract Domains.}
It is well known since \cite{CC79} that abstract domains can be
equivalently specified either as Galois insertions or as 
closures.  These two
approaches are completely equivalent. On the one hand, if $\mu\in\uco(C)$ and
$A$ is a complete lattice which is isomorphic to $\img(\mu)$, where 
$\iota : \img(\mu)\ra A$ and $\iota^{-1}:
A\ra\img(\mu)$ provide the isomorphism, then
$(\iota\circ\mu,C,A,\iota^{-1} )$ is a GI. On the other hand, if $\gdca$ is a GI then
$\mu_A \ud \gamma \circ \alpha\in\uco(C)$ is the closure associated with $A$
such that $\tuple{\img(\mu_A),\leq_C}$ is a complete lattice which is isomorphic to
$\tuple{A,\leq_A}$. Furthermore, these two constructions
are inverse of each other.  Let us also remark that an abstract domain
$A$ is disjunctive iff $\mu_A$ is additive. Given an abstract domain $A$ specified
by a GI $\gdca$, its associated closure $\gamma\circ\alpha$ on $C$ can
be thought of as the ``logical meaning'' of $A$ in $C$, since this is
shared by any other abstract representation for the objects of $A$.
Thus, the closure operator approach is particularly convenient when
reasoning about properties of abstract domains independently from the
representation of their objects.

Abstract domains specified by GIs can be pre-ordered w.r.t.\ precision
as follows: if $A_1,A_2 \in \Abs(C)$ then $A_1$ is more precise (or
concrete) than
$A_2$ (or $A_2$ is an abstraction of $A_1$), denoted by $A_1 \preceq
A_2$, when $\mu_{A_1} \sqsubseteq \mu_{A_2}$.  The pointwise
ordering $\sqsubseteq$ between uco's corresponds therefore to the standard ordering used to
compare abstract domains with respect to their precision. Also, $A_1$
and $A_2$ are equivalent, denoted by $A_1 \simeq A_2$, when their
associated closures coincide, i.e.\ $\mu_{A_1}=\mu_{A_2}$.  Hence, the
quotient $\Abs(C)_{/\simeq}$ gives rise to a poset that, by a slight
abuse of notation,  is simply denoted by 
$\tuple{\Abs(C),\sqsubseteq}$. Thus, when we write $A\in\Abs(C)$ we
mean that $A$ is any representative of an equivalence class in 
$\Abs(C)_{/\simeq}$
and is 
specified by a Galois insertition $\gdca$.  
It turns out that 
$\tuple{\Abs(C),\sqsubseteq}$ is a complete lattice, called the
lattice of abstract interpretations of $C$
\cite{CC77,CC79},  because it is
isomorphic to the complete lattice $\tuple{\uco(C),\sqsubseteq}$.
Lub's and glb's in $\Abs(C)$ have therefore the following reading as
operators on domains.  Let $\{ A_i\}_{i\in I}\subseteq\Abs(C)$:
(i)~$\sqcup_{i\in I}A_i $ is the most concrete among the domains which
are abstractions of all the $A_i$'s; (ii)~$\sqcap_{i\in I} A_i $ is
the most abstract among the domains which are more concrete than every
$A_i$~---~this latter domain is also known as reduced product of all
the $A_i$'s.

\subsubsection{Completeness}\label{secco}
Let $C$ be a concrete domain, $f:C\ra C$ be a concrete semantic function\footnote{For
simplicity of notation we consider here unary functions since the extension
to generic $n$-ary functions is straightforward.}  
and let $f^\sharp:A \ra A$ be a corresponding abstract
function on an abstract domain $A\in \Abs(C)$ specified by a GI 
$(\alpha,C,A,\gamma)$.   Then,
$\tuple{A,f^\sharp}$ is a sound abstract interpretation when $\ok{\alpha
\circ f \sqsubseteq f^\sharp\circ \alpha}$ holds. The abstract
function $\ok{f^\sharp}$ is called a correct approximation on $A$ of
$f$. 
This means that a concrete
computation  $f(c)$
can be correctly approximated in $A$ by $\ok{f^\sharp (\alpha (c))}$, namely $\alpha(f(c))
\leq_A \ok{f^\sharp (\alpha (c))}$. An abstract function 
$\ok{f_1^\sharp}:A \ra A$ is more precise than 
$\ok{f_2^\sharp}:A \ra A$ when 
$\ok{f_1^\sharp} \sqsubseteq \ok{f_2^\sharp}$. 
Since  $\ok{\alpha
\circ f \sqsubseteq f^\sharp\circ \alpha}$ holds iff  $\ok{\alpha
\circ f \circ \gamma \sqsubseteq f^\sharp}$ holds, 
the abstract function
$\ok{f^A \ud \alpha \circ f \circ \gamma: A\rightarrow A}$ is called the best correct
approximation of $f$ in $A$.  

Completeness in abstract interpretation
corresponds to requiring that, in addition to soundness, no loss of
precision occurs when $f(c)$ is approximated in $A$  
by $\ok{f^\sharp(\alpha(c))}$. Thus, completeness of $\ok{f^\sharp}$
for $f$ is encoded by the equation $\ok{\alpha \circ f = f^\sharp \circ
\alpha}$. This is also called backward completeness because a dual
form of forward completeness may be considered. 
As a very simple example, let us consider the abstract domain
$\mathit{Sign}$ representing the sign of an integer variable, namely
$\mathit{Sign} = \{\bot, \ok{\bZ_{< 0}}, 0, \ok{\bZ_{>0}}, \top\}\in
\Abs(\wp(\bZ)_\subseteq )$. Let us consider the binary concrete operation of
integer addition on sets
of integers, that is 
$X+Y \ok{\ud} \{x+y~|~ x\in X,\, y\in Y\}$, and the square operator on sets
of integers, that is $\ok{X^2} \ok{\ud} \ok{\{x^2 ~|~ x\in X\}}$.  It turns out that the
best correct approximation
$\ok{+^\mathit{Sign}}$ of integer addition in
$\mathit{Sign}$ is sound but not complete~---~because $\alpha(\{-1\} +
\{1\}) = \ok{0 <_{\mathit{Sign}}}  \top = \alpha(\{-1\})
\ok{+^{\mathit{Sign}}} \alpha(\{1\})$~---~ while it is easy to check
that the best correct approximation of
the square operation in $\mathit{Sign}$ is instead complete. 

A dual form of completeness may be considered.  The
soundness condition $\ok{\alpha \circ f \sqsubseteq f^\sharp\circ
\alpha}$ can be equivalently formulated as $\ok{f\circ \gamma
\sqsubseteq \gamma \circ f^\sharp}$. Forward completeness for $\ok{f^\sharp}$
corresponds to requiring that the equation $\ok{f\circ \gamma = \gamma
\circ f^\sharp}$ holds,
and therefore means that no loss of precision occurs when a concrete computation
$f(\gamma(a))$, for some abstract value $a\in A$, is approximated in
$A$ by $\ok{f^\sharp (a)}$. Let us
notice that backward and forward completeness are orthogonal
concepts. In fact: (1) as observed above, we have that $\ok{+^\mathit{Sign}}$ is
not backward complete while it is forward complete because for any $a_1,a_2 \in
\mathit{Sign}$, $\gamma(a_1) + \gamma(a_2) = 
\gamma( a_1 \ok{+^\mathit{Sign}} a_2)$; (2)
the best correct approximation $\ok{(\cdot)^{2_{\mathit{Sign}}}}$ of
the square operator on $\mathit{Sign}$ is
not forward complete because $\ok{\gamma( \bZ_{>0} )^2} \subsetneq
\gamma(\mathbb{Z}_{>0})=\gamma( \ok{(\bZ_{>0})^{2_{\mathit{Sign}}}})$
while, as observed above, it is instead backward complete.

Giacobazzi et al.~\cite{jacm} observed that completeness uniquely
depends upon the abstraction map, i.e.\ upon the abstract domain: this
means that if $\ok{f^\sharp}$ is backward complete for $f$ then the best correct
approximation $\ok{f^A}$ of $f$ in $A$ is
backward complete as well, and, in this case, $\ok{f^\sharp}$ indeed coincides
with $\ok{f^A}$.  Hence, for any abstract domain $A$, one can define
a backward complete abstract operation $\ok{f^\sharp}$  on $A$ if and only if $\ok{f^A}$ is
backward complete. Thus, an abstract domain $A \in \Abs(C)$ is defined to be
backward complete for $f$ iff the equation $\alpha\circ f = \ok{f^A}\circ \alpha$ holds. This
simple observation makes backward completeness an abstract domain property,
namely an intrinsic characteristic of the abstract domain.  Let us
observe that $\alpha\circ f = \ok{f^A} \circ \alpha$ holds iff $\gamma \circ
\alpha \circ f = \gamma \circ \ok{f^A} \circ \alpha = \gamma \circ \alpha
\circ f \circ \gamma \circ \alpha$ holds, so that $A$ is
backward complete for $f$ when $\mu_A \circ f = \mu_A \circ f \circ
\mu_A$. Thus, a closure $\mu\in \uco(C)$, that defines some abstract
domain, is backward complete for $f$ when $\mu \circ f = \mu \circ f \circ
\mu$ holds.
Analogous observations apply to 
forward completeness, which is also an abstract domain property:
$A\in \Abs(C)$ is forward complete for $f$ (or forward $f$-complete) 
when $f \circ \mu_A = \mu_A \circ f \circ
\mu_A$, while a closure $\mu\in \uco(C)$
is forward complete for $f$ when $f \circ \mu = \mu \circ f \circ
\mu$ holds.

Let us also recall that, by a well-known result (see, e.g.,
\cite[Theorem~7.1.0.4]{CC79}, \cite[Fact~2.3]{ap86} and
\cite[Lemma~4.3]{db83}),  backward 
complete abstract domains are ``fixpoint complete'' as well. This
means that if $A\in \Abs(C)$ is backward complete for a concrete monotone
function $f:C\ra C$ then 
$\alpha(\lfp(f)) = \ok{\lfp(f^A)}$. Moreover, if $\alpha$ and $f$ are
both co-continuous then this
also holds for greatest fixpoints, namely $\alpha(\gfp(f)) =
\ok{\gfp(f^A)}$. 
As far as forward completeness is concerned, the following result
holds.  
\begin{lemma}\label{fixrem2}
If $A\in \Abs(C)$ is forward complete
for a monotone $f$ then $\alpha(\gfp(f))= \gfp(f^A)$. 
Moreover, if $\gamma$ and $f$ are
both continuous and $\gamma(\bot_A)=\bot_C$
then $\alpha(\lfp(f))= \lfp(f^A)$.
\end{lemma}
\begin{proof}
Let us show that $\alpha(\gfp(f)) = \gfp(f^A)$.
On the one hand, since $\gfp(f) \leq \gamma(\alpha (\gfp(f)))$, we have that
$\gfp(f)=f(\gfp(f)) \leq f(\gamma(\alpha (\gfp(f))))$, therefore, by using
forward completeness, $\gfp(f) \leq \gamma(f^A (\alpha (\gfp(f))))$. Thus,
$\alpha(\gfp(f)) \leq f^A (\alpha(\gfp(f)))$, from which follows that
$\alpha(\gfp(f)) \leq \gfp (f^A)$. On the other hand, by using forward
completeness, $f(\gamma(\gfp(f^A))) = \gamma (f^A(\gfp(f^A))) =
\gamma(\gfp(f^A))$, so that $\gamma(\gfp(f^A)) \leq \gfp(f)$, and
therefore, by applying $\alpha$, we obtain that $\gfp(f^A) =
\alpha(\gamma(\gfp(f^A))) \leq \alpha(\gfp(f))$. \\
Assume now that $\gamma$ and $f$ are
both continuous and $\gamma(\bot_A)=\bot_C$.
Let us show
by induction on $k$ that for any $k\in \mathbb{N}$, 
$\gamma ((f^A)^{k,\uparrow} (\bot_A)) =
f^{k,\uparrow} (\bot_C)$. 
\\
($k=0$): By hypothesis, $\gamma ((f^A)^{0,\uparrow} (\bot_A)) = \gamma (\bot_A) =\bot_C = 
f^{0,\uparrow} (\bot_C)$.\\ 
($k+1$): 
\begin{align*}
\gamma ((f^A)^{k+1,\uparrow} (\bot_A)) &=   \\
\gamma (f^A ((f^A)^{k,\uparrow} (\bot_A))) &= \text{~~~[by forward
completeness]}\\ 
f(\gamma ((f^A)^{k,\uparrow}
(\bot_A))) &= \text{~~~[by inductive
hypothesis]}\\ 
f(f^{k,\uparrow} (\bot_C)) &=  \\
f^{k+1,\uparrow} (\bot_C)). &
\end{align*}
Thus, by applying $\alpha$, we obtain that for any $k\in \mathbb{N}$, 
$$(f^A)^{k,\uparrow} (\bot_A) =
\alpha (f^{k,\uparrow} (\bot_C)).\eqno(\dagger)$$
Since $\gamma$ and $f$ are continuous and $\alpha$ is always additive,
we have that $f^A = \alpha \circ f \circ \gamma$ is continuous because
it is a composition of continuous functions. Hence:
\begin{align*}
\lfp(f^A) &=  \text{~~~[by Knaster-Tarski's theorem]} \\
\vee_{k\in \bN} (f^A)^{k,\uparrow}(\bot_A) & = \text{~~~[by
$(\dagger)$]} \\
\vee_{k\in \bN} \alpha (f^{k,\uparrow}(\bot_C)) & = \text{~~~[as
$\alpha$ is additive]} \\
\alpha(\vee_{k\in \bN} f^{k,\uparrow}(\bot_C)) & = \text{~~~[by
Knaster-Tarski's theorem]} \\
\alpha(\lfp(f)) &
\end{align*}
and this concludes the proof. 
\end{proof}

It is worth noting that concretization maps of abstract domains which
satisfies the ascending chain conditions (i.e., every ascending chain
is eventually stationary) are always trivially continuous. 

\subsubsection{Shells}
Refinements of abstract domains have been studied from the
beginning of abstract interpretation \cite{CC77,CC79} and led to the
notion of shell of an abstract domain \cite{fgr96,GR97,jacm}.   
Given a generic poset $P_\leq$ of semantic objects~---~where $x\leq y$
intuitively means  that $x$ is a ``refinement'' of $y$~---~and a property $\cP
\subseteq P$ of these objects, 
the generic 
notion of \emph{shell} goes as follows:
the $\cP$-shell of an object $x \in P$ is defined to be 
an object $s_x \in P$  such that:
\begin{itemize}
\item[{\rm (i)}] $s_x$ satisties the property $\cP$, 
\item[{\rm (ii)}] $s_x$ is a refinement  of $x$, and
\item[{\rm (iii)}] $s_x$ is the greatest among the objects satisfying (i)
and (ii). 
\end{itemize}
Note that if a $\cP$-shell exists then it is unique. Moreover, if the
$\cP$-shell exists for any object in $P$ then it turns out that the operator mapping
$x\in P$ to its $\cP$-shell is a lower closure operator on $\cP$,
being monotone, idempotent and reductive: this operator will be called
the \emph{$\cP$-shell refinement}. 
We will be particularly interested in shells of abstract domains and
partitions, namely shells in the complete lattices of abstract domains and
partitions. 
Given a state space $\Sigma$ and a partition property
$\cP\subseteq \Part(\Sigma)$, the $\cP$-shell of $P\in \Part(\Sigma)$
is the coarsest refinement of $P$ satisfying $\cP$, when this exists. 
Also, given a concrete domain $C$ and a domain property
$\cP\subseteq \Abs(C)$, the $\cP$-shell of $A\in \Abs(C)$, when this
exists, is the
most abstract domain that satisfies $\cP$ and refines $A$.  
Giacobazzi et al.~\cite{jacm} gave a constructive characterization of
backward complete abstract domains, under the assumption of dealing
with continuous concrete functions.  As a consequence, they showed 
that backward complete shells always exist when the concrete
functions are continuous. 
In Section~\ref{csp} we will follow
this same idea for forward completeness and this will provide the
link between strongly preserving abstract models and complete
abstract interpretations.

\subsection{Abstract Model Checking and Strong Preservation}\label{amc}

Standard temporal languages like $\CTL$, $\CTLS$, $\ACTL$, the
$\mu$-calculus, $\LTL$, etc., are interpreted on models specified as
Kripke structures.  Given a set $\mathit{AP}$ of atomic propositions
(of some language), a Kripke structure $\cK= (\Sigma ,\sra,\ell)$ over
$\mathit{AP}$ consists of a transition system $(\Sigma ,\sra)$ together
with a state labeling function $\ell:\Sigma \ra \wp(\mathit{AP})$. We
use the following notation: for any $s\in \Sigma$, $[s]_\ell \ud
\{s'\in \Sigma~|~ \ell(s)=\ell(s')\}$, while $P_\ell \ud \{[s]_\ell~|~
s\in \Sigma\}\in \Part(\Sigma)$ denotes the state partition that is induced by
$\ell$.  The notation $s \ok{\models^\cK} \varphi$ means
that a state $s\in \Sigma$ satisfies in $\cK$ a state formula $\varphi$ of
some language $\fL$, where the specific definition of the satisfaction
relation $\ok{\models^\cK}$ depends on the language $\fL$
(interpretations of standard logical/temporal operators can be found
in \cite{cgp99}).

Standard abstract model checking \cite{cgl94,cgp99} 
relies on abstract Kripke structures that are defined
over partitions of the concrete state space $\Sigma$.
A set $A$ of abstract states is related to $\Sigma$ 
by a surjective abstraction $h:\Sigma \ra A$ that maps concrete states into
abstract states and thus gives rise to a state partition
$P_h \ud \{h^{-1}(a)~|~a\in A\} \in
\Part(\Sigma)$. 
Thus, in standard abstract model checking, formulae are interpreted on
an abstract Kripke structure $\cA = (A, \ok{\sra^\sharp},
\ok{\ell^\sharp})$ whose states are an abstract
representation in $A$ of some block of the partition $P_h$.  Given a
specification language $\fL$ of state formulae, a weak preservation result for $\fL$
guarantees that if a formula in $\fL$ holds on an abstract Ktipke structure
$\cA$ then it also holds on the corresponding concrete structure
$\cK$: for any $\varphi\in \fL$, $a\in A$ and $s\in \Sigma$ such that
$h(s)=a$, if $a
\ok{\models^{\cA}} \varphi$ then
$s \ok{\models^{\cK}} \varphi$.  Moreover, 
strong preservation (s.p.\ for short) for $\fL$ encodes the equivalence of
abstract and concrete validity for formulae in $\fL$:
for any $\varphi\in \fL$, $a\in A$ and $s\in \Sigma$ such that
$h(s)=a$, $a
\ok{\models^{\cA}} \varphi$ if and only if
$s \ok{\models^{\cK}} \varphi$.
 
The definition of weakly/strongly preserving  
abstract Kripke structures depends on the language $\fL$. 
Let us recall some well-known examples \cite{cgl94,cgp99,gl94}.
Let $\cK= (\Sigma ,\sra,\ell)$ be a concrete Kripke
structure
$h: \Sigma \ra A$ be a surjection. 
\begin{itemize}
\item[{\rm (i)}] Consider the language $\ACTLS$. If $P_h\preceq
P_\ell$ then the abstract Kripke
structure $\cA = (A, \ok{\sra_h^{\exists\exists}}, \ell_h)$
weakly preserves $\ACTLS$,
where $\ell_h (a) =\cup \{\ell(s)~|~s\in \Sigma,\: h(s)=a\}$ and
$\ok{\sra_h^{\exists\exists}}\subseteq A\times A$ is defined as: 
$h(s_1)\, \ok{\sra_h^{\exists\exists}}\, h(s_2)  \;\,\Lra\;\,\exists
s_1',s_2'.\: h(s_1')=h(s_1) \:\;\&\;\: h(s_2')=h(s_2) \:\;\&\;\:
s_1' \sra s_2'$.
\item[{\rm (ii)}] Let
$P_{\mathrm{sim}}\in \Part(\Sigma)$ be the partition induced by 
simulation equivalence on
$\cK$. If $P_h =
P_{\mathrm{sim}}$ (this also holds when $P_h \preceq P_{\mathrm{sim}}$)
then the abstract Kripke
structure $\cA = (A, \ok{\sra_h^{\forall\exists}},\ell_h)$ strongly preserves
$\ACTLS$, where 
$h(s_1)\,\ok{\sra_h^{\forall\exists}} \, h(s_2) \;\,\Lra\;\,\forall
s_1'.\: h(s_1')=h(s_1).\: \exists s_2'.\: h(s_2')=h(s_2) \;\&\;
s_1'\sra s_2'$.
\item[{\rm (iii)}] Let
$P_{\mathrm{bis}}\in \Part(\Sigma)$ 
be the partition induced by bisimulation equivalence on $\cK$. 
If $P_h = P_{\mathrm{bis}}$ (this also holds when $P_h \preceq P_{\mathrm{bis}}$)
then the abstract Kripke
structure $\cA = (A, \ok{\sra_h^{\exists\exists}},
\ell_h)$ strongly preserves $\CTLS$. 
\end{itemize}

Following Dams \cite[Section~6.1]{dams96} 
and Henzinger et al.~\cite[Section~2.2]{hmr05},
the notion of strong preservation  can be also given w.r.t.\ a mere
state partition  rather than w.r.t.\ an abstract Kripke structure. 
Let $\ok{\grasse{\cdot}_\cK}: \fL \ra \wp(\Sigma)$ be the semantic
function of state formulae in $\fL$ w.r.t.\ a Kripke structure
$\cK = (\Sigma ,\sra,\ell)$, i.e., $\ok{\grasse{\varphi}_\cK} \ud \{s\in \Sigma~|~
s\ok{\models^\cK} \varphi\}$.  
Then,  the semantic interpretation of $\fL$ on $\cK$ induces
the following logical equivalence $\ok{\equiv_\fL^\cK}\: \subseteq \Sigma 
\times \Sigma $:~ $$s\,\ok{\equiv^\cK_\fL}\, s' \text{~~iff~~} 
\forall \varphi \in \fL.\, s\in
\ok{\grasse{\varphi}_\cK} \: \Lra \: s' \in \ok{\grasse{\varphi}_\cK}.$$
Let
$\ok{P_\fL} \in \Part(\Sigma)$ be the partition induced by
$\ok{\equiv^\cK_\fL}$  (the
index $\cK$ denoting the Kripke structure is omitted). Then, a
partition $P\in \Part(\Sigma)$ is strongly preserving\footnote{Dams 
\cite{dams96} uses the term ``fine'' instead of ``strongly
preserving''.} for $\fL$
(when interpreted on $\cK$) if $P \preccurlyeq \ok{P_\fL}$. Thus,
$\ok{P_\fL}$ is the coarsest partition that is strongly preserving for
$\fL$. 
For a number of well known temporal languages, like 
$\ACTLS$, $\CTLS$ (see, respectively, the above points~(ii) and (iii)), $\CTLSX$ and
the fragments of the $\mu$-calculus described by Henzinger et al.~\cite{hmr05}, 
it turns out that if $P$ is
strongly preserving for $\fL$ then the abstract
Kripke structure $(P, \ok{\sra^{\exists\exists}},
\ell_\fL)$ is strongly
preserving for $\fL$, where, for any $B\in P$, $\ell_\fL (B) =
\cup_{s\in B} \ell(s)$.
In particular,  $(P_\fL, \ok{\sra^{\exists\exists}},
\ell_\fL)$  is strongly preserving for $\fL$ and, additionally, $P_\fL$ is the
smallest possible abstract state space, namely   
if $\cA = (A, \ok{\sra^\sharp}, \ok{\ell^\sharp})$ is an abstract Kripke structure that
strongly preserves $\fL$ then $|P_\fL| \leq |A|$. 

However, given a language $\fL$ and a  Kripke structure 
$\cK$ where formulae of $\fL$ are interpreted, 
the following example shows that  it is not always possible to
define an abstract Kripke structure $\cA$ on the partition
$P_\fL$ such that $\cA$ strongly preserves $\fL$.

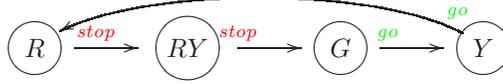
\begin{figure}[t]
  \centering
  \mbox{
    \xymatrix  { 
     *+[o][F-]{\phantom{Y}R\phantom{Y}}\ar[r] ^(0.40){\color{red}\mathit{stop}} &  	    
      *+[o][F-]{\phantom{Y}RY\phantom{Y}} \ar[r] ^(0.33){\color{red}\mathit{stop}} & 
      *+[o][F-]{\phantom{Y}G\phantom{Y}}\ar[r] ^(0.33){\color{green}\mathit{go}}  &
      *+[o][F-]{\phantom{Y}Y\phantom{Y}}\ar@(ul,ur)[lll] _(0.12){\color{green}\mathit{go}}
	\\
    }
  }
\caption{A U.K.\ traffic light.}
  \label{figsem}
\end{figure}

\begin{example2}\label{semaforo}
Consider the following simple language $\fL$:
$$\fL\ni \varphi ::= \mathit{stop} ~|~ \mathit{go} ~|~ \mathrm{AXX} \varphi$$
and the Kripke structure $\cK$ depicted in Figure~\ref{figsem}, where
superscripts determine the labeling function.
$\cK$ models a four-state traffic light controller (like in the U.K.\ and in
Germany): Red $\rightarrow$ RedYellow $\rightarrow$ Green
$\rightarrow$ Yellow. 
According to the standard semantics of $\mathrm{AXX}$, 
we have that $s\ok{\models^\cK} \mathrm{AXX}\varphi$ iff for any path
$s_0 s_1 s_2\ldots $
starting from $s_0=s$, it happens that $s_2 \ok{\models^\cK} \varphi$. 
It turns out that
$\ok{\grasse{\mathrm{AXX}\mathit{stop}}_\cK}=\{G,Y\}$ and 
$\ok{\grasse{\mathrm{AXX}\mathit{go}}_\cK}=\{R,RY\}$. Thus, we have
that $\ok{P_\fL} =
\ok{\{\{R,RY\},\{G,Y\}\}}$. 
However, let us show that there exists no abstract transition relation $\ok{\sra^\sharp}
\subseteq \ok{P_\fL \times P_\fL}$ such that the abstract Kripke structure
$\ok{\cA}= (\ok{P_\fL},
\ok{\sra^\sharp} ,\ok{\ell_\fL})$  strongly preserves $\ok{\fL}$. Assume
by contradiction that such an abstract Kripke structure $\cA$
exists. Let $B_1 = \{R,RY\}\in \ok{P_\fL}$ and $B_2 = \{G,Y\}\in \ok{P_\fL}$.
Since $R \ok{\models^\cK}
\mathrm{AXX}\mathit{go}$ and
$G \ok{\models^\cK} \mathrm{AXX}\mathit{stop}$,   by strong preservation,
it must be that $B_1 \ok{\models^\cA} \mathrm{AXX}\mathit{go}$ 
and $B_2 \ok{\models^\cA} \mathrm{AXX}\mathit{stop}$. Hence, necessarily,
$B_1 \ok{\sra^\sharp} B_2$ and $B_2 \ok{\sra^\sharp} B_1$. This leads to the
contradiction
$B_1 \ok{{\not\models}^\cA}
\mathrm{AXX}\mathit{go}$. In fact, 
if $\ok{\sra^\sharp} = 
\{(B_1,B_2),(B_2,B_1)\}$ then we would have that $B_1 \ok{{\not\models}^\cA}
\mathrm{AXX}\mathit{go}$. On the other hand, if, instead, $B_1
\ok{\sra^\sharp} B_1$ (the case $B_2 \ok{\sra^\sharp} B_2$ is
analogous), then  we would still
have that 
$B_1 \ok{{\not\models}^\cA}
\mathrm{AXX}\mathit{go}$. Even more, along the same lines it is not hard
to show that no proper abstract Kripke structure that strongly preserves
$\fL$ can
be defined,  because even if either $B_1$ or $B_2$ is split
we still cannot define an abstract transition relation that is strongly
preserving for $\fL$. 
\qed
\end{example2}

\section{Partitions as Abstract Domains}\label{paa}

\begin{figure}[t]
\begin{center}
\begin{tabular}{ccc}
\mbox{
\setlength{\unitlength}{.65cm}
\begin{picture}(6,5)
\linethickness{0.1mm}
\multiput(0,0)(1,0){7}
{\line(0,1){5}}
\multiput(0,0)(0,1){6}
{\line(1,0){6}}
\thicklines
{\color{magenta}
\qbezier(1.5,1.5)(1.2,3)(2,4.5)
\qbezier(2,4.5)(2.5,4.75)(3.2,4.2)
\qbezier(3.2,4.2)(3,2.5)(4.5,2.5)
\qbezier(4.5,2.5)(5,2.5)(5.5,2)
\qbezier(5.5,2)(4,1.5)(1.5,1.5)
\color{blue}
\thicklines
\put(1,1){\line(0,1){4}}
\put(1,5){\line(1,0){3}}
\put(4,5){\line(0,-1){2}}
\put(4,3){\line(1,0){2}}
\put(6,3){\line(0,-1){2}}
\put(1,1){\line(1,0){5}}
}
\end{picture}
}
&
\mbox{~~~~~~~~~~~~~}
&
\mbox{
\setlength{\unitlength}{.67cm}
\begin{picture}(6,5)
\linethickness{0.1mm}
\multiput(0,0)(1,0){7}
{\line(0,1){5}}
\multiput(0,0)(0,1){6}
{\line(1,0){6}}
\thicklines
{\color{magenta}
\qbezier(1.5,1.5)(1.2,3)(2,4.5)
\qbezier(2,4.5)(2.5,4.75)(3.2,4.2)
\qbezier(3.2,4.2)(3,2.5)(4.5,2.5)
\qbezier(4.5,2.5)(5,2.5)(5.5,2)
\qbezier(5.5,2)(4,1.5)(1.5,1.5)
\color{red}
\thicklines
\put(2,2){\line(0,1){2}}
\put(2,4){\line(1,0){1}}
\put(3,4){\line(0,-1){2}}
\put(3,2){\line(-1,0){1}}
}
\end{picture}
}
\end{tabular}
\end{center}
\caption{Partitions as abstract domains: over-approximation on
the left and under-approximation on the right.}\label{figuco}
\end{figure}
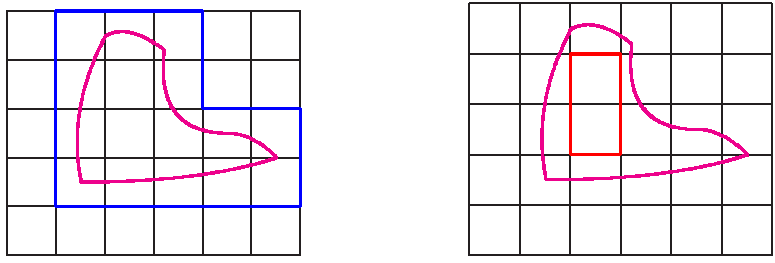

Let $\Sigma$ be any (possibly infinite) set of states.  Following
\cite[Section~5]{CC94}, a  partition
$P\in \Part(\Sigma)$ can be viewed as an abstraction of
$\wp(\Sigma)_\subseteq$ as follows:  
any $S\subseteq
\Sigma$ is over approximated by the unique minimal cover of $S$ in $P$, 
namely by the 
union of all the  blocks $B\in P$ such that $B\cap S\neq
\varnothing$. A graphical example is depicted on the 
left-hand side of Figure~\ref{figuco}. This abstraction is formalized
by a GI $(\alpha_P,\wp(\Sigma)_\subseteq,\wp(P)_\subseteq, \gamma_P)$
where: 
$$\alpha_P(S) \ud \{B\in P~|~ B \cap S\neq \varnothing\}~~~~~~~
\gamma_P(\mathcal{B})\ud \cup_{B\in \mathcal{B}} B.$$
Hence, any partition $P\in \Part(\Sigma)$ induces an abstract domain
$\adp(P)\in \Abs(\wp(\Sigma))$, and an 
abstract domain $A\in \Abs(\wp(\Sigma))$ is called \emph{partitioning} when $A$ is
equivalent to $\adp(P)$ for some partition $P$. 
Observe that the closure 
$\adp(P)=\gamma_P\circ
\alpha_P$
associated to a partitioning abstract domain is
defined as $\adp(P)
= \lambda S.\cup \{B\in P~|~ B \cap S\neq \varnothing\}$. 
Accordingly, a closure $\mu\in \uco(\wp(\Sigma))$ that coincides with 
$\gamma_P \circ \alpha_P$, for some
partition $P$, is called partitioning. 
  We denote by $\Absp(\wp(\Sigma))$ and $\ucop(\wp
(\Sigma ))$  the sets of, respectively, partitioning abstract
domains and closures on $\wp(\Sigma)$. As noted in
\cite{CC99}, a 
surjective abstraction $h:\Sigma \ra A$ used in standard abstract
model checking  that maps concrete states into
abstract states (cf.\ Section~\ref{amc}) gives rise to a partitioning Galois insertion
$(\alpha_h, \wp(\Sigma)_\subseteq,\wp(A)_\subseteq,\gamma_h)$  where $\alpha_h \ud \lambda
S\subseteq \Sigma. \{h(s)\in A~|~ s\in S\}$ and $\gamma_h \ud \lambda X \subseteq
A. \{s \in \Sigma~|~h(s)\in X\}$.

Partitions can be also viewed as dual abstractions
when a set $S$ is under approximated by the union of all the blocks
$B\in P$ such that $B\subseteq S$. A graphical example of this under approximation is
depicted on the right-hand side of
Figure~\ref{figuco}. This dual abstraction is formalized
by the GI
$(\widetilde{\alpha}_P,\wp(\Sigma)_\supseteq,\wp(P)_\supseteq,
\widetilde{\gamma}_P)$ 
where the ordering on the concrete domain $\wp(\Sigma)$ is given by
the subset relation and  
$$\widetilde{\alpha}_P(S) \ud \{B\in P~|~ B \subseteq S\}~~~~~~~
\widetilde{\gamma}_P(\mathcal{B})\ud \cup_{B\in \mathcal{B}} B.$$
In the following, we will be interested in viewing partitions as over
approximations, that is partitions as abstract domains of
$\wp(\Sigma)_\subseteq$.

Thus, partitions can be viewed as representations of abstract domains. On the other hand,
it turns out that abstract domains can be abstracted to partitions. 
An abstract domain $A\in \Abs(\wp(\Sigma)_\subseteq)$ 
induces a state equivalence $\equiv_A$ on $\Sigma$ by
identifying those states that cannot be distinguished by $A$:
$$s \equiv_A s' \text{~~~iff~~~} \alpha(\{s\}) = \alpha(\{s'\}).$$
For any $s\in \Sigma$, $[s]_A \ud \{s'\in \Sigma~|~
\alpha(\{s\}) = \alpha(\{s'\})\}$ is a block of the state partition $\pr(A)$
induced by $A$:
$$\pr(A) \ud \{[s]_A ~|~ s\in \Sigma\}.$$ 
Thus, $\pr:
\Abs(\wp(\Sigma)) \ra \Part(\Sigma)$ is a mapping from abstract domains to
partitions.

\begin{example2}\label{simple}
Let $\Sigma =\{1,2,3,4\}$ and let us specify  abstract domains as
uco's on $\wp(\Sigma)$. The uco's
$\mu_1 = \{\varnothing, 12, 3,4,1234\}$, $\mu_2 = \{\varnothing, 12,
3, 4, 34, 1234\}$, $\mu_3 = \{\varnothing, 12, 3, 4, 34, 123, 124,
1234\}$, 
$\mu_4 = \{ 12,
123, 124, 1234\}$ and $\mu_5 = \{\varnothing, 12,
123, 124, 1234\}$ all  induce the same
partition $P=\pr(\mu_i) = \{12,3,4\}\in \Part(\Sigma)$. For example, 
$\mu_5(\{1\})=\mu_5 (\{2\})=\{1,2\}$, $\mu_5(\{3\})=\{1,2,3\}$ and
$\mu_5 (\{4\})=\{1,2,3,4\}$ so that $\pr(\mu_5)=P$. 
Observe that $\mu_3$ is the
only partitioning abstract domain because $\adp(P)=\mu_3$.  \qed
\end{example2}

Abstract domains of $\wp(\Sigma)$ carry
additional information other than the underlying state
partition and this additional information allows us to distinguish
them. It turns out that this can be precisely stated by
abstract interpretation since the above mappings $\pr$ and $\adp$ allows us to show that
the whole lattice of partitions of
$\Sigma$ can be viewed as a (``higher-order'') abstraction of the lattice of abstract
domains of $\wp(\Sigma)$.

\begin{theorem}\label{teoGI}
$(\pr, \Abs(\wp(\Sigma))_\sqsupseteq,\Part(\Sigma)_\succeq,
\adp)$ is a Galois insertion.
\end{theorem}
\begin{proof}
Let $A \in \Abs(\wp(\Sigma))$ and $P\in
\Part(\Sigma)$ and let $\mu_A\in \uco(\wp(\Sigma))$ be the closure
associated with the abstract domain $A$. Let us prove that  
$P\preceq \pr(A) \; \Lra\; \adp(P) \sqsubseteq
\mu_A$. \\
$(\Rightarrow)$~For $S\in \wp(\Sigma)$ we have to prove that
$\adp(P)(S) \subseteq \mu_A(S)$. Consider $s\in \adp(P)(S)$. Hence,
there exists some $B\in P$ such that $s\in B$ and $B\cap S\neq
\varnothing$. Let $q\in B\cap S$. Since $P\preceq \pr(A)$, there
exists some block $[r]_A \in \pr(A)$ such that $B\subseteq
[r]_A$. Thus, for any $x,y\in B$,
$\alpha(\{x\})=\alpha(\{r\})=\alpha(\{y\})$, in particular,
$\alpha(\{s\})=\alpha(\{q\})$. Consequently, 
since $q\in S$ and therefore 
$\mu_A(\{q\})\subseteq \mu_A(S)$, we have that $\mu_A(\{s\}) =
\mu_A(\{q\}) \subseteq \mu_A(S)$, so that $s\in \mu_A(S)$. \\
$(\Leftarrow)$~ Consider a block $B\in P$ and some $s\in B$. We show
that $B\subseteq [s]_A$, namely if $s',s''\in B$ then $\alpha(\{s'\}) =
\alpha(\{s''\})$. Since $\adp(P) \sqsubseteq \mu_A$,
if $s',s''\in B$ then $\adp(P)(\{s'\})=B \subseteq \mu_A(\{s'\})$ so that
$s''\in \mu_A(\{s'\})$ and therefore $\mu_A(\{s''\})\subseteq
\mu_A(\{s'\})$. Likewise, $\mu_A(\{s'\})\subseteq
\mu_A(\{s''\})$ so that $\mu_A(\{s'\}) = \mu_A(\{s''\})$ and in turn
$\alpha(\{s'\}) =
\alpha(\{s''\})$.
\\
Finally, observe that $\adp$ is 1-1 so that the above adjunction
is indeed a Galois insertion. 
\end{proof}

Let us observe that, as recalled in Section~\ref{aic}, the
adjoint maps $\pr$ and $\adp$ give rise to an order isomorphism
between the lattices $\tuple{\Part(\Sigma),\preceq}$ and
$\tuple{\Absp(\wp(\Sigma)),\sqsubseteq}$.

\begin{corollary}\label{coropart} Let $A \in \Abs(\wp(\Sigma))$. The
following statements are equivalent:\\
{\rm (1)} $A$ is partitioning.\\
{\rm (2)} $\gamma$ is
additive and $\{\gamma (\alpha (\{s\}))\}_{s\in \Sigma }$ is a partition of $\Sigma$.
In this case, $\pr(A) =
\{\gamma(\alpha(\{s\}))\}_{s\in \Sigma }$. \\
{\rm (3)} $A$ is
forward complete for the complement operator
$\complement$.
\end{corollary}
\begin{proof} 
Let $A\in \Abs(\wp(\Sigma))$ and let $\mu_A =\gamma\circ \alpha \in
\uco(\wp(\Sigma))$ be the corresponding uco.\\
(1) $\Ra$ (2) By Theorem~\ref{teoGI},  $A\in \Absp(\wp (\Sigma ))$ iff
$\adp(\pr(A)) =A$. Thus, if $\adp(\pr(A))=A$ then $\mu_A=\gamma\circ
\alpha$ is obviously additive. Moreover, $s\equiv_A s'$ iff
$\alpha(\{s\}) = \alpha(\{s'\})$ iff $\gamma(\alpha(\{s\})) =
\gamma(\alpha(\{s'\}))$, so that, for any $s\in \Sigma$,
$[s]_A=\gamma(\alpha(\{s\}))$ and therefore $\pr(A) =
\{\gamma(\alpha(\{s\}))\}_{s\in \Sigma }$.\\
(2) $\Ra$ (1) Since
$\{\gamma (\alpha (\{s\}))\}_{s\in \Sigma } = P\in \Part(\Sigma)$ we
have that
for any $s\in \Sigma$,
$[s]_A=\gamma(\alpha(\{s\}))$: in fact, if $s'\in
\gamma(\alpha(\{s\}))$ then $\alpha(\{s'\}) \leq \alpha(\{s\})$, hence
$\gamma(\alpha(\{s'\})) \subseteq \gamma(\alpha(\{s\}))$ and therefore
$\gamma(\alpha(\{s'\})) = \gamma(\alpha(\{s\}))$. Thus, 
$\pr(A)=P$. Moreover, since $\gamma$ is additive, for any $S\subseteq
\Sigma$, $\cup_{s\in S}
\gamma(\alpha(\{s\}))= \gamma(\vee_{s\in S}
\alpha(\{s\}))=\gamma(\alpha(S))\in \mu_A$. Hence, since
$\adp(P)= \{ \cup_{s\in S}
\gamma(\alpha(\{s\}))~|~ S \subseteq \Sigma \}$ we have that 
$\adp(\pr (A))=A$. \\
(1) $\Ra$ (3) Assume that $A \in \Absp(\wp(\Sigma))$. It is
enough to prove that for any $s\in \Sigma $, $\complement(\mu_A(\{s\}))\in \mu_A$:
  in fact, by (1) $\Ra$ (2), $\gamma$ is additive and
therefore $\mu_A$ is additive (because it is a composition of additive
maps) and therefore
 if $S\in\mu_A$ then $S=\cup_{s\in S}\mu_A(\{s\})$ so that
  $\complement (S)=\cap_{s\in S}\complement(\mu_A(\{s\}))$.
Let us observe the following fact~$(*)$: for any $s,s'\in \Sigma $, $s\not\in
\mu_A(\{s'\}) \Leftrightarrow \mu_A(\{s\})\cap \mu_A(\{s'\})=\varnothing$;
this is a consequence of the fact that, by (1) $\Ra$ (2),
$\{\mu_A(\{s\})\}_{s\in \Sigma }$ is a partition. 
For any $s\in \Sigma $, we have that
$\complement(\mu_A(\{s\}))\in\mu_A$ because:
  \begin{align*}
    \mu_A(\complement(\mu_A(\{s\})))
    & = \mu_A(\{s'\in \Sigma ~|~s'\not\in\mu_A(\{s\})\}  & \text{[by
additivity of  $\mu_A$]}  \\
    &=\cup \{ \mu_A(\{s'\})~|~s'\not \in \mu_A(\{s\})\}  & 
\text{[by the above fact $(*)$]}\\
    & = \cup \{ \mu_A(\{s'\})~|~\mu_A(\{s'\})\cap \mu_A(\{s\})=\varnothing\}   &\\
    &= \cup \{ \mu_A(\{s'\})~|~\mu_A(\{s'\})\subseteq \complement(
\mu_A(\{s\})) \} & \\
   &\subseteq \complement(\mu_A(\{s\}))
  \end{align*}
(3) $\Ra$ (1) 
Assume that $\mu_A$ is  
  forward complete for $\complement$, i.e.\ $\mu_A$ is closed under
complements.  By (2)~$\Ra$~(1),
it is enough to
prove that $\gamma$ is additive and that
  $\{\mu_A(\{s\})\}_{s\in \Sigma }\in\Part(\Sigma)$. \\
(i) $\gamma$ is additive. Observe that $\gamma$ is additive iff
$\mu_A$ is additive iff $\mu_A$ is closed under arbitrary unions. 
If $\{S_i\}_{i\in I}\subseteq \mu_A$ then
    $\cup_i S_i = \complement(\cap_i \complement(S_i))\in\mu_A$,
    because, $\mu_A$ is closed under complements (and arbitrary intersections). \\
(ii) $\{\mu_A(\{s\})\}_{s\in \Sigma }\in\Part(\Sigma)$.
Clearly, we have that $\cup_{s\in \Sigma }
\mu_A(\{s\})=\Sigma $. Consider now $s,r\in \Sigma $ such that 
$\mu_A(\{s\})\cap\mu_A(\{r\})\neq \varnothing$. Let us show that
$\mu_A(\{s\})=\mu_A(\{r\})$. In order to show this, let us prove that
$s\in\mu_A(\{r\})$. Notice that
$\mu_A(\{s\})\smallsetminus\mu_A(\{r\})=\mu_A(\{s\})\cap\complement(\mu_A(\{r\}))\in\mu_A$,
because $\mu_A$ is closed under complements.  If
$s\not\in\mu_A(\{r\})$ then we would have that $s\in
\mu_A(\{s\})\smallsetminus \mu_A(\{r\})\in \mu_A$, and this would imply 
$\mu_A(\{s\})\subseteq \mu_A(\{s\})\smallsetminus \mu_A(\{r\}) \subseteq
\mu_A(\{s\})$, namely $\mu_A(\{s\}) =  \mu_A(\{s\})\smallsetminus
\mu_A(\{r\})$. Thus, we would obtain the contradiction
$\mu_A(\{s\})\cap
\mu_A(\{r\}) = \varnothing$. 
Hence, we have that $s\in \mu_A(\{r\})$ and therefore
$\mu_A(\{s\})\subseteq \mu_A(\{r\})$. By swapping the roles of $s$ and
$r$, we also
obtain that $\mu_A(\{r\})\subseteq \mu_A(\{s\})$, so that
$\mu_A(\{s\})=\mu_A(\{r\})$. 
\end{proof}

Let us remark that $\bP \ud \adp \circ \pr$ is a lower closure operator on
$\tuple{\Abs(\wp(\Sigma)),\sqsubseteq}$ and that for any $A \in
\Abs(\wp(\Sigma))$, $A$ is partitioning iff $\bP(A)=A$. 
Hence, $\bP$ is exactly the partitioning-shell refinement, namely
$\bP(A)$ is the most abstract refinement of $A$ that is partitioning.

\section{Abstract Semantics of Languages}
\subsection{Concrete Semantics}\label{cs}
We consider temporal specification languages $\fL$ whose
state formulae $\varphi$ are inductively defined by: 
$$\fL\ni \varphi ::= p ~|~ f(\varphi_1, ...,\varphi_n) $$
where $p$ ranges over a (typically finite) set of atomic propositions $\AP$, while $f$
ranges over a finite set $\mathit{Op}$ of operators. $\AP$ and $\Op$
are also denoted, respectively, by $\AP_\fL$ and $\Op_\fL$. Each
operator $f\in \mathit{Op}$ has an arity\footnote{It would be
possible to consider generic operators whose arity is any
possibly infinite ordinal, thus allowing, for example, infinite
conjunctions or disjunctions.} 
$\sharp(f) >0$. 

Formulae in $\fL$ are interpreted on a \emph{semantic structure} $\cS
= (\Sigma, I)$ where $\Sigma$ is 
any (possibly
infinite) set of states and $I$ is an interpretation function $I: \AP
\cup \mathit{Op} \ra \Fun(\wp(\Sigma))$ that
maps $p\in \AP$ to $I(p)\in \wp(\Sigma)$ and $f\in \mathit{Op}$ to 
$I(f): \wp(\Sigma)^{\sharp(f)} \ra \wp(\Sigma)$. 
$I(p)$ and $I(f)$ are also denoted by, respectively,
$\boldsymbol{p}$ and $\boldsymbol{f}$. Moreover, $\boldsymbol{AP}\ud
\{\boldsymbol{p} \in \wp(\Sigma)~|~ p \in \AP\}$ and $\boldsymbol{Op}
\ud \{ \boldsymbol{f}:\wp(\Sigma)^{\sharp(f)}\ra \wp(\Sigma)~|~ f\in \Op\}$.
Note that the interpretation $I$ induces
a state labeling $\ell_I: \Sigma \ra \wp(\mathit{AP})$ by
$\ell_I(s)\ud \{ p\in \AP~|~ s \in I(p)\}$.  
The 
\emph{concrete state semantic function} $\grasse{\cdot}_\cS: \fL\ra
\wp(\Sigma)$ evaluates a formula $\varphi\in \fL$ to the set of states making
$\varphi$ true w.r.t.\ the semantic structure $\cS$:
$$\grasse{p}_\cS = \boldsymbol{p} \mbox{{\rm ~~~and~~~ }}
\grasse{f(\varphi_1,...,\varphi_n)}_\cS =
\boldsymbol{f}(\grasse{\varphi_1}_\cS,...,\grasse{\varphi_n}_\cS).$$
Semantic structures generalize the role of Kripke
structures. In fact, in standard model checking a semantic structure
is usually defined through a Kripke structure $\cK$ so that the
interpretation of logical/temporal operators is defined in terms of
standard logical operators and paths in $\cK$. 
In the following, we freely use standard logical and temporal
operators together with their corresponding usual interpretations: for
example, $I(\wedge)= \cap$, $I(\vee)= \cup$, $I(\neg)=
\complement$, $I(\mathrm{EX})= \pre_R$, $I(\mathrm{AX})=\pret_R$,
etc. As an example, consider the
standard semantics of $\CTL$: 
$$
\CTL \ni \varphi ::=  p ~|~ \varphi_1 \wedge \varphi_2 ~|~ \neg \varphi ~|~ \mathrm{AX}\varphi ~|~ \mathrm{EX}\varphi
~|~ \mathrm{AU}(\varphi_1,\varphi_2) ~|~
\mathrm{EU}(\varphi_1,\varphi_2) ~|~ \mathrm{AR}(\varphi_1,\varphi_2) ~|~
\mathrm{ER}(\varphi_1,\varphi_2) 
$$
with respect to a Kripke structure $\cK= (\Sigma
,R,\ell)$. Hence, $\cK$ determines a corresponding
interpretation $I$ for atoms in $\AP$ and
operators of $\Op_{\CTL}$, namely $I(\mathrm{AX})=\pret_R$,
$I(\mathrm{EX})=\pre_R$, etc., and this defines the concrete semantic
function  $\grasse{\cdot}_\cK: \CTL \ra
\wp(\Sigma)$. 

If $g$ is any operator with arity $\sharp(g)=n>0$ whose
interpretation is given by $\boldsymbol{g}:\wp(\Sigma)^{n} \ra \wp
(\Sigma )$ and $\cS=(\Sigma,I)$ is a semantic structure then we say
that a language $\fL$ is \emph{closed under} $g$ for $\cS$ when for any
$\varphi_1, ...,\varphi_{n} \in \fL$ there exists some $\psi\in\fL$
such that $\boldsymbol{g}(\grasse{\varphi_1}_\cS,...,\grasse{\varphi_{n}}_\cS)
= \grasse{\psi}_\cS$. For instance, if $\Op_\fL$ includes 
$\mathrm{EX}$ and negation with their standard
interpretations, i.e.\
$I(\mathrm{EX})=\pre_R$ and
$I(\neg)=
\complement$, then
$\fL$ is closed under $\mathrm{AX}$ with its standard interpretation $\pret_R$ 
because $\pret_R = \complement \circ \pre_R \circ \complement$. 
This notion can be extended in a straightforward
way to infinitary operators: for instance, $\fL$ is closed under
infinite logical conjunction for $\cS$ iff for any $\Phi\subseteq \fL$, there
exists some $\psi\in\fL$ such that $\bigcap_{\varphi\in
\Phi}\grasse{\varphi}_\cS = \grasse{\psi}_\cS$. In particular, let us
remark that if $\fL$ is
closed under infinite logical conjunction then it must exist some
$\psi\in \fL$ such that $\cap \varnothing
= \Sigma  = \grasse{\psi}_\cS$, namely $\fL$ is able to express the tautology
$\mathit{true}$. Let us remark that 
if the state space $\Sigma$ is finite and 
$\fL$ is closed under logical conjunction then we always
mean that there exists some $\psi\in \fL$ such that $\cap \varnothing
= \Sigma  = \grasse{\psi}_\cS$. Finally, note that $\fL$ is
closed  under negation and infinite logical conjunction if and only if
$\fL$ includes propositional logic.  

\subsection{Abstract Semantics}\label{as}
In the following, we apply the standard abstract interpretation approach for
defining abstract semantics \cite{CC77,CC79}. Let $\fL$ be a language
and $\cS=(\Sigma,I)$ be a semantic structure for $\fL$.  An
\emph{abstract semantic structure} $\cS^\sharp = (A,I^\sharp)$ is
given by an abstract domain $A\in 
\Abs(\wp(\Sigma)_\subseteq)$ and by an abstract interpretation function
$I^\sharp: \AP \cup \Op \ra \Fun(A)$. An abstract
semantic structure $\cS^\sharp$ therefore induces an \emph{abstract
semantic function} $\grasse{\cdot}_{\cS^\sharp}: \fL\ra A$
that evaluates formulae in $\fL$ to
abstract values in $A$.
The abstract interpretation $I^\sharp$  is a
correct over-approximation (respectively, under-approximation) of $I$ on $A$ when 
for any $p\in \AP$, $\gamma(I^\sharp(p)) \supseteq 
I (p)$ (respectively, $\gamma(I^\sharp(p)) \subseteq 
I (p)$) 
and for any $f\in \Op$,  $\gamma \circ I^\sharp (f) \sqsupseteq 
I(f) \circ \gamma$ (respectively, $\gamma \circ I^\sharp (f) \sqsubseteq 
I(f) \circ \gamma$).
If $I^\sharp$  is a
correct over-approximation (respectively, under-approximation) 
of $I$ and the semantic operations in
$\boldsymbol{Op}$ are monotone then the
abstract
semantics is an over-approximation (respectively, under-approximation)
of the concrete semantics, namely
for any $\varphi\in \fL$, 
 $\gamma(\grasse{\varphi}_{\cS^\sharp}) \supseteq
\grasse{\varphi}_\cS$
(respectively, $\gamma(\grasse{\varphi}_{\cS^\sharp}) \subseteq
\grasse{\varphi}_\cS$).  

In particular, the abstract domain $A$ always induces an abstract
semantic structure $\cS^A=(A,I^A)$ where $I^A$ is the best correct
approximation of $I$ on $A$, i.e.\ $I^A$ interprets atoms $p$ and
operators $f$ as best correct approximations on $A$ of,
respectively, 
$\boldsymbol{p}$ and $\boldsymbol{f}$: for any $p\in \AP$ and $f\in \Op$, 
$$I^A (p)\ud \alpha(\boldsymbol{p}) ~~~\text{ and }~~~ I^A (f) \ud
\boldsymbol{f}^A.$$
Thus, the abstract domain $A$ systematically induces an abstract
semantic function $\grasse{\cdot}_{\cS^A}: \fL\ra A$,   
also denoted by $\grasse{\cdot}_\cS^A$, which is therefore defined by:
$$\grasse{p}_\cS^A = \alpha(\boldsymbol{p})
~~~\text{ and }~~~
\grasse{f(\varphi_1,...,\varphi_n)}_\cS^A =
\boldsymbol{f}^A(\grasse{\varphi_1}^A_\cS,...,\grasse{\varphi_n}^A_\cS).$$

\noindent
As usual in abstract interpretation, observe that the concrete semantics
is a particular abstract semantics, namely it is the abstract
semantics induced by the ``identical abstraction'' $(\id,\wp(\Sigma),\wp(\Sigma),\id)$.


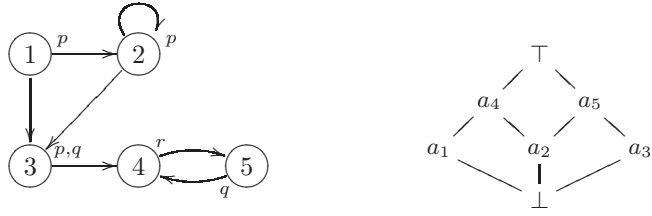
\begin{figure}[t]
\begin{center}
\begin{tabular}{ccc}
 \mbox{\xymatrix{
      *++[o][F]{1} \ar[d] \ar[r]^(0.3){p} ^(1.3){p} &
      *++[o][F]{2} \ar@(ul,ur)[] \ar[ld] &  \\ 
      *++[o][F]{3}\ar[r] ^(0.35){p,q} & *++[o][F]{4}\ar@/^/[r] ^(0.24){r}
& *++[o][F]{5} \ar@/^/[l] ^(0.24){q} 
    }
  }
&
\mbox{~~~~~~~~~~~~~}
&
\mbox{\small
    \xymatrix@C=4pt@R=7pt{
      &   & \top & &  \\
        &   a_4\ar@{-}[ru]   &  & a_5\ar@{-}[lu] &\\
     a_1 \ar@{-}[ru] & &   a_2\ar@{-}[ru]\ar@{-}[lu]      & & a_3\ar@{-}[lu] \\
         && \bot\ar@{-}[llu]\ar@{-}[u]\ar@{-}[rru] &&  
    }
  }
\end{tabular}
\end{center}
\caption{A Kripke structre on the left and an abstract domain on the right.}\label{figuco2}
\end{figure}

\begin{example2}
  \label{ex1}
Let $\fL \ni
\varphi::=p~|~q~|~r~|~\varphi_1\wedge\varphi_2~|~\EX\varphi$.
Let us consider the Kripke structure
$\cK=(\Sigma,\shortrightarrow,\ell)$ 
and the lattice $A$ both
depicted in Figure~\ref{figuco2}.
Let $\cS$ be the
semantic structure induced by the Kripke structure $\cK$ so
that $\boldsymbol{\EX}=\pre_{\shortrightarrow}$. Let us consider the
formulae $\EX r$ and $\EX (p \wedge q)$, whose concrete semantics are
as follows: $\grasse{\EX r}_\cS = \{3,5\}$ and $\grasse{\EX (p \wedge
q)}_\cS= \{1,2\}$. 
$A$ is an abstract domain of $\wp(\Sigma)$ where the 
Galois insertion $\gdcas$ is determined by the following
concretization map: 
\begin{align*}
&\gamma(\bot)=\varnothing;~~\gamma(a_1)=\{1,2\};~~\gamma(a_2)=\{3\};
~~\gamma(a_3)=\{3,4\};\\ 
&\gamma(a_4)=\{1,2,3\};~~ \gamma(a_5
)=\{3,4,5\};~~ \gamma(\top)=\{1,2,3,4,5\}.
\end{align*}
Note that, by Corollary~\ref{coropart}, 
$A$ is not partitioning because $\gamma$ is not additive: $\gamma(a_2)
\cup \gamma(a_3) =\{3,4\} \subsetneq \{3,4,5\} = \gamma(a_2
\vee a_3)$. It turns out that:
$$ 
\begin{array}{ll}
\grasse{\EX r}_\cS^A
    &= \alpha (\pre_{\shortrightarrow} ( \gamma(\grasse{r}_\cS^A)) 
    = \alpha (\pre_{\shortrightarrow} ( \gamma (\alpha(\boldsymbol{r})))) 
    = \alpha (\pre_{\shortrightarrow} ( \gamma( a_3) )) \\ 
    & = \alpha (\pre_{\shortrightarrow} (\{3,4\}))  
    = \alpha (\{1, 2, 3, 5\} )  = \top; \\[7.5pt]
\grasse{\EX (p \wedge q)}_\cS^A
    & = \alpha (\pre_{\shortrightarrow} ( \gamma( \grasse{p}_\cS^A \wedge
\grasse{q}_\cS^A))) 
    = \alpha (\pre_{\shortrightarrow} ( \gamma (\alpha(\boldsymbol{p}) \wedge
\alpha(\boldsymbol{q}))))\\ 
    & = \alpha (\pre_{\shortrightarrow} ( \gamma( a_4 \wedge a_5 ))) 
= \alpha (\pre_{\shortrightarrow} ( \gamma( a_2 ))) 
= \alpha (\pre_{\shortrightarrow} (3)) 
= \alpha (\{1,2\}) = a_1 . 
\end{array}
$$
Observe that 
the abstract semantics $\grasse{\EX r}_\cS^A$ is 
  a proper over-approximation of
$\grasse{\EX r}_\cS$ because $\grasse{\EX r}_\cS \subsetneq 
\gamma(\grasse{\EX r}_\cS^A)$. On the other hand,
the concrete semantics  
$\grasse{\EX (p \wedge q)}_\cS$ is precisely represented in $A$
because $\gamma( \grasse{\EX (p
\wedge q)}_\cS^A ) = \grasse{\EX (p \wedge q)}_\cS$.
\qed
\end{example2}

\section{Generalized Strong Preservation} 

We showed in Section~\ref{paa} how a state partition
$P$ can be viewed as a partitioning abstract domain  $\adp(P)$
specified by the GI $(\alpha_P,
\wp(\Sigma)_\subseteq,\wp(P)_\subseteq,\gamma_P)$.  
Thus, given a language $\fL$ and a corresponding
semantic structure $\cS=(\Sigma,I)$, it turns out that 
any partition  $P\in \Part(\Sigma)$ systematically induces
a corresponding abstract semantics $\ok{\grasse{\cdot}_\cS^P}\ud 
\ok{\grasse{\cdot}_\cS^{\adp(P)}}
:\fL \ra \adp(P)$ that evaluates a formula in $\fL$ to a (possibly
empty) union of blocks of $P$.  
Strong preservation for a partition $P$ can be characterized in terms of the corresponding
abstract domain $\adp(P)$ as follows. 

\begin{lemma}\label{char}
$P\in \Part(\Sigma)$ is s.p.\ for $\fL$ iff $\forall \varphi \in
\fL$ and $S\subseteq \Sigma $, $\alpha_P(S)
\subseteq \grasse{\varphi}_\cS^P \: \Lra \: S\subseteq\grasse{\varphi}_\cS$.  
\end{lemma}
\begin{proof}
$(\Rightarrow)$: 
  Let us first observe that for any $\varphi\in \fL$, 
$\gamma_P(\alpha_P(\grasse{\varphi}_\cS)) = \grasse{\varphi}_\cS$: in fact, 
for any $s\in \grasse{\varphi}_\cS$, $\alpha_P(\{s\})$ is the block of 
$P$ containing $s$; since $P\preceq P_\fL$, we have that $\alpha_P(\{s\})\subseteq
\grasse{\varphi}_\cS$, and from this $\alpha_P(\grasse{\varphi}_\cS) \subseteq
\grasse{\varphi}_\cS$ and in turn $\gamma_P(\alpha_P(\grasse{\varphi}_\cS)) =
\grasse{\varphi}_\cS$.\\ 
Let us now prove by 
structural induction on $\varphi\in \fL$ that
$\grasse{\varphi}_\cS=\gamma_P(\grasse{\varphi}^P_\cS)$:

\begin{itemize}
\item[--] $\varphi\equiv p\in \AP_\fL$: by using the above observation, 
  $\grasse{p}_\cS=\gamma_P(\alpha_P(\grasse{p}_\cS))=\gamma_P(\grasse{p}_\cS^P)$. 

\item[--] $\varphi\equiv f(\varphi_1,\ldots,\varphi_n)$: 
  \begin{align*}
    \grasse{f(\varphi_1,\ldots,\varphi_n)}_\cS = & \text{~~~[by
the above observation]}\\
    \gamma_P(\alpha_P(\grasse{f(\varphi_1,\ldots,\varphi_n)}_\cS)) = &
\text{~~~[by definition]}  \\ 
    \gamma_P(\alpha_P(\boldsymbol{f}
(\grasse{\varphi_1}_\cS,\ldots,\grasse{\varphi_n}_\cS))) = &
\text{~~~[by inductive hypothesis]} \\
    \gamma_P(\alpha_P(\boldsymbol{f}
(\gamma_P(\grasse{\varphi_1}_\cS^P),\ldots,\gamma_P(\grasse{\varphi_n}_\cS^P))))
= &  \text{~~~[by definition]} \\
\gamma_P( \grasse{f(\varphi_1,\ldots,\varphi_n)}_\cS^P). \phantom{=}&
  \end{align*}
\end{itemize}
Now, consider any $\varphi\in
\fL$. If $S\subseteq \grasse{\varphi}_\cS$ then 
  $\alpha_P(S) \subseteq \alpha_P(\grasse{\varphi}_\cS) =
\alpha_P(\gamma_P (\grasse{\varphi}_\cS^P)) = \grasse{\varphi}_\cS^P$.
  Conversely, if
$\alpha_P(S)\subseteq\grasse{\varphi}_\cS^P$ 
then 
  $S\subseteq \gamma_P(\grasse{\varphi}_\cS^P)= \grasse{\varphi}_\cS$. \\
$(\Leftarrow)$:
   Consider a block $B\in P$ and $s,s'\in B$ so that
$\alpha_P(\{s\})=B=\alpha_P(\{s'\})$. By hypothesis, for any $\varphi\in
\fL$, we have that  
  $s\in\grasse{\varphi}_\cS$ iff $\alpha_P(\{s\})
  \subseteq\grasse{\varphi}_\cS^P$ iff $\alpha_P(\{s'\})
  \subseteq\grasse{\varphi}_\cS^P$ iff 
  $s'\in\grasse{\varphi}_\cS$. Thus,
  $s\equiv_\fL s'$. 
\end{proof}

This
states that a partition $P\in \Part(\Sigma)$ is s.p.\ for $\fL$ 
if and only if to check whether some set $S$ of states satisfies some
formula $\varphi\in \fL$, i.e.\ $S\subseteq\grasse{\varphi}_\cS$, is equivalent to
check whether the abstract state $\alpha_P(S)$ is more precise than the abstract semantics
$\grasse{\varphi}_\cS^P$, that is $S$ is over-approximated by $\grasse{\varphi}_\cS^P$.
The key observation here is that in our
abstract interpretation-based framework partitions are particular
abstract domains. This allows us to generalize
the notion of strong preservation from partitions to generic abstract
semantic functions as follows.

\begin{definition2}\label{spd}
  Let $\fL$ be a language, $\cS=(\Sigma,I)$ be a semantic structure for
$\fL$ and $\cS^\sharp =(A,I^\sharp)$ be a corresponding abstract
semantic structure. The abstract semantics
  $\ok{\grasse{\cdot}_{\cS^\sharp}}$ is \emph{strongly preserving} for $\fL$ (w.r.t.\
$\cS$) if for any $\varphi\in \fL$ and $S\subseteq \Sigma $,
  $$\alpha(S)\leq_A \ok{\grasse{\varphi}_{\cS^\sharp}} \;\; \Lra \;\; S\subseteq
  \grasse{\varphi}_\cS. \qed$$
\end{definition2}

Definition~\ref{spd} generalizes standard strong preservation from partitions, as
characte\-rized by Lem\-ma~\ref{char}, both to an arbitrary abstract
domain $A\in \Abs(\wp(\Sigma))$ and to a corresponding abstract
interpretation function $I^\sharp$. 
Likewise, standard weak preservation can be generalized as follows. Let
$\cK=(\Sigma,R,\ell)$ be a concrete Kripke structure that induces the
concrete semantics
$\ok{\grasse{\varphi}_\cK} = \{ s\in
\Sigma~|~ s \ok{\models^\cK } \varphi\}$.
Let $h:\Sigma
\ra A$ be a surjective abstraction and let
$(\alpha_h,\wp(\Sigma),\wp(A),\gamma_h)$ be the corresponding
partitioning abstract domain. Let $\cA = (A,\ok{R^\sharp},
\ok{\ell^\sharp})$ be an abstract Kripke structure on
$A$ that gives rise to  the abstract semantics $\ok{\grasse{\varphi}_\cA} = \{ a\in
A~|~ a \ok{\models^\cA } \varphi\}$. Then, $\cA$ weakly preserves  $\fL$
when 
$$\forall \phi\in \fL.\forall S\subseteq \Sigma.\; \alpha_h(S)
\subseteq \ok{\grasse{\varphi}_\cA} \;\; \Ra \;\; S\subseteq
\ok{\grasse{\varphi}_\cK}.$$ 
Hence, weak preservation can be generalized to generic abstract domains and
abstract semantics accordingly to Definition~\ref{spd}. 

\subsection{Strong Preservation is an Abstract Domain Property}
Definition~\ref{spd} is a direct and natural generalization of the standard
notion of strong preservation in abstract model checking. It can be
equivalently stated as follows.   
\begin{lemma}\label{charsp}
$\grasse{\cdot}_{\cS^\sharp}$ is s.p.\ for $\fL$ iff for any $\varphi
\in \fL$, $\grasse{\varphi}_\cS =
\gamma(\grasse{\varphi}_{\cS^\sharp})$. 
\end{lemma}
\begin{proof}
$(\Rightarrow)$ On the one hand,  $\gamma(\grasse{\varphi}_{\cS^\sharp})
\subseteq \grasse{\varphi}_\cS$ iff 
$\alpha(\gamma(\grasse{\varphi}_{\cS^\sharp}))
\leq \grasse{\varphi}_{\cS^\sharp}$ iff 
$\grasse{\varphi}_{\cS^\sharp} \leq \grasse{\varphi}_{\cS^\sharp}$,
which is trivially true. On the other hand,  
$\grasse{\varphi}_\cS \subseteq \gamma(\grasse{\varphi}_{\cS^\sharp})$
iff $\alpha(\grasse{\varphi}_\cS) \leq \grasse{\varphi}_{\cS^\sharp}$
iff $\grasse{\varphi}_\cS \subseteq \grasse{\varphi}_\cS$, that
is trivially true. \\
$(\Leftarrow)$ We have that $S\subseteq \grasse{\varphi}_\cS$ iff $S\subseteq
\gamma(\grasse{\varphi}_{\cS^\sharp})$ iff $\alpha(S) \leq
\grasse{\varphi}_{\cS^\sharp}$. 
\end{proof}

\noindent
In particular, it is worth noting that if   $\ok{\grasse{\cdot}_{\cS^\sharp}}$ is s.p.\ for
$\fL$ then $\ok{\grasse{\cdot}_{\cS^\sharp}} = \alpha \circ
\grasse{\cdot}_\cS$ holds. 

\begin{lemma}\label{spadp} Let $A\in \Abs(\wp(\Sigma))$.\\
{\rm (1)} Let $\cS^\sharp_1 =(A,I^\sharp_1)$ and $\cS^\sharp_2
=(A,I^\sharp_2)$ be abstract semantic structures on $A$.  If
$\ok{\grasse{\cdot}_{\cS_1^\sharp}}$ and
$\ok{\grasse{\cdot}_{\cS_2^\sharp}}$ are both s.p.\ for $\fL$ then
$\ok{\grasse{\cdot}_{\cS_1^\sharp}}=
\ok{\grasse{\cdot}_{\cS_2^\sharp}}$. \\
{\rm (2)} Let $\cS^\sharp =(A,I^\sharp)$ 
be an abstract semantic structure on $A$. 
If $\ok{\grasse{\cdot}_{\cS^\sharp}}$
is s.p.\ for $\fL$ then
$\ok{\grasse{\cdot}_\cS^A}$ 
is s.p.\ for $\fL$. 
\end{lemma}
\begin{proof}
(1) By Lemma~\ref{charsp}, for any $\varphi\in \fL$, 
$ \gamma(\ok{\grasse{\varphi}_{\cS_1^\sharp}}) = \grasse{\varphi}_\cS 
= \gamma(\ok{\grasse{\varphi}_{\cS_2^\sharp}})$, so that, by applying
$\alpha$, 
$\ok{\grasse{\varphi}_{\cS_1^\sharp}} = \alpha(
\gamma(\ok{\grasse{\varphi}_{\cS_1^\sharp}}))  = \alpha(\grasse{\varphi}_\cS) 
= \alpha(\gamma(\ok{\grasse{\varphi}_{\cS_2^\sharp}})) 
=\ok{\grasse{\varphi}_{\cS_2^\sharp}}$.
\\
(2) Let us first observe that 
for any $\varphi\in\fL$, $\gamma(\alpha(\grasse{\varphi}_\cS)) =
\grasse{\varphi}_\cS$. In fact, 
$\gamma(\alpha(\grasse{\varphi}_\cS)) \subseteq
\grasse{\varphi}_\cS \Lra \alpha(\gamma(\alpha(\grasse{\varphi}_\cS)))
\leq \ok{\grasse{\varphi}_{\cS^\sharp}} \Lra 
\alpha(\grasse{\varphi}_\cS)
\leq \ok{\grasse{\varphi}_{\cS^\sharp}} \Lra \grasse{\varphi}_\cS \subseteq
\grasse{\varphi}_\cS$.  
As a consequence of this
fact, by structural
induction  on $\varphi\in \fL$, analogously to the proof
of Lemma~\ref{char}, it is easy to prove that 
$\gamma(\grasse{\varphi}_\cS^A)=\grasse{\varphi}_\cS$. Thus, by
Lemma~\ref{charsp}, $\grasse{\cdot}_\cS^A$ is s.p.\ for $\fL$.
\end{proof}

Thus, it turns out that strong preservation is an
\emph{abstract domain property}. This means that given any abstract domain $A\in
\Abs(\wp(\Sigma))$, it is possible to define an abstract semantic
structure $\cS^\sharp=(A,I^\sharp)$ on $A$ such that the corresponding
abstract
semantics $\ok{\grasse{\cdot}_{\cS^\sharp}}$ is s.p.\ for $\fL$ if and
only if the induced abstract semantics $\ok{\grasse{\cdot}^A_\cS}:\fL
\ra A$ is s.p.\ for $\fL$.  In
particular, this also holds for the standard approach: if $\cA =
(A,\ok{R^\sharp}, 
\ok{\ell^\sharp})$ is an abstract Kripke structure for $\fL$, 
where $h:\Sigma \ra A$ is the corresponding surjection, then the
standard abstract semantics $\grasse{\cdot}_\cA$ 
strongly preserves $\fL$ if and only if the abstract semantics induced
by the partitioning abstract domain $(\alpha_h, \wp(\Sigma),
\wp(A),\gamma_h)$ strongly preserves $\fL$, and in this case this
abstract semantics coincides with 
the standard abstract semantics $\ok{\grasse{\cdot}_\cA}$.  
Strong preservation is an abstract domain property and therefore can be 
defined without loss of generality as follows. 
\begin{definition2}
An abstract domain  $A\in
\Abs(\wp(\Sigma))$ is \emph{strongly preserving for} $\fL$ (w.r.t.\ a
semantic structure $\cS$) when
$\grasse{\cdot}_\cS^A$ is s.p.\ for $\fL$ (w.r.t.\ $\cS$). 
We denote by $\SP_\fL\subseteq \Abs(\wp(\Sigma))$ the set of abstract
domains that are s.p.\ for $\fL$. 
\qed
\end{definition2}

\begin{example2}
Let us consider Example~\ref{ex1}. It turns out that the abstract
domain $A$ is not s.p.\ for $\fL$ because, by Lemma~\ref{charsp}, $\grasse{\EX
r}_\cS = \{3,5\} \subsetneq \{1,2,3,4,5\} =\gamma(\top)=
\gamma(\grasse{\EX r}_\cS^A)$. \qed 
\end{example2}

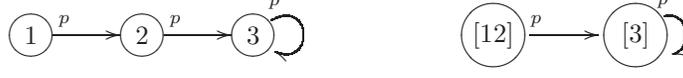
\begin{figure}[t] 
\begin{center}
\begin{tabular}{ccc}
    \mbox{
      \xymatrix@C=25pt@R=15pt{
       *++[o][F]{1}\ar[r]^(0.3){p} & *++[o][F]{2}\ar[r]^(0.3){p} &
*++[o][F]{3} \ar@(ur,dr)[]^(0.15){p} 
       }
     }
&
\mbox{~~~~~~~~~~~~~~~}
&
    \mbox{
      \xymatrix@C=10pt@R=10pt{
       *++[o][F]{[12]}\ar[rr]^(0.3){p} & & *++[o][F]{\;[3]\;} \ar@(ur,dr)[]^(0.18){p}  
       }
     }
\end{tabular}
\end{center}
\caption{A Kripke structure $\cK$ on the left and an abstract Kripke
structure $\cA$ on the right.}\label{figk} 
\end{figure}

\begin{example2}\label{esnu}
Let us consider the simple language
  $\cL\ni \varphi ::= p ~|~ \EX \varphi$ and the Kripke 
structure $\cK$ depicted in Figure~\ref{figk}.
 The Kripke structure $\cK$
induces the semantic structure $\cS=(\{1,2,3\},I)$ such that
$I(p)=\{1,2,3\}$ and $I(\EX)=\pre_\sra$. Hence, we
have that $\grasse{p}_\cS=\{1,2,3\}$, $\grasse{\EX p}_\cS
=\{1,2,3\}$ and, for $k>1$, $\grasse{\EX^k p}_\cS
=\{1,2,3\}$. Let us  consider the
partitioning abstract domain 
$A$ induced by the partition $P=\{[12],[3]\}$ and related to
$\wp(\Sigma)$ by $\alpha$ and $\gamma$.
Let us  consider two different abstract semantic structures on $A$.
\begin{itemize}
\item[--] The abstract semantic structure $\cS^A=(A,I^A)$ is induced
as best correct approximation of $I$ by $A$.
\item[--] The abstract semantic structure $\cS^\cA=(A,I^\cA)$ is instead
induced by the abstract Kripke structure
$\cA=(A,\sra^\sharp,\ell^\sharp)$ in Figure~\ref{figk}. Hence, 
$I^\cA(p) = \{[12],[3]\}$ and $I^\cA(\EX)=\pre_{\sra^\sharp}$. 
\end{itemize}
$\cS^A$ is different from $\cS^\cA$ because  $I^A(\EX)\neq I^\cA(\EX)$. In fact, 
$I^A(\EX)(\{[12]\}) = 
\alpha (\pre_\sra (\gamma(\{[12]\}))) =
\alpha (\pre_\sra (\{1,2\})) 
= \ok{\alpha (\{1\})} = \ok{\{[12]\}}$, while
$I^\cA(\EX)(\{[12]\}) = 
\pre_{\sra^\sharp}(\{[12]\}) = \varnothing$. \\
Let us show  that both the abstract semantics 
$\grasse{\cdot}_{\cS}^A$ and $\grasse{\cdot}_{\cS^\sharp}$ are s.p.\
for $\fL$. 
\begin{itemize}
\item[--]
We have that  
$\grasse{p}_\cS^A=\{[12],[3]\}$, $\grasse{\EX p}_\cS^A =\alpha
(\pre_\sra (\{1,2,3\})) = \alpha(\{1,2,3\})=
\{[12],[3]\}$ and, for
$k>1$, $\grasse{\EX^k p}_\cS^A =\{[12],[3]\}$. Thus,
for any $\varphi \in \cL$, $\grasse{\varphi}_\cS = \gamma
(\grasse{\varphi}_\cS^A)$. 
\item[--] We have that  
$\grasse{p}_{\cS^\cA}=\{[12],[3]\}$, $\grasse{\EX p}_{\cS^\cA}
=\pre_{\sra^\sharp} (\{[12],[3]\})=\{[12],[3]\}$ and, for
$k>1$, $\grasse{\EX^k p}_{\cS^\cA} =\{[12],[3]\}$. Thus,
for any $\varphi \in \cL$, $\grasse{\varphi}_\cS = \gamma
(\grasse{\varphi}_{\cS^\cA})$. 
\end{itemize}
Consequently, by Lemma~\ref{charsp}, 
both abstract semantics are s.p.\ for $\fL$. 
\qed 
\end{example2}

\subsection{The Most Abstract Strongly Preserving Domain}
As recalled in Section~\ref{amc}, a language $\fL$ and a
semantic structure $\cS$ for $\fL$ induce a corresponding logical partition
$P_\fL\in \Part(\Sigma)$.  By Lemma~\ref{char}, it turns out that
$P_\fL$ is the coarsest strongly preserving partitioning abstract
domain for $\fL$. This can be generalized to arbitrary abstract domains
as follows. Let us define  $\AD_\fL$ by:
$$\AD_\fL \ud  \cM(\{\grasse{\varphi}_\cS~|~\varphi\in\fL\}).$$
Hence, $\AD_\fL$ is the closure under arbitrary intersections of the set
of concrete semantics of formulae in $\fL$.
Observe that $\AD_\fL \in \Abs(\wp(\Sigma))$ because it 
is a Moore-family of $\wp(\Sigma)$. 

\begin{theorem}\label{spchar}
For any 
$A \in \Abs(\wp(\Sigma))$, $A \in \SP_\fL$ iff $A \sqsubseteq
  \AD_\fL$.
\end{theorem}
\begin{proof}
Let $\mu=\gamma\circ \alpha\in \uco(\wp(\Sigma))$ and let $\mu_\fL\in \uco(\wp(\Sigma))$ be
the uco associated to $\AD_\fL$, that is $\mu_\fL (S) = \cap \{
\grasse{\varphi}_\cS~|~ \varphi \in \fL,\: S \subseteq
\grasse{\varphi}_\cS\}$. Recall that $A\sqsubseteq \AD_\fL$ iff for any
$\varphi\in \fL$, $\grasse{\varphi}_\cS \in \mu$.  \\
($\Ra$)~ For any $\varphi\in \fL$, we have that 
$\gamma(\alpha(\grasse{\varphi}_\cS))= \grasse{\varphi}_\cS$ because, 
by Lemma~\ref{charsp},
$\gamma(\alpha(\grasse{\varphi}_\cS)) 
= \gamma(\alpha(\gamma(\grasse{\varphi}_{\cS}^A))) =
\gamma(\grasse{\varphi}_{\cS}^A) = \grasse{\varphi}_\cS$.\\
($\Leftarrow$)~By hypothesis, $\gamma(\alpha(\grasse{\varphi}_\cS))=
\grasse{\varphi}_\cS$ for any $\varphi$. Let us show  
by structural induction on $\varphi\in \fL$ that
$\grasse{\varphi}_\cS=\gamma(\grasse{\varphi}_\cS^A)$.
\begin{itemize}
\item[--] $\varphi\equiv p\in \AP_\fL$: by using the hypothesis, 
  $\grasse{p}_\cS=\gamma_P(\alpha_P(\grasse{p}_\cS))=\gamma_P(\grasse{p}_\cS^A)$. 

\item[--] $\varphi\equiv f(\varphi_1,\ldots,\varphi_n)$: 
  \begin{align*}
    \grasse{f(\varphi_1,\ldots,\varphi_n)}_\cS = & \text{~~~[by
hypothesis]}\\
    \gamma(\alpha(\grasse{f(\varphi_1,\ldots,\varphi_n)}_\cS)) = &
\text{~~~[by definition]}  \\ 
    \gamma(\alpha(\boldsymbol{f}
(\grasse{\varphi_1}_\cS,\ldots,\grasse{\varphi_n}_\cS))) = &
\text{~~~[by inductive hypothesis]} \\
    \gamma(\alpha(\boldsymbol{f}
(\gamma(\grasse{\varphi_1}_\cS^A),\ldots,\gamma(\grasse{\varphi_n}_\cS^A))))
= &  \text{~~~[by definition]} \\
\gamma( \grasse{f(\varphi_1,\ldots,\varphi_n)}_\cS^A). \phantom{=}&
  \end{align*}
\end{itemize}
Thus, by Lemma~\ref{charsp}, $A \in \SP_\fL$.  
\end{proof}

Thus, $\ok{\AD_\fL}$ is the  most abstract domain that is s.p.\ for
$\fL$ w.r.t.\ $\cS$. As a consequence, it turns out  that
$A$ is s.p.\ for $\fL$  if and only if $A$ represents with no loss of precision
the concrete semantics of any formula in $\fL$, that is
$\forall \varphi\in \fL.\; \gamma(\alpha(\grasse{\varphi}_\cS)) =
\grasse{\varphi}_\cS$. Lemma~\ref{spadp} states that if a s.p.\ abstract
semantics on a given abstract domain exists then this is
unique. Nevertheless, 
Example~\ref{esnu} shows that this unique s.p.\ abstract semantics may
be induced from different abstract semantic structures, i.e.\
different abstract interpretation functions. However, 
when $\fL$ is closed
under conjunction,
it turns out
that on the most abstract s.p.\ domain  $\AD_\fL$,  the abstract interpretation
function is unique and is given by the best correct approximation
$I^{\AD_\fL}$.  
\begin{theorem}\label{unique}
Let $\fL$ be closed under infinite logical conjunction and let 
$\cS^\sharp = (\AD_\fL, I^\sharp)$ be an abstract semantic structure
on $\AD_\fL$. If $\grasse{\cdot}_{\cS^\sharp}$ is s.p.\ for $\fL$ then
$I^\sharp = I^{\AD_\fL}$. 
\end{theorem}
\begin{proof}
Since $\fL$ is closed under arbritrary logical conjunctions 
we have that $\AD_\fL = \{\grasse{\varphi}_\cS ~|~ \varphi \in
\fL\}$. Thus, for any $a\in \AD_\fL$, there exists some
$\varphi\in \fL$ such that
$a=\grasse{\varphi}_{\cS^\sharp}=\grasse{\varphi}_\cS^{\AD_\fL}$. In fact, if
$a\in \AD_\fL$ then $a= \grasse{\varphi}_\cS$, for some $\varphi\in
\fL$, so that, by Lemmata~\ref{charsp} and~\ref{spadp}, $a=\grasse{\varphi}_\cS =
\gamma(\grasse{\varphi}_{\cS^\sharp})=
\grasse{\varphi}_{\cS^\sharp} = \grasse{\varphi}_\cS^{\AD_\fL}$. \\
Let $p\in \AP$. Then, by Lemma~\ref{spadp}, $\grasse{p}_{\cS^\sharp}
=\grasse{p}_{\cS}^{\AD_\fL}$ so that
$I^\sharp(p)=  I^{\AD_\fL}(p)$.\\
Let $f\in \Op$. Then, 
\begin{align*}
I^\sharp (f) (a_1,...,a_n) = & \text{~~~~~~[by
the observation above]}\\
    I^\sharp (f) (\grasse{\varphi_1}_{\cS^\sharp},...,\grasse{\varphi_n}_{\cS^\sharp}) = &
\text{~~~~~~[by definition]}  \\ 
   \grasse{f(\varphi_1,...,\varphi_n)}_{\cS^\sharp}  = &
\text{~~~~~~[by Lemma~\ref{spadp}]} \\
   \grasse{f(\varphi_1,...,\varphi_n)}_{\cS}^{\AD_\fL}  = &\text{~~~~~~[by definition]}  \\ 
I^{\AD_\fL} (f)
(\grasse{\varphi_1}_{\cS}^{\AD_\fL},...,\grasse{\varphi_n}_{\cS}^{\AD_\fL})
= & \text{~~~~~~[by
the observation above]}\\
I^{\AD_\fL} (f) (a_1,...,a_n).\phantom{=}&
  \end{align*}
Thus, $I^\sharp=I^{\AD_\fL}$.
\end{proof}

\noindent
Hence, in the most abstract s.p.\ domain $\AD_\fL$ there is a unique
choice for interpreting  
atoms and operations of $\fL$. 

In our generalized framework, strong preservation for partitions
becomes 
a particular instance through the Galois insertion $\pr\!/\!\adp$. 
Moreover, when $\fL$ is closed under infinite conjunction,
it turns out that the most abstract s.p.\ domain $\AD_\fL$ is 
partitioning if and only if $\fL$ is also closed under negation. 

\begin{proposition}\label{onetwo}\ \\ \
{\rm (1)} $P_\fL = \pr(\AD_\fL)$ and $\adp(P_\fL)=\bP(\AD_\fL)$. \\
{\rm (2)} $P$ is strongly preserving
for $\fL$ iff $P \preccurlyeq \pr(\AD_\fL)$ iff $\adp(P)
\sqsubseteq \AD_\fL$.\\
{\rm (3)} Let $\fL$ be closed under conjunction. Then, 
$\AD_\fL$ is partitioning iff $\fL$ is closed under logical negation. 
\end{proposition}
\begin{proof}
(1)~ Let $\mu_\fL\in \uco(\wp(\Sigma))$ be
the uco associated to $\AD_\fL$.
We have that $\pr(\AD_\fL) =\{[s]_{\AD_\fL}~|~s\in \Sigma\}$, where
$[s]_{\AD_\fL} = \{s'\in \Sigma~|~ \mu_\fL(\{s'\}) = \mu_\fL
(\{s\})\}$. We also have that 
$s\equiv_{\fL} s'$ iff $\forall \varphi\in
\fL. s\in  \grasse{\varphi}_\cS \Lra s'\in
\grasse{\varphi}_\cS$ iff 
$\mu_\fL(\{s\}) = \mu_\fL(\{s'\})$, so that
$P_\fL = \pr(\AD_\fL)$. Moreover, 
$\adp(P_{\fL}) = \adp(\pr(\AD_\fL))=\bP(\AD_\fL)$. \\
(2)~ $P$ is s.p.\ for $\fL$ iff $P\preccurlyeq P_\fL$ iff, by 
Point~(1), $P\preccurlyeq \pr(A_{\fL})$ iff, by Theorem~\ref{teoGI},
  $\adp(P)\sqsubseteq \AD_\fL$. \\
(3)~   Since $\fL$ is closed under infinite logical conjunction, 
$\AD_\fL
=\{\grasse{\varphi}_\cS~|~\varphi\in\fL\}$.  Thus, 
$\fL$ is closed under logical
  negation iff $\AD_\fL$ is closed under complementation
$\complement$ and this exactly means that
  $\AD_\fL$ is forward complete for the complement $\complement$.  By
  Corollary~\ref{coropart}, 
this latter fact happens iff $\AD_\fL$ is partitioning. 
\end{proof}

In particular, when $\fL$ is closed under conjunction but not under
negation, it turns out that $\adp(P_\fL) \sqsubset \AD_\fL$, i.e.\ a
proper loss of information occurs when the domain $\AD_\fL$ is
abstracted 
to the partition $\pr(\AD_\fL)=P_\fL$. On the other hand, when 
$\fL$ is closed under conjunction and negation, we have that
$\adp(P_\fL)=\AD_\fL$ and therefore, by Theorem~\ref{unique}, the
abstract interpretation function on the partitioning abstract domain
$\adp(P_\fL)$ is uniquely determined.

\begin{example2}\label{exefex}
Let us consider the traffic light controller $\cK$ in
Example~\ref{semaforo}. As already observed, 
formulae of $\fL$ have the following semantics in $\cK$:
\begin{align*}
\ok{\grasse{\mathit{stop}}_\cK} = \{ R,RY\};~~~
\ok{\grasse{\mathit{go}}_\cK} = \{ G,Y\};~~~
\ok{\grasse{\mathrm{AXX}\mathit{stop}}_\cK}=\{G,Y\};~~~
\ok{\grasse{\mathrm{AXX}\mathit{go}}_\cK}=\{R,RY\}
\end{align*}
so that $$\AD_\fL =\cM(\{
\ok{\grasse{\varphi}_\cK}~|~ \varphi\in \fL\}) 
= \{ \varnothing, \{R,RY\}, \{G,Y\}, \{R,RY,G,Y\} \}$$ 
and $P_\fL = \pr(\AD_\fL)=
\{ \{R,RY\},\{G,Y\}\}$. We denote by $\mu_\fL$ the uco associated to
$\AD_\fL$.  As shown in Example~\ref{semaforo}, it turns
out  that 
no abstract Kripke structure that properly abstracts $\cK$ and
strongly preserves $\fL$ can be defined. In our approach, the
abstract domain $\AD_\fL$ induces a corresponding  strongly preserving
abstract semantics $\ok{\grasse{\cdot}_\cK^{\AD_\fL}}: \fL \ra
\AD_\fL$, where the best correct approximation
of the operator $\mathbf{AXX}:\wp(\Sigma)\ra \wp(\Sigma)$ on $\AD_\fL$
is:
\begin{align*}
\mu_\fL \circ \mathbf{AXX} =&~ \{\varnothing \mapsto \varnothing, \{R,RY\} \mapsto \{G,Y\}, 
\{G,Y\} \mapsto \{R,RY\},  \\
&~~~\{R,RY,G,Y\}  \mapsto \{R,RY,G,Y\}\}. \\[-35pt]
\end{align*}
\mbox{~~}\qed
\end{example2}

\begin{figure}[t]
\begin{center}
\begin{tabular}{ccc}
  \mbox{\xymatrix{
      *++[o][F]{1} \ar[d] \ar[r]^(0.3){p} ^(1.3){p} &
      *++[o][F]{2} \ar@(ul,ur)[] \ar[ld] &  \\ 
      *++[o][F]{3}\ar[r] ^(0.3){p} & *++[o][F]{4}\ar@/^/[r] ^(0.24){p}
& *++[o][F]{5} \ar@/^/[l] ^(0.24){q} \\
    }
  }
& \mbox{~~~~~~~~~~~~~~~~~~~~} &
  \mbox{\xymatrix  {
      *++[o][F]{\![12]\!} \ar[d] ^(0.32){p} \ar@(ul,ur)[] & &  \\ 
      *++[o][F]{\,[3]\,}\ar[r] ^(0.28){p} & *++[o][F]{\,[4]\,}\ar@/^/[r] ^(0.26){p}
& *++[o][F]{\,[5]\,} \ar@/^/[l] ^(0.26){q} \\
    }
  }
\end{tabular}
\end{center}
\caption{Concrete (on the left) and abstract (on the right) Kripke structures.}\label{fig1}
\end{figure}
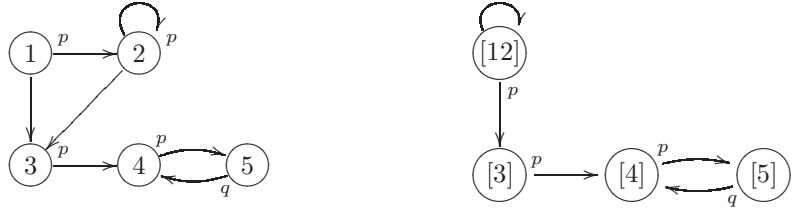

\begin{example2}\label{exef}
Consider the language $\CTL$ and the Kripke structure 
$\cK=(\Sigma,R, \ell)$ depicted in
Figure~\ref{fig1}, where
the interpretation of temporal 
operators of $\CTL$ on $\cK$ is standard. 
It is well known that the coarsest s.p.\
partition $P_{\mathrm{CTL}}$ can
be obtained by refining the initial partition $P=\{1234,5\}$ induced
by the labeling $\ell$ through the Paige-Tarjan \cite{pt87}
algorithm, since $P_{\mathrm{CTL}}$
coincides with bisimulation equivalence on $\cK$.  
It is easy to check that
$P_{\mathrm{CTL}}= \{12,3,4,5\}$. This partition
determines (see point~(2) in Section~\ref{amc}) 
the s.p.\ abstract Kripke structure depicted in
Figure~\ref{fig1}. 
Since $\CTL$ is closed under conjunction and negation, 
by Proposition~\ref{onetwo}~(1) and~(3), it turns out that  the most abstract s.p.\
domain $A_{\mathrm{CTL}}$ is partitioning and 
coincides with the following partitioning closure: 
$$\adp (P_{\mathrm{CTL}}) = \{\varnothing, 12,3,4,5,34, 35, 45, 122,
124, 125, 345, 1234, 1235, 1245, 12345\}.$$

Let us now consider the following language 
$\fL\ni \varphi ::= p ~|~ q~|~ \varphi_1\wedge \varphi_2~|~
\mathrm{EF}_{[0,2]} \varphi$, 
where $\mathrm{EF}_{[0,2]}$ is a time bounded reachability
operator that is useful for quantitative temporal analysis \cite{emss90},
e.g., of discrete real-time systems \cite[Chapter~16]{cgp99}. The standard 
interpretation of $\mathrm{EF}_{[0,2]}$ is as follows:
$s\ok{\models^\cK} \mathrm{EF}_{[0,2]} \varphi$ iff there exists a path
$s_0 s_1 s_2 s_3\ldots$ in $\cK$
starting from $s=s_0$ and some $n\in [0,2]$ such that  
$s_n \ok{\models^\cK} \varphi$. 
Let us characterize the semantics of formulae in $\fL$:
\begin{align*}
& \ok{\grasse{p}_\cK} = \{ 1,2,3,4\};~~~
 \ok{\grasse{q}_\cK} = \{ 5\};~~~
\ok{\grasse{\mathrm{EF}_{[0,2]} p}_\cK} = \{ 1,2,3,4,5\};\\
& \ok{\grasse{\mathrm{EF}_{[0,2]}q}_\cK} = \{ 3,4,5\};~~~
\ok{\grasse{\mathrm{EF}_{[0,2]}(\mathrm{EF}_{[0,2]}q)}_\cK} =
\{1,2,3,4,5\};\\
& \ok{\grasse{p \wedge \mathrm{EF}_{[0,2]}q }_\cK} =
\{3,4\};~~~
\ok{\grasse{\mathrm{EF}_{[0,2]}(p \wedge \mathrm{EF}_{[0,2]}q) }_\cK} =
\{1,2,3,4,5\}.
\end{align*}
Thus, $\AD_\fL =\cM(\{
\ok{\grasse{\varphi}_\cK}~|~ \varphi\in \fL\}) = \{\varnothing, 5, 34,
345,
1234, 12345\}$. On the other hand, by Proposition~\ref{onetwo}~(1),
$P_\fL = \pr(\AD_\fL)= \{12, 34, 5\}$.
In this case, it turns out that $\adp(P_\fL) \sqsubset
\AD_\fL$. Moreover, analogously to Example~\ref{semaforo}, 
let us show that there exists no abstract transition relation $\sra^\sharp
\subseteq P_\fL \times P_\fL$ that determines an abstract Kripke structure
$\cA= (P_\fL,
\sra^\sharp,\ell_\fL)$  which strongly preserves $\fL$. 
Let $B=\{1,2\}$, $B'=\{3,4\}$ and $B''=\{5\}$ be the blocks in
$P_\fL$. Assume
by contradiction that such an abstract Kripke structure $\cA$
exists.  
\begin{itemize}
\item[{\rm (i)}] 
On the concrete model $\cK$ we have that 
$3 \ok{{\models^\cK}} \mathrm{EF}_{[0,2]}q$. Thus, by strong
preservation, it must be that $B' \ok{\models^\cA}
\mathrm{EF}_{[0,2]}q$.
On the other hand, if $B' \sra^\sharp B$ and 
$B \sra^\sharp B''$
then $B \ok{\models^\cA} \mathrm{EF}_{[0,2]}q$ and therefore, by weak preservation, we
would have that
$ 1\ok{\models^\cK} \mathrm{EF}_{[0,2]}q$, which is a contradiction. 
Thus, necessarily,
 $B' \sra^\sharp B''$. 
\item[{\rm (ii)}] 
Let us observe that  $1 \ok{{\models^\cK}}
\mathrm{EF}_{[0,2]}\mathrm{EF}_{[0,2]}q$.  Hence, by strong
preservation, $B \ok{\models^\cA}
\mathrm{EF}_{[0,2]}\mathrm{EF}_{[0,2]}q$. If $B \sra^\sharp B''$
then, as in point~(i), we would still have that  
$ 1\ok{\models^\cK} \mathrm{EF}_{[0,2]}q$, i.e.\ a
contradiction. Hence, necessarily, $B \sra^\sharp B'$.
\item[{\rm (iii)}] From $B \sra^\sharp  B'$ and $B'
\sra^\sharp  B''$, we would obtain that 
$B \ok{\models^\cA}\mathrm{EF}_{[0,2]}q$ that, as observed in
point~(ii), is a contradiction. 
\end{itemize}

\noindent 
Thus, this shows that it is not possible to define an abstract Kripke structure on the
abstract state space $P_\fL$ that strongly preserves $\fL$. The
abstract domain $\AD_\fL$ induces a corresponding  
abstract semantics $\ok{\grasse{\cdot}_\cK^{\AD_\fL}}$ that
instead strongly preserves $\fL$. In this case, the best correct approximation
of the operator $\mathbf{EF}_{[0,2]}$ on $\AD_\fL$
is:
\begin{align*}
\mu_\fL \circ \mathbf{EF}_{[0,2]} =&~ \{\varnothing \mapsto \varnothing,\, 5 \mapsto 345,\, 
34 \mapsto 12345,\, 345 \mapsto 12345, \\ 
&~~~1234  \mapsto 12345,\, 12345 \mapsto 12345\}. \\[-35pt]
\end{align*}
\mbox{~~}\qed
\end{example2}

\section{Strong Preservation and Completeness}\label{csp}
In this section we establish a precise correspondence between 
generalized strong preservation of abstract models  and
completeness of abstract interpretations, so that the problem of
minimally refining an abstract model in order to get strong
preservation can be formulated as a complete domain refinement in abstract
interpretation. 

\subsection{Forward Complete Shells}\label{fcs-section}
Let us consider forward completeness of abstract domains $A\in \Abs(C)$ for
generic $n$-ary concrete operations $f:C^n \ra C$, with $n\geq
0$. Hence, $A$  is forward complete for $f$, or simply $f$-complete,
when $f\circ \tuple{\mu_A,...,\mu_A} =\mu_A \circ f \circ
\tuple{\mu_A,...,\mu_A}$, that is, for any
$\vec{x}\in C^n$, $f(\mu_A(x_1),...,\mu_A(x_n)) = 
\mu_A(f(\mu_A(x_1),...,\mu_A(x_n)))$. Equivalently, $A$ is
$f$-complete when  
for any
$\vec{a}\in A^n$, $f(\gamma(a_1),...,\gamma(a_n)) = 
\gamma(\alpha(f(\gamma(a_1),...,\gamma(a_n))))$.
For a set of operations $F\subseteq
\Fun(C)$, $A$ is $F$-complete when $A$ is $f$-complete for each
$f\in F$. Observe that $F$-completeness for an abstract domain $A$ means that
the associated closure $\mu_A$ is
closed under the image of functions in $F$, namely  
$F(\mu_A)\subseteq \mu_A$. 
Also note that when $k:C^0\ra C$, i.e.\ $k\in C$ is a constant, $A$ is 
$k$-complete iff $k$ is precisely represented in $A$, i.e.\
$\gamma(\alpha(k)) = k$. 
Let us also note that an abstract domain $A\in \Abs(C)$ is always
forward meet-complete
because any uco is Moore-closed.  

Let us first note that forward $F$-complete shells
always exist. Let $\sS_F:\Abs(C)\ra \Abs(C)$ be defined as 
$\sS_F (A)\ud \sqcup \{ X\in \Abs(C)~|~ X \sqsubseteq A ,\,
X \text{ is $F$-complete}\}$.

\begin{lemma}\label{fcs}
$\sS_F(A)$ is the $F$-complete shell of $A$. 
\end{lemma}
\begin{proof}
Let $\eta = \sqcup \{\rho  \in \uco(C)~|~ \rho \sqsubseteq \mu_A,\,
\rho \text{ is $F$-complete}\} = \cap \{\rho  \in \uco(C)~|~ \rho \sqsubseteq \mu_A,\,
\rho \text{ is $F$-complete}\}$.
  Let $f\in F$, with $\sharp(f)=n >0$ (if $\sharp(f)=0$ then, trivially, $f\in \eta$) 
and $\ok{\vec{c}}\in \ok{\eta^{n}}$.
  Consider any $\rho \in\uco(C)$ that is $F$-complete and such that
  $\rho \sqsubseteq \mu$. Since $\eta\subseteq \rho$, we have that $\ok{\vec{c}}\in
  \ok{\rho^{n}}$ and therefore $\ok{f(\vec{c})}\in\rho$ because
  $\rho$ is $F$-complete.  Thus, $\ok{f(\vec{c})}\in\eta$, i.e., $\eta$ is $F$-complete.
\end{proof}

A forward complete shell $\sS_F(A)$ is a more concrete abstraction
than $A$. How to characterize $\sS_F(A)$? 
It is here useful to view abstract domains as closure
operators on the concrete domain, i.e.\ as subsets of $C$.  Hence, $A$
is viewed as the
subset $\img(\mu_A)=\gamma(A)$ of the concrete domain $C$ so that 
$\sS_F(A)$ can be characterized as the least Moore-closed subset of $C$ that 
contains $\img(\mu_A)$ and is forward $F$-complete. 
We need to characterize the least amount of
concrete information that must be added to $\gamma(A)$ in order to get
forward completeness. 
It turns out that forward complete shells admit a constructive
fixpoint characterization. Let $\ok{F^{\uco}}: \uco(C)\ra \uco(C)$ be defined as follows:
$\ok{F^{\uco}}(\rho)\ud\cM(F(\rho))$, namely $\ok{F^{\uco}} (\rho)$ is the most
abstract domain that contains the image of $F$ on $\rho$. 
Observe that the operator $\lambda \rho .\mu_A\sqcap
\ok{F^{\uco}}(\rho):\uco(C)\ra \uco(C)$ is monotone.

\begin{lemma}\label{itera}
$\sS_F (A) =
\gfp (\lambda \rho. \mu_A \sqcap \ok{F^{\uco}} (\rho))$. 
\end{lemma}
\begin{proof}
Observe that a uco $\rho$ is $F$-complete iff $F(\rho) \subseteq \rho$
iff $\cM(F(\rho))=\ok{F^{\uco}} (\rho)\subseteq \rho$ iff $\rho
\sqsubseteq \ok{F^{\uco}}(\rho)$. Thus, we have that
$\sS_F (A) = \sqcup \{ \rho \in \uco(C)~|~ \rho \sqsubseteq \mu_A ,\:
\rho \text{ is $F$-complete}\} = \sqcup \{ \rho \in \uco(C)~|~ \rho \sqsubseteq \mu_A ,\:
\rho \sqsubseteq \ok{F^{\uco}}(\rho) \} = 
\sqcup \{ \rho \in \uco(C)~|~ \rho \sqsubseteq \mu_A \sqcap \ok{F^{\uco}}(\rho) \}
= \gfp ( \lambda \rho. \mu_A \sqcap \ok{F^{\uco}} (\rho))$. 
\end{proof}

\noindent
Thus, it turns
out that the lower iteration sequence of $\lambda \rho. \mu_A \sqcap
\ok{F^{\uco}} (\rho)$ 
in $\uco(C)$
converges to the complete shell $\sS_F(\mu_A)$.

\begin{example2}
Let us consider the square operator on sets of integers
$\sq:\wp(\mathbb{Z})\ra \wp(\mathbb{Z})$, i.e.\ $\sq(X)=X^2=\{x^2~|~
x\in X\}$, and the abstract domain
$\mathit{Sign}=\{\varnothing,\bZ_{<0},\{0\},\bZ_{>0},\bZ\}$. As observed in
Section~\ref{secco}, $\mathit{Sign}$ is not forward complete for the
square operator. Let us apply Lemma~\ref{itera} in order to 
compute the forward complete shell
$\sS_{\sq}(\mathit{Sign})$.  Observe that
$$\varnothing^2 =\varnothing \in\Sign;~~~ \{0\}^2 = \{0\}\in \Sign;~~~
\bZ_{<0}^2=\bZ_{>0}^2 =\bZ^2
\not\in \Sign.$$
Thus, the first step of iteration refines $\Sign$ to $\Sign\cup
\ok{\{\bZ^2\}}$ (notice that this is an abstract domain because it is
Moore-closed).  Then, $\ok{(\bZ^2)^2} =\ok{\bZ^{2^2}} \not\in \Sign
\cup \{\ok{\bZ^2}\}$, so that on
the second step of iteration we obtain $\Sign \cup
\{\ok{\bZ^2},\ok{\bZ^{2^2}}\}$. In general, for $n\geq 1$,  the $n$-th step of
iteration provides $\Sign \cup \{\ok{\bZ^{2^k}}~|~ k\in [1,n]\}$, so that the complete
shell $\sS_{\sq}(\Sign)$ coincides with the least fixpoint $\Sign\cup \{\ok{\bZ^{2^n}}~|~
n\geq 1\}$.
\qed
\end{example2}

Finally, the following easy observation will be useful later on. 
\begin{lemma}\label{shellequiv}
Let $F,G\subseteq \Fun(C)$. Then, $\sS_F = \sS_G$ if and only if for any $A \in
\Abs(C)$, $A$ is $F$-complete $\Lra$ $A$ is $G$-complete. 
\end{lemma}
\begin{proof}
($\Ra$) If $A$ is $F$-complete then $A=\sS_F(A) = \sS_G(A)$
and therefore $A$ is $G$-complete as well. \\
($\Leftarrow$) This follows from
$\sS_F (A)= \sqcup \{ X \in \Abs(C)~|~ X \sqsubseteq A ,\,
X \text{ is $F$-complete}\} = \sqcup \{ X \in \Abs(C)~|~ X \sqsubseteq A ,\,
X \text{ is $G$-complete}\} = \sS_G (A)$.   
\end{proof}

\subsection{Strong Preservation and Complete Shells}
Let $\fL$ be a language with atoms in $\AP_{\fL}$ and operators in
$\Op_{\fL}$ and let $\cS=(\Sigma,I)$ be a semantic structure for
$\fL$ so that $\APb_{\fL}$ and $\Opb_\fL$ denote, respectively, the
corresponding sets of semantic interpretations of atoms and operators.
It turns out that forward completeness
for $\APb_{\fL}$ and $\Opb_\fL$
implies strong preservation for $\fL$.
\begin{lemma}
If $A\in \Abs(\wp(\Sigma))$ is forward complete for $\APb_{\fL}$ and
$\Opb_\fL$ then $A$ is s.p.\ for $\fL$.
\end{lemma}
\begin{proof}
By Theorem~\ref{spchar}, we show that $A\sqsubseteq \AD_\fL$. Let
us show by
induction that
for any $\varphi\in \fL$,
$\grasse{\varphi}_\cS=\gamma(\alpha(\grasse{\varphi}_\cS))$. 
\begin{itemize}
\item[--] $\varphi\equiv p\in \AP_\fL$: since $A$ is forward
complete for $\boldsymbol{p}$, 
  $\grasse{p}_\cS=\boldsymbol{p}= \gamma(\alpha(\boldsymbol{p})) =
\gamma(\alpha(\grasse{p}_\cS))$. 

\item[--] $\varphi\equiv f(\varphi_1,\ldots,\varphi_n)$ with $f\in \Op_\fL$: 
  \begin{align*}
    \grasse{f(\varphi_1,...,\varphi_n)}_\cS = & \text{~~~[by
definition]}\\
    \boldsymbol{f}
(\grasse{\varphi_1}_\cS,...,\grasse{\varphi_n}_\cS) = &
\text{~~~[by inductive hypothesis]} \\
    \boldsymbol{f}
(\gamma(\alpha(\grasse{\varphi_1}_\cS)),...,\gamma(\alpha(\grasse{\varphi_n}_\cS)))
= &  \text{~~~[since $A$ is forward complete for $\boldsymbol{f}$]} \\
    \gamma(\alpha(\boldsymbol{f}
(\gamma(\alpha(\grasse{\varphi_1}_\cS)),...,\gamma(\alpha(\grasse{\varphi_n}_\cS)))))
= &  \text{~~~[by inductive hypothesis and by definition]} \\
    \gamma(\alpha(\grasse{f(\varphi_1,...,\varphi_n)}_\cS)).\phantom{=} &
  \end{align*}
\end{itemize}
\end{proof}

On the other hand, the converse is not true, that is 
strong preservation does not imply forward
completeness, as shown by the following example. 

\begin{example2}
Let us consider again Example~\ref{esnu} where we showed that the partitioning abstract
domain $A=\wp(P)_\subseteq$ is s.p.\ for $\fL$. However, $A$ is not
forward complete for $\boldsymbol{Op}_\fL = \{\pre_\sra\}$.  In fact: 
$\gamma(\alpha ( \pre_\sra (\gamma (\alpha (\{3\}))))) =
\gamma(\alpha ( \pre_\sra (\{3\}))) 
= \gamma (\alpha (\{2,3\})) = \{1,2,3\}$ while 
$\pre_\sra (\gamma (\alpha (\{3\})))= \pre_\sra (\{3\}) =
\{2,3\}$. 
\qed
\end{example2}

Instead, it turns out that most abstract s.p.\ domains can be
characterized as forward complete shells.  
\subsubsection{Complete Shells as Strongly Preserving Abstract Domains} \label{cssp}

Partition refinement
algorithms for computing behavioural equivalences like bisimulation
\cite{pt87}, simulation equivalence \cite{bg03,hhk95,TC01} and (divergence
blind) stuttering equivalence \cite{gv90} are used
in standard abstract model checking to
compute the coarsest strongly preserving partition of temporal languages
like $\CTLS$ or the
$\mu$-calculus for the case of bisimulation equivalence, $\ACTLS$ for simulation
equivalence and $\CTLSX$ for stuttering equivalence.

Given a language $\fL$ and a concrete state space $\Sigma$, 
these partition refinement algorithms work by
iteratively refining an initial partition $P$ within the lattice of
partitions $\Part(\Sigma)$ until the fixpoint $P_\fL$ is reached.
The input partition $P$ determines the set
$\AP_P$ 
of atoms
and
their interpretation $I_P$ as follows: $\AP_P\ud \{p_B~|~B\in P\}$ and
$I_P(p_B) \ud B$. 
More in general, any $\cX\subseteq \wp(\Sigma)$ determines a 
set $\{p_X\}_{X\in \cX}$ of atoms with interpretation
$I_\cX( p_X) =X$. In particular, this can be done for an abstract domain
$A \in \Abs(\wp(\Sigma))$ by considering its concretization 
$\gamma(A)\subseteq \Sigma$, namely $A$ is viewed as a set of atoms 
with interpretation 
$I_A (a) = \gamma(a)$. Thus, an abstract domain $A\in
\Abs(\wp(\Sigma))$ 
together with a set of functions $F\subseteq
\Fun(\wp(\Sigma))$ determine a language $\fL_{A,F}$, with atoms in
$A$, operations in $F$ and endowed with a semantic structure
$\cS_{A,F}=(\Sigma,I_A \cup I_F)$ such that  for any $a\in A$, $I_A(a)=\gamma(a)$
and for any $f\in F$, $I_F(f)=f$.   
Therefore, the most abstract s.p.\ domain $\AD_{\fL_{A,F}}$ generalizes in
our framework the
output of a partition refinement algorithm for some language. Accordingly, 
we aim at
characterizing $\AD_{\fL_{A,F}}$ as the output of a
refinement process of the initial domain $A$ within the lattice 
$\Abs(\wp(\Sigma))$
of abstract domains. The following 
result shows that forward
completeness for the operations in $F$ is the right
notion of refinement to be used for the case of abstract
domains. 

\begin{theorem}\label{main2} Let
$A \in \Abs(\wp (\Sigma ))$, $F\subseteq \Fun(\wp(\Sigma))$ 
and assume that $\fL_{A,F}$ is closed under infinite logical conjunction. Then,
$\AD_{\fL_{A,F}} = \sS_{F}(A)$. 
\end{theorem}
\begin{proof}
Since $\fL_{A,F}$ is closed under conjunction we 
 have that $\AD_{\fL_{A,F}} =
\{\grasse{\varphi}_{\cS_{A,F}}~|~\varphi\in\fL_{A,F}\}$.
  Let us first prove that
  $\{\grasse{\varphi}_{\cS_{A,F}}~|~\varphi\in\fL_{A,F}\} 
\subseteq\sS_{F}(A)$ by structural
induction on $\varphi\in \fL_{A,F}$:
  \begin{itemize}
  \item[--] $\varphi\equiv a\in A$: $\grasse{a}_{\cS_{A,F}} = I_A(a)=\gamma(a)
\in \gamma(A) \subseteq
    \sS_{F}(A)$.
  \item[--] $\varphi\equiv f(\varphi_1,...,\varphi_n)$ with $f\in F$: 
     $\grasse{f(\varphi_1,...,\varphi_n)}_{\cS_{A,F}}=
    f(\grasse{\varphi_1}_{\cS_{A,F}},...,
\grasse{\varphi_n}_{\cS_{A,F}})$,  where,
by inductive hypothesis, 
$\grasse{\varphi_i}_{\cS_{A,F}}\in \sS_{F}(A)$. 
Therefore, since $\sS_{F}(A)$ is 
forward $f$-complete, we have that
$f(\grasse{\varphi_1}_{\cS_{A,F}},...,\grasse{\varphi_n}_{\cS_{A,F}})\in
\sS_{F}(A)$.   
  \end{itemize}

\noindent
Let us now prove the opposite inclusion. 
Let us first observe that $\AD_{\fL_{A,F}}$
is forward
$F$-complete. For simplicity of notation, 
consider  $f\in F$ with
$\sharp(f)=1$. If $\grasse{\varphi}_{\cS_{A,F}} \in
\AD_{\fL_{A,F}}$, where $\varphi\in \fL_{A,F}$, then, $f(\varphi)\in
\fL_{A,F}$ and $f(\grasse{\varphi}_{\cS_{A,F}}) =
\grasse{f(\varphi)}_{\cS_{A,F}} \in \AD_{\fL_{A,F}}$. 
By Lemma~\ref{itera}, we know that 
  $\sS_{A}(A)=\sqcap_{\alpha \in\Ord} (\lambda \rho. \mu_A
\sqcap \cM(F(\rho)))^{\alpha,\downarrow} (\top_{\uco(\wp(\Sigma))})$, so that
  it is sufficient to prove by transfinite induction on $\alpha \in
\Ord$ that
  $$(\lambda \rho. \mu_A
\sqcap \cM(F(\rho)))^{\alpha,\downarrow} 
(\top_{\uco(\wp(\Sigma))}) \subseteq \AD_{\fL_{A,F}}.$$
  \begin{itemize}

  \item[--] $\alpha = 0$:  $(\lambda \rho. \mu_A
\sqcap \cM(F(\rho)))^{0,\downarrow} (\top_{\uco(\wp(\Sigma))})
=\top_{\uco(\wp(\Sigma))}=\{\Sigma \} \in \gamma(A) \subseteq \AD_{\fL_{A,F}}$.
  \item[--] $\alpha +1$: By inductive
hypothesis, $(\lambda \rho. \mu_A
\sqcap \cM(F(\rho)))^{\alpha,\downarrow} (\top_{\uco(\wp(\Sigma))}) 
\subseteq\AD_{\fL_{A,F}}$. Moreover, $\AD_{\fL_{A,F}}$ is Moore-closed
and forward
$F$-complete (hence closed under $F$).
Thus, $\cM(F ( (\lambda \rho. \mu_A
\sqcap \cM(F(\rho)))^{\alpha,\downarrow} (\top_{\uco(\wp(\Sigma))}))) \subseteq
\AD_{\fL_{A,F}}$,  namely 
$(\lambda \rho. \mu_A
\sqcap \cM(F(\rho)))^{\alpha+1,\downarrow} (\top_{\uco(\wp(\Sigma))})\subseteq
\AD_{\fL_{A,F}}$.
\item[--] limit ordinal $\alpha$: 
This follows from 
$$ (\lambda \rho. \mu_A
\sqcap \cM(F(\rho)))^{\alpha,\downarrow}
(\top_{\uco(\wp(\Sigma))}) =
\sqcap_{\beta < \alpha} (\lambda \rho. \mu_A
\sqcap \cM(F(\rho)))^{\beta,\downarrow}
(\top_{\uco(\wp(\Sigma))}) $$
because, by inductive hypothesis, $(\lambda \rho. \mu_A
\sqcap \cM(F(\rho)))^{\beta,\downarrow} (\top_{\uco(\wp(\Sigma))}) 
\subseteq \AD_{\fL_{A,F}}$, for any
$\beta< \alpha$.
  \end{itemize}
\end{proof}

\subsubsection{Strongly Preserving Abstract Domains as Complete Shells} 

Let us consider a language $\fL$, with atoms in $\AP_\fL$ and
operators in $\Op_\fL$, and a semantic structure $\cS=(\Sigma,I)$.  
As an immediate consequence of Theorem~\ref{main2}, 
the most abstract s.p.\ domain 
$\AD_{\fL}$ for $\fL$ w.r.t.\ $\cS$ can be characterized as the
forward $\boldsymbol{AP}_\fL \cup \boldsymbol{Op}_\fL$-complete shell
of the most abstract domain $\{\Sigma\}$.

\begin{corollary}\label{strai}
  Let $\fL$ be closed under infinite
logical conjunction.
Then, $\AD_\fL=\sS_{\boldsymbol{AP}_\fL\cup\boldsymbol{Op}_\fL}(\{\Sigma\})$.
\end{corollary}

Let us also observe that
$\AD_\fL$ can be equivalently characterized as 
the
forward $\boldsymbol{Op}_\fL$-complete shell
of an initial abstract
domain $\cM(\boldsymbol{AP}_\fL)$ induced by atoms: 
$\AD_\fL=\sS_{\boldsymbol{Op}_\fL}(\cM(\boldsymbol{AP}_\fL))$.

\subsubsection{Strongly Preserving Partitions}


Theorem \ref{main2} and Corollary~\ref{strai} 
provide an elegant generalization of 
partition refinement algorithms for strong
preservation from an
abstract interpretation perspective. 

Given a language $\fL$ with operators in $\Op_\fL$ and a corresponding
semantic
structure $\cS=(\Sigma,I)$, as recalled in Section~\ref{cssp}, 
an input partition $P\in \Part(\Sigma)$ for a partition refinement
algorithm determines the set
$\AP_\fL = \{p_B~|~B\in P\}$ 
of atoms of $\fL$ 
and
their interpretation $I (p_B)=B$. Thus, $\cM(\boldsymbol{\AP}_\fL) =
\cM(P)=P\cup \{\varnothing,\Sigma\}$. 
It turns out that the coarsest s.p.\ partition $P_\fL$ for $\fL$ 
can be characterized in our abstract domain-based approach as
follows.

\begin{corollary}\label{spcb}
  Let $\fL$
be closed under infinite logical conjunction. \\
  {\rm (1)}  $P_\fL = \pr(\sS_{\boldsymbol{Op}_\fL}(\cM(P)))$.\\
  {\rm (2)} Let $\fL$ be closed under logical negation. Then, 
  $\adp(P_{\fL}) =
\sS_{\boldsymbol{Op}_\fL}(\cM(P))$.
\end{corollary}
\begin{proof}
  \noindent {\rm (1)} 
By Corollary~\ref{strai}, 
$\AD_\fL=\sS_{\boldsymbol{Op}_\fL}(\cM(P))$ and by
Proposition~\ref{onetwo}~(1), $P_\fL=\pr(\AD_\fL)=
\pr(\sS_{\boldsymbol{Op}_\fL}(\cM(P)))$.\\ 
{\rm (2)} By Proposition~\ref{onetwo}~(1) and~(3),
Corollary~\ref{strai} and point~(1), 
$\adp(P_\fL)=\adp(\pr(\AD_\fL))=\AD_\fL=\sS_{\boldsymbol{Op}_\fL}(\cM(P))$. 
\end{proof}

It is worth remarking that when $\cL$ is not closed under negation, by
Proposition~\ref{onetwo}~(3) and Corollary~\ref{spcb}~(2), it turns out
that $\adp(P_\fL) \sqsubset \sS_{\boldsymbol{Op}_\fL}(\cM(P))$.  This
means that when $\fL$ is not closed under negation the output
partition $P_\fL$ of any partition refinement algorithm for achieving
strong preservation for $\fL$ is not optimal within the lattice of
abstract domains.

\begin{example2}
Let us consider the language $\fL$ and 
the concrete Kripke structure $\cK$ in Example~\ref{exef}.
The labeling determines the initial partition
$P=\{\boldsymbol{p}=1234, \boldsymbol{q}=5\}\in \Part(\Sigma)$, so
that $\cM(P)= \{\varnothing, 1234,5,12345\}\in \Abs(\wp(\Sigma))$. 
Here, $\Op_\fL = \{\wedge, \mathrm{EF}_{[0,2]}\}$. Abstract domains
are Moore-closed so that $\sS_{\boldsymbol{Op}_\fL}=
\sS_{\mathbf{EF}_{[0,2]}}$. Let us compute
$\sS_{\mathbf{EF}_{[0,2]}}(\cM(P))$. 
\begin{align*} 
A_0 & = \cM(P)= \{\varnothing, 1234,5,12345\}\\
A_1 & = A_0 \sqcap \cM(\mathbf{EF}_{[0,2]}(A_0))= \cM(A_0 \cup \mathbf{EF}_{[0,2]}(A_0)) \\ 
& =
\cM(\{\varnothing, 1234,5,12345\} \cup \{
\mathbf{EF}_{[0,2]}(\{5\})=345\}) = \{\varnothing, 5, 34,
1234,12345\}\\
A_2 & = A_1 \text{~~~~(fixpoint)}
\end{align*}
As already observed in Example~\ref{exef}, $P_\fL = \{12,34,5\}$ is
such that $\adp(P_\fL)\sqsubset \mu_\fL$ and it is not possible
to define a strongly
preserving abstract Kripke
structure on the abstract space $P_\fL$.
\qed
\end{example2}

\section{An Application to some Behavioural Equivalences}\label{abse}
It is well known that some temporal
languages like $\CTL$, $\ACTL$ and $\CTLX$ induce state logical equivalences that
coincide with standard behavioural equivalences 
like bisimulation equivalence for $\CTL$, 
(divergence blind) stuttering equivalence for $\CTLX$ and
simulation equivalence for
$\ACTL$. 
We derive here
a novel characterization of these behavioural equivalences
in terms of forward completeness of abstract
interpretations. 

\subsection{Bisimulation Equivalence}
Let $\cK = (\Sigma ,\sra,\ell)$ be a Kripke structure over some set
$\AP$ of atomic propositions. A
relation $R  \subseteq \Sigma \times \Sigma $ is a bisimulation on
$\cK$ if for any
$s,s'\in \Sigma $ such that $s R s'$:
\begin{itemize}
\item[{\rm (1)}] $\ell(s) =\ell (s')$; 
\item[{\rm (2)}] For
any $t\in \Sigma $ such that $\ok{s \sra t}$, there exists $t'\in \Sigma $
such that $\ok{s'\sra t'}$
and $t R t'$;
\item[{\rm (3)}] $s' R s$, i.e.\ $R$ is symmetric. 
\end{itemize}
Since the empty
relation is a bisimulation and bisimulations are closed under union,
it turns out that the 
largest (as a set) bisimulation relation exists. This largest
bisimulation is an equivalence relation called bisimulation
equivalence and is 
denoted by $\sim_{\mathrm{bis}}$ while $P_{\mathrm{bis}}\in \Part(\Sigma)$ denotes
the corresponding partition. 
Thus,  a partition $P\in \Part(\Sigma)$ is a bisimulation on $\cK$
when $P\preceq P_{\mathrm{bis}}$.

It is well known~\cite{bcg88} that
when $\cK$ is finitely branching, bisimulation
equivalence coincides with the state
equivalence induced by $\CTL$, i.e.,
$P_{\mathrm{bis}}= P_\CTL$ (the same holds
for $\CTLS$ and the $\mu$-calculus, see e.g.\
\cite[Lemma~6.2.0.5]{dams96}). 
Moreover, it is known (see e.g.\ \cite[Section~12]{vg90}) that it is enough
to consider finitary Hennessy-Milner logic \cite{hm85}, i.e.\ a language
$\fL_1$ including propositional logic 
and the existential next operator in order to have that
$P_{\fL_1} = P_{\mathrm{bis}}$: 
$$\fL_1 \ni \varphi ::= p ~|~ \varphi_1 \wedge \varphi_2 ~|~
\neg \varphi ~|~
\mathrm{EX}\varphi 
$$ where, as usual, 
the interpretation $\mathbf{EX}$ of $\mathrm{EX}$ in $\cK$ is 
$\pre_\sra$.  A number of algorithms for computing bisimulation
equivalence exists \cite{bfh90,dpp04,ly92,pt87}. The
Paige-Tarjan algorithm \cite{pt87}  runs in $O(|\sra|\log(|\Sigma
|))$-time and 
is the most time-efficient
algorithm that computes bisimulation equivalence.  

We recalled above that $P_{\fL_1} = P_\CTL$. In our framework,
this can be obtained as a consequence of the fact 
that the most
abstract s.p.\ domains for $\CTL$ and $\fL_1$ coincide.

\begin{lemma}\label{ctlequiv}
Let $\cK$ be finitely branching. Then, $\AD_\CTL  = \AD_{\fL_1}=\adp(P_{\mathrm{bis}})$. 
\end{lemma}
\begin{proof}
Let $\boldsymbol{Op}_{\CTL} =   
\{\cap,  \complement, \mathbf{AX}, \mathbf{EX},
\mathbf{AU},\mathbf{EU}, \mathbf{AR},  \mathbf{ER}\}$ be the
set of standard interpretations of the operators of $\CTL$ on $\cK$, 
so that $\mathbf{AX}=\pret_\sra$ and
$\mathbf{EX}=\pre_\sra$.   
We show
that $\mu\in \uco(\wp(\Sigma))$ is forward complete  for 
$\boldsymbol{Op}_{\CTL}$ 
 iff
$\mu$ is forward complete  for $\{\complement , \pre_\sra\}$. 
Assume that $\mu$ is forward complete  for $\{\complement ,
\pre_\sra\}$. Let us first prove that $\mu$ is forward
complete for $\pret_\sra = \mathbf{AX}$: 
\begin{align*}
\mu \circ \pret_\sra \circ \mu &= \text{~~~~~[by definition of $\pret_\sra$]}\\
\mu \circ \complement \circ \pre_\sra \circ \complement \circ \mu & = 
\text{~~~~~[as $\mu$ is complete for $\complement$]}\\
\mu \circ \complement \circ \pre_\sra \circ \mu \circ \complement \circ \mu & = 
\text{~~~~~[as $\mu$ is complete for $\pre_\sra$]}\\
\mu \circ \complement \circ \mu \circ \pre_\sra \circ \mu \circ \complement \circ \mu & = 
\text{~~~~~[as $\mu$ is complete for $\complement$]}\\
\complement \circ \mu \circ \pre_\sra \circ \mu \circ \complement \circ \mu & = 
\text{~~~~~[as $\mu$ is complete for $\pre_\sra$]}\\
\complement \circ \pre_\sra \circ \mu \circ \complement \circ \mu & = 
\text{~~~~~[as $\mu$ is complete for $\complement$]}\\
\complement \circ \pre_\sra \circ \complement \circ \mu & = 
\text{~~~~~[by definition of $\pret_\sra$]}\\
\pret_\sra \circ \mu &
\end{align*}
The following fixpoint
characterizations are well known~\cite{cgp99}: 
\begin{itemize}
\item[--] $\mathbf{AU}(S_1,S_2) = \lfp(\lambda Z. S_2 \cup (S_1 \cap
\pret_\sra(Z)))$; 
\item[--] $\mathbf{EU}(S_1,S_2) = \lfp(\lambda Z. S_2 \cup (S_1 \cap
\pre_\sra(Z)))$;
\item[--] $\mathbf{AR}(S_1,S_2) = \gfp(\lambda Z. S_2 \cap (S_1 \cup
\pret_\sra(Z)))$; 
\item[--] $\mathbf{ER}(S_1,S_2) = \gfp(\lambda Z. S_2 \cap (S_1 \cup
\pre_\sra(Z)))$.
\end{itemize}
Let us show that $\mu$ is forward complete for $\mathbf{AU}$. The proofs
for the remaining operators in $\boldsymbol{Op}_{\CTL}$
are analogous. 
We need to show that $\mu (\lfp(\lambda Z. \mu(S_2) \cup (\mu(S_1) \cap
\pret_\sra(Z)))) = \lfp(\lambda Z. \mu(S_2) \cup (\mu(S_1) \cap
\pret_\sra(Z)))$. Let us  show that $\mu$ is forward
complete for the function $\lambda Z. \mu(S_2) \cup (\mu(S_1) \cap
\pret_\sra(Z))$: 
\begin{align*}
\mu( \mu(S_2) \cup (\mu(S_1) \cap \pret_\sra (\mu (Z)))) &= 
\text{~~~~~[as $\mu$ is complete for $\pret_\sra$]}\\
\mu( \mu(S_2) \cup (\mu(S_1) \cap \mu (\pret_\sra (\mu (Z))))) &= 
\text{~~~~~[as $\mu$ is  complete for $\cap$]}\\
\mu( \mu(S_2) \cup \mu(\mu(S_1) \cap \mu (\pret_\sra (\mu (Z))))) &=
\text{~~~~~[as $\mu$ is complete for $\cup$]}\\
\mu(S_2) \cup \mu(\mu(S_1) \cap \mu (\pret_\sra (\mu (Z)))) &=
\text{~~~~~[as $\mu$ is  complete for $\cap$]}\\
\mu(S_2) \cup (\mu(S_1) \cap \mu (\pret_\sra (\mu (Z)))) &=
 \text{~~~~~[as $\mu$ is  complete for $\pret_\sra$]}\\
\mu(S_2) \cup (\mu(S_1) \cap \pret_\sra (\mu (Z))).&
\end{align*}
Observe that since $\mu$ is additive (and therefore continuous)
we have that 
$\mu(\varnothing)=\varnothing$. 
Moreover, let us show that from the hypothesis that 
$\cK$ is finitely branching it follows  that
$\pret_\sra$ is continuous. First, notice that
$\pret_\sra$ is continuous iff $\pre_\sra$ is co-continuous. Hence,
let us check that $\pre_\sra$ is co-continuous. Let
$\{X_i\}_{i\in \mathbb{N}}$ be a decreasing chain of subsets of $\Sigma$ and let
$x\in \cap_{i\in \mathbb{N}} \pre_\sra (X_i)$. Since $\cK$ is
finitely branching, $\post_\sra(\{x\})$ is finite so that 
there exists some $k\in \mathbb{N}$ such that for any
$j>0$, $\post_\sra(\{x\})\cap X_k = \post_\sra(\{x\}) \cap
X_{k+j}$. Hence, there exists some $z\in \cap_{i\in \mathbb{N}} X_i
\cap \post_\sra(\{x\})$, so that $x\in \pre_\sra (\cap_{i\in \mathbb{N}}X_i)$.
Therefore, since $\pret_\sra$ is continuous we also have that    
$\lambda Z. \mu(S_2) \cup (\mu(S_1) \cap
\pret_\sra(Z))$ is continuous. 
We can therefore apply Lemma~\ref{fixrem2} so that
$\mu (\lfp(\lambda Z. \mu(S_2) \cup (\mu(S_1) \cap
\pret_\sra(Z)))) = \lfp(\lambda Z. \mu(S_2) \cup (\mu(S_1) \cap
\pret_\sra(Z)))$.\\
Thus, 
by Lemma~\ref{shellequiv}, $\sS_{\{\complement ,
\pre_\sra\}} = \sS_{\boldsymbol{Op}_{\CTL}}$, so that, by
Corollary~\ref{strai}, $\AD_{\fL_1} = \AD_\CTL$. Finally, since
$\cK$ is finitely branching and 
$\fL_1$ is closed under conjunction and negation, $\adp(P_{\fL_1})=
\adp(P_{\mathrm{bis}}) = \adp(P_{\fL_1})= \AD_{\fL_1}$.
\end{proof}

As a consequence of this and of the results in Section~\ref{csp}
(in particular of Corollary~\ref{spcb}), any partition refinement
algorithm $\Algbis$ for computing bisimulation equivalence on a
finitely branching Kripke structure, like those in
\cite{bfh90,dpp04,ly92,pt87}, can be characterized as a complete shell
refinement as follows:
$$\Algbis(P) = \pr (\sS_{\{\complement , \pre_\sra\}}
(\cM(P))).$$
Thus, $\Algbis$ is viewed as an algorithm for computing a particular
abstraction, that is $\pr$, of a particular complete shell, that is  
$\sS_{\{\complement , \pre_\sra\}}$. In particular, this holds for the
Paige-Tarjan algorithm~\cite{pt87} and leads to design a generalized
Paige-Tarjan-like procedure for computing most abstract strongly preserving
domains~\cite{rt05}.


Finally, our abstract intepretation-based approach allows us to give the
following nice
characterization of 
bisimulation for a partition $P$ in terms of forward completeness for
the corresponding partitioning abstract domain $\adp(P)$. 
\begin{theorem}\label{thbis}
Let $P\in \Part(\Sigma)$. Then, $P$ is a bisimulation on $\cK$ iff $\adp (P)$
is forward complete for $\{\boldsymbol{p}~|~p\in \mathit{AP}\} \cup
\{\pre_\sra\}$.
\end{theorem}
\begin{proof}
We view $\adp(P)$ as a uco so that $\adp(P)= \{\cup_i B_i \in \wp(\Sigma)~|~
\{B_i\} \subseteq P\}$. Let
us first observe that $P\preceq P_\ell$ iff $\adp(P)$ is forward complete for 
$\{\boldsymbol{p}\subseteq \Sigma~|~p\in AP\}$.  On the one hand,
since $\boldsymbol{p}=\{s\in \Sigma ~|~p\in \ell (s)\}$, 
if $s\in \boldsymbol{p}$ and $s\in B$, for some $B\in P$, then
$B\subseteq [s]_\ell \subseteq \boldsymbol{p}$. Hence, 
$\boldsymbol{p}$ is a union of some blocks of $P$ and therefore
$\boldsymbol{p}\in \adp(P)$. On the other hand, 
if $\adp(P)$ contains $\{\boldsymbol{p}\subseteq \Sigma~|~p\in AP\}$ then,
for any $p\in AP$, $\boldsymbol{p}$ is a union of some blocks in
$P$. Thus, for any $B\in P$, either $B\subseteq \boldsymbol{p}$ or $B\cap
\boldsymbol{p}=\varnothing$. Consequently, if 
$s \in B$ then  $B\subseteq [s]_\ell \in P_\ell$.
\\
Let us now note that $\ok{\adp(P)}$ is forward complete for $\ok{\pre_\sra}$ iff
    for any block $B\in P$, $\ok{\pre_\sra(B)}$ is a (possibly empty) union of
blocks of $P$:
this holds because $\ok{\pre_\sra}$ is additive, and therefore if $\ok{\{B_i\}}\subseteq
P$ then $\ok{\pre_\sra(\cup_i B_i)}=\cup_i \ok{\pre_\sra(B_i)}$. 
The fact that, for some  $B\in P$, $\ok{\pre_\sra(B)}= \ok{\cup_i
B_i}$, for some blocks $\ok{\{B_i\}}\subseteq P$, 
implies that if $s\in
\ok{\pre_\sra(B)}$, i.e., $s \sra t$ for some $t\in B$, then
$s\in \ok{B_j}$, for some $j$, and if $s'\in \ok{B_j}$ then $s'\in \ok{\pre_\sra(B)}$,
i.e., $s' \sra t'$ for some $t'\in B$, namely
condition~(2) of bisimulation for $P$ holds. On the other hand, if
condition~(2) of bisimulation for $P$ holds then if $s,s'\in B'$ and
$s\in \ok{\pre_\sra(B)}$, for some $B,B'\in P$, then $s'\sra t'$ 
for some $t\in B$, i.e., $s'\in \ok{\pre_\sra(B)}$,
and therefore $\ok{\pre_\sra(B)}$ is a union of blocks of $P$. This closes
the proof.
\end{proof}

\subsubsection{On the Smallest Abstract Transition Relation}\label{str1} 
As recalled in Section~\ref{amc},
the abstract Kripke structure $\cA = (P_{\mathrm{bis}},\ok{\sra^{\exists \exists}},
\ok{\ell^\exists})$ strongly
preserves $\CTL$,  where 
$B_1 \ok{{\sra^{\exists\exists}}} B_2$ iff there exist
$s_1\in B_1$ and $s_2 \in B_2$ such that $s_1 \sra s_2$, 
and $\ok{\ell^\exists (B)} = \cup_{s\in B} \ell(s)$. 
As a simple and  elegant 
consequence of our approach, it is easy to show that $\ok{\sra^{\exists
\exists}}$ is the \emph{unique} (and therefore the smallest)
abstract transition relation  on $P_{\mathrm{bis}}$
that induces strong preservation for $\CTL$. 

Let $\cK=(\Sigma,\sra,\ell)$ be finitely branching so that, by
Lemma~\ref{ctlequiv}, $\AD_{\fL_1}=\adp(P_{\mathrm{bis}})=\wp(P_{\mathrm{bis}})$. Recall that
the concrete interpretation $I$ induced by $\cK$ is such that
$I(\EX)=\pre_\sra$. 
By
Theorem~\ref{unique}, the unique interpretation of atoms
and operations in $\fL_1$ on the
abstract domain $\wp(P_{\mathrm{bis}})$ that gives rise to a s.p.\
abstract semantics is the best correct approximation
$I^{\wp(P_{\mathrm{bis}})}$.  
Hence, if $\ok{\cA} =
(P_{\mathrm{bis}},\ok{\sra^{\sharp}}, 
\ok{\ell^\sharp})$ is strongly preserving for $\CTL$ then the
interpretation $\ok{\pre_{\sra^{\sharp}}}$ of $\EX$ 
induced by $\cA$ must coincide with $\ok{I^{\wp(P_{\mathrm{bis}})}(\EX)}$.  
Consequently, $\ok{\pre_{\sra^{\sharp}}} = \ok{\alpha \circ \pre_\sra \circ
\gamma}$ so that for any $B_1,B_2 \in \ok{P_{\mathrm{bis}}}$, we have that 
$\ok{B_1 \sra^\sharp B_2}$ iff $B_1 \in \ok{\alpha(\pre_\sra
(\gamma(\{B_2\})))}$. Therefore, we conclude by observing that 
$B_1 \in \ok{\alpha(\pre_\sra
(\gamma(\{B_2\})))}$ iff $B_1 \ok{{\sra^{\exists\exists}}} B_2$. 

We believe that a similar reasoning could be also useful for other
languages $\fL$ in order to prove that the smallest abstract
transition relation on $P_{\fL}$ that induces strong preservation
exists. For example, this has been proved for the case of
$\ACTL$ by Bustan and Grumberg~\cite{bg03}.

\subsection{Stuttering Equivalence}
Lamport's criticism \cite{lam83} of the next-time operator $\mathrm{X}$
in $\CTL$/$\CTLS$ is well known. This motivated the study of  temporal
logics $\CTLX$/$\CTLSX$ 
obtained  from $\CTL$/$\CTLS$ by removing the next-time operator and this led to
study notions of behavioural \emph{stuttering}-based equivalences
\cite{bcg88,dnv95,gv90}. We are interested here in
\emph{divergence blind stuttering} (dbs for short) equivalence. 
Let $\cK = (\Sigma ,\sra ,\ell)$ be a Kripke structure over a set
$\AP$ of atoms. 
A
relation $R \subseteq \Sigma \times \Sigma $ is a divergence blind 
stuttering relation on $\cK$ if for any
$s,s'\in \Sigma $ such that $sR s'$: 
\begin{itemize}
\item[{\rm (1)}] $\ell(s) =\ell (s')$;
\item[{\rm (2)}] If
$\ok{s\sra t}$ 
then there exist $t_0,...,t_k\in \Sigma$, with $k\geq 0$, such that: (i)
$t_0=s'$; (ii)  for all $i\in [0,k-1]$, $\ok{t_i
\sra  t_{i+i}}$ and $s R t_i$; 
(iii) $t R t_k$;
\item[{\rm (3)}] $s'Rs$, i.e.\ $R$ is symmetric. 
\end{itemize}
Observe that condition~(2) allows the case $k=0$
and this simply boils down to requiring that $t R s'$.  
Since the empty
relation is a dbs relation and dbs relations are closed under union,
it turns out that the 
largest dbs  relation relation exists. It turns out that this largest
dbs relation  is an equivalence relation called dbs equivalence  and is 
denoted by $\sim_{\mathrm{dbs}}$ while $P_{\mathrm{dbs}}\in \Part(\Sigma)$ denotes
the corresponding partition.  In particular, a partition $P\in \Part(\Sigma)$ is a
dbs relation on $\cK$ when 
when $P\preceq P_{\mathrm{dbs}}$.

De Nicola and Vaandrager \cite[Theorem~3.2.5]{dnv95} showed that for
finite Kripke structures and for an interpretation of 
universal/existential path quantifiers over all the, possibly finite,
prefixes, dbs equivalence coincides with 
the state equivalence induced from the language $\CTLX$
(this also holds for $\CTLSX$), that is $P_{\mathrm{dbs}} =
P_{\CTLX}$. This is not true with the standard interpretation of path
quantifiers over infinite paths, since this requires a
divergence sensitive notion of stuttering (see the details in
\cite{dnv95}). Groote and Vaandrager \cite{gv90} presented a partition
refinement algorithm that computes the partition $P_{\mathrm{dbs}}$
in $O(|\Sigma| |\sra|)$-time.

We provide a characterization of divergence blind stuttering equivalence
as the state equivalence induced by 
the following language $\fL_2$ that includes propositional logic and 
the existential until operator $\mathrm{EU}$, where the interpretation
of the existential path quantifier is standard, i.e.\ over infinite paths:
$$\fL_2 \ni \varphi ::= 
~p ~|~ \varphi_1 \wedge \varphi_2 ~|~ \neg \varphi ~|~ 
\mathrm{EU}(\varphi_1,\varphi_2)
$$  
Since the transition relation $\sra$ is assumed to be total, 
let us recall that the standard semantics $\beu:\wp(\Sigma)^2\ra
\wp(\Sigma)$ of the existential until operator
is as
follows:
\begin{tabbing} 
~~~~$\beu(S_1,S_2) = S_2 \cup \{s\in S_1 ~|~$\=$\exists s_0,...,s_n \in
\Sigma, \text{ with } n\geq 0, \text{ such that~(i)~} s_0=s,$\\
\>${\rm ~(ii)~} \forall i\in [0,n-1].\,
s_i\in S_1 \text{ and } s_i  \sra  s_{i+1},$
${\rm ~(iii)~} s_n\in S_2\}$.
\end{tabbing}
\noindent
The following result characterizes
a dbs partition $P$ in terms of forward completeness for the
corresponding partitioning abstract domain $\adp(P)$. 

\begin{theorem}\label{chardbs}
Let $P\in \Part(\Sigma)$. Then, $P\in\Part(\Sigma)$ is a dbs partition on $\cK$ iff 
$\adp(P)$ is forward complete for $\{ \boldsymbol{p}~|~ p\in \AP\}
\cup \{ \beu\}$. 
\end{theorem}
\begin{proof}
As already shown in the proof of Theorem~\ref{thbis}, it
turns out that $P\preceq P_\ell$ iff $\adp(P)$ is forward complete for
$\{\boldsymbol{p} \subseteq \Sigma~|~ p\in \AP\}$. 
Thus, it remains to
show $P\in \Part(\Sigma)$ satisfies condition~(2) of the definition of
dbs relation iff
$\adp(P)$ is forward complete for $\beu$. 
    Let us first observe that $P\in\Part(\Sigma)$ satisfies this 
condition~(2) 
  iff for any $B_1,B_2\in P$, $\beu(B_1,B_2) = B_1 \cup B_2$.\\
$(\Rightarrow)$ 
   If $B_1=B_2$ then $\beu(B_1,B_1)=B_1$. 
  Otherwise, assume that $B_1\neq B_2$. If $B_2 \subsetneq
\beu(B_1,B_2) \subseteq B_1 \cup B_2$ then there exists $s\in
\beu(B_1,B_2)$ such that $s\in B_1$. Thus, if $s'\in B_1$ then, by
condition~(2), $s'\in \beu(B_1,B_2)$. This implies that $\beu(B_1,B_2)
= B_1 \cup B_2$.    \\
 $(\Leftarrow)$
  Let $B\in P$, $s,s'\in B$ and $s \sra  t$. If $t\in B$
then condition~(2) is satisfied. Otherwise, $t\in B'$, for some $B'\in
P$, with $B\neq B'$. Thus, $s\in \beu(B,B')$ and therefore
$\beu(B,B')=B\cup B'$. This means that condition~(2) is
satisfied for $P$. \\
  To complete the proof it is now sufficient to show that if, 
 for any $B_1,B_2\in P$, $\beu(B_1,B_2)= B_1 \cup B_2$ then
$\adp(P)$ is forward complete for $\beu$, i.e., for any
$\{B_i\}_{i\in I},\{B_j\}_{j\in J} \subseteq P$, $\beu(\cup_i B_i,
\cup_j B_j) = \cup_k B_k$, for some $\{B_k\}_{k\in K}\subseteq P$. 
The function $\beu$ is additive in its second argument, thus we only need to show
that, for any $B\in P$, 
$\beu(\cup_i B_i,B)=\cup_k B_k$, namely if $s\in \beu(\cup_i
B_i,B)$ and $s\in B'$, for some $B'\in P$, then $B'\subseteq
\beu(\cup_i B_i, B)$. 
  If $s\in\beu(\cup_i B_i, B)$ and $s\in B'$, for some $B'\in \{B_i\}_i$, then there exist
  $n\geq 0$ and   $s_0,...,s_n \in
\Sigma $ such that $s_0=s$, $\forall
j\in [0,n-1]. s_j\in \cup_i B_i$ and  $s_j  \sra s_{j+1}$, and   $s_n\in B$. 
  Let us prove by induction on $n\in\mathbb{N}$ that if $s'\in B'$ then 
  $s'\in\beu(\cup_i B_i, B)$.\\[5pt]
($n=0$):
    In this case $s\in \cup_i B_i$ and $s\in B=B'$. Hence, for some
$k$, $s\in B_k=B=B'$ and therefore $s\in \beu(B,B)$. By hypothesis,
$\beu(B,B)=B$. Moreover, $\beu$ is monotone on its first component and
therefore $B'=B=\beu(B,B)\subseteq \beu(\cup_i B_i,B)$. \\[5pt]
($n+1$):
    Suppose that 
there exist
  $s_0,...,s_{n+1} \in
\Sigma $ such that $s_0=s$, $\forall
j\in [0,n]. s_j\in \cup_i B_i$ and  $s_j  \sra 
s_{j+1}$, and   $s_{n+1}\in B$. Let $s_n \in B_k$, for some $B_k\in
\{B_i\}_{i\in I}$. Then, $s\in \beu(\cup_i B_i,B_k)$ and $s=s_0 \sra
s_1 \sra  ... \sra 
s_n$. Since this finite path has length $n$, 
by inductive hypothesis, $s'\in \beu(\cup_i B_i,B_k)$. Hence,
there exist $r_0,...,r_m \in \Sigma $, with $m\geq 0$, such that 
$s'=r_0$, $\forall
j\in [0,m-1]. r_j\in \cup_i B_i$ and  $r_j  \sra 
r_{j+1}$, and   $r_{m}\in B_k$. Moreover, since $s_n \sra 
s_{n+1}$, we have that $s_n\in \beu(B_k,B)$. By
hypothesis, $\beu(B_k,B)=B_k\cup B$, and therefore $r_m\in
\beu(B_k,B)$. Thus, there exist $q_0,...,q_l \in \Sigma $, with $l\geq 0$, such that 
$r_m=q_0$, $\forall
j\in [0,l-1]. q_j\in B_k$ and  $q_j  \sra 
q_{j+1}$, and   $q_{l}\in B$. We have thus found the following finite path: 
$s'=r_0 \sra  r_1 \sra 
... \sra  r_m = q_0 \sra  q_1 \sra  ... \sra  q_l$, where all the
states in the sequence but the last one $q_l$  belong to $\cup_i B_i$,
while $q_l\in B$. This means that $s'\in \beu(\cup_i B_i,B)$.  
\end{proof}

As a consequence, we obtain a characterization of dbs equivalence as
the state equivalence induced by the standard interpretation of the
language $\fL_2$.

\begin{corollary}\label{coro7}
Let $\Sigma$ be finite. Then, $P_{\mathrm{dbs}}= P_{\fL_2}$.
\end{corollary}
\begin{proof}
By definition, $P_{\mathrm{dbs}} = 
\curlyvee_{\Part(\Sigma)} \{P\in
\Part(\Sigma)~|~ P$ is a dbs relation on  $\cK \}$. 
By Theorem~\ref{chardbs}, $P_{\mathrm{dbs}} = 
\curlyvee_{\Part(\Sigma)} \{P\in
\Part(\Sigma)~|~ \adp(P)$ is complete for  
$\{ \boldsymbol{p}~|~ p\in \AP\}
\cup \{ \beu\}\}$. 
By Theorem~\ref{teoGI}, $\adp$ is co-additive on
$\Part(\Sigma)_\succeq$, that is $\adp$ preserves lub's in
$\Part(\Sigma)_\preceq$. Hence, $\adp(P_{\mathrm{dbs}}) = 
\sqcup_{\Abs(\wp(\Sigma))}  \{\adp(P)\in
\Abs(\wp(\Sigma))~|~ P\in \Part(\Sigma),\, 
\adp(P)$ is complete for  $\{ \boldsymbol{p}~|~ p\in \AP\}
\cup \{ \beu\}\}$. By Theorem~\ref{teoGI}, 
$\Absp(\wp(\Sigma)) = \{\adp(P)~|~P\in \Part(\Sigma)\}$ so that
 $\adp(P_{\mathrm{dbs}}) = 
\sqcup_{\Abs(\wp(\Sigma))}  \{A \in
\Absp(\wp(\Sigma))~|~ 
A$  is complete for  $\{ \boldsymbol{p}~|~ p\in \AP\}
\cup \{ \beu\}\}$. By Corollary~\ref{coropart}, $A \in
\Absp(\wp(\Sigma))$ iff $A$ is forward complete for $\complement$,
so that $\adp(P_{\mathrm{dbs}}) = 
\sqcup_{\Abs(\wp(\Sigma))}  \{A \in
\Abs(\wp(\Sigma))~|~ 
A$ is complete for  $\{ \boldsymbol{p}~|~ p\in \AP\}
\cup \{\complement, \beu\}\}$. Then, we note that $A$ is forward
complete for $\{ \boldsymbol{p}~|~ p\in \AP\}$ iff $A \sqsubseteq \cM(\{
\boldsymbol{p}~|~ p\in \AP\})$. Hence, 
 $\adp(P_{\mathrm{dbs}}) = 
\sqcup_{\Abs(\wp(\Sigma))}  \{A \in
\Abs(\wp(\Sigma))~|~ A \sqsubseteq \cM(\{ \boldsymbol{p}~|~ p\in \AP\}), \, 
A$ is complete for 
$\{\complement, \beu\}\} = \sS_{\{\complement, \beu\}} (\cM(\{
\boldsymbol{p}~|~ p\in \AP\}))$. Finally, since $\Sigma$ is finite and
therefore closure under infinite conjunction boils down to closure
under finite conjunction,  by Corollary~\ref{strai}, 
$\sS_{\{\complement, \beu\}} (\cM(\{
\boldsymbol{p}~|~ p\in \AP\}))=\AD_{\fL_2}$. Thus,  by Proposition~\ref{onetwo}~(1),
$\adp(P_{\mathrm{dbs}}) = \AD_{\fL_2}$, so that 
$P_{\mathrm{dbs}} = \pr(\adp( P_{\mathrm{dbs}})) = \pr(\AD_{\fL_2}) =
P_{\fL_2}$.   
\end{proof}

As a consequence of Corollary~\ref{spcb}, the Groote-Vaandrager
algorithm~\cite{gv90} $\GV$ for computing dsb equivalence on a finite
Kripke structure  can be characterized as a complete shell
refinement as follows:
$$\GV(P) = \pr (\sS_{\{\complement , \beu\}}
(\cM(P))).$$

\subsection{Simulation Preorder and Equivalence}\label{simeq}
Simulations are possibly nonsymmetric bisimulations, that is
$R\subseteq \Sigma \times \Sigma $ is a
simulation on a Kripke structure $\cK=(\Sigma,\sra,\ell)$ 
if for any $s,s'\in \Sigma $ such that $s R s'$: 
\begin{itemize}
\item[{\rm (1)}] $\ell(s') \subseteq \ell (s)$; 
\item[{\rm (2)}] For
any $t\in \Sigma $ such that $\ok{s \sra t}$, there exists $t'\in \Sigma $
such that $\ok{s'\sra t'}$
and $t R t'$.
\end{itemize}
The empty
relation is a simulation and simulation relations are closed under
union, so that the 
largest simulation relation exists. It turns out that the largest
simulation is a preorder relation called similarity preorder and
denoted by $R_{\mathrm{sim}}\in \PreOrd(\Sigma)$. 
Therefore,  a preorder relation $R\in \PreOrd(\Sigma)$ is a simulation on $\cK$
when $R \subseteq  R_{\mathrm{sim}}$. Simulation equivalence 
$\sim_{\mathrm{simeq}}\,\subseteq
\Sigma \times \Sigma $ is the symmetric closure of $R_{\mathrm{sim}}$:
$s\sim_{\mathrm{simeq}}s'$ iff there exist two simulation
relations $R_1$ and $R_2$ such that $s R_1 s'$ and $s' R_2s$. 
$P_{\mathrm{simeq}}\in \Part(\Sigma)$ denotes
the partition corresponding to $\sim_{\mathrm{simeq}}$. 

A number of algorithms for computing simulation equivalence
have been proposed
\cite{bp95,bg03,cps93,gpp03,hhk95} and some of them like
\cite{bp95,hhk95} first compute the similarity preorder and then
from it they obtain simulation equivalence.    The problem of 
computing simulation equivalence is important in model checking
because, as recalled in Section~\ref{amc}, simulation equivalence
strongly preserves $\ACTL$ so that
$P_{\mathrm{simeq}} = P_\ACTL$  (see \cite[Section~4]{gl94}).
Recall that $\ACTL$ is 
obtained by restricting $\CTL$, as defined in Section~\ref{cs}, to
universal quantifiers and by allowing negation on atomic propositions only:
$$
\ACTL \ni \varphi ::=  p ~|~ \neg p ~|~ \varphi_1 \wedge \varphi_2 ~|~
\varphi_1 \vee \varphi_2 ~|~ \mathrm{AX}\varphi 
~|~ \mathrm{AU}(\varphi_1,\varphi_2) ~|~
\mathrm{AR}(\varphi_1,\varphi_2)$$

It turns out that the most abstract s.p.\ domain for $\ACTL$
can be obtained as the most abstract s.p.\ domain for 
the following sublanguage $\fL_3$:
$$\fL_3 \ni \varphi ::= p ~|~ \neg p ~|~ 
\varphi_1 \wedge \varphi_2 ~|~ \varphi_1 \vee \varphi_2 ~|~
\mathrm{AX}\varphi $$
\begin{lemma}\label{lem2}
Let $\cK$ be finitely branching. Then, $\AD_{\ACTL} = \AD_{\fL_3}$.
\end{lemma}
\begin{proof}
Let $\boldsymbol{Op}_{\ACTL} =  
\{\cap,\cup,  \mathbf{AX}, 
\mathbf{AU}, \mathbf{AR} \}$ be the
set of standard interpretations of the operators of $\ACTL$ on $\cK$,
so that $\mathbf{AX}=\pret_\sra$.  
Analogously to the proof of Lemma~\ref{ctlequiv}, as a consequence of
the least/greatest fixpoint characterizations of $\mathbf{AU}$ and
$\mathbf{AR}$, it turns out that
for any $A \in \Abs(\wp(\Sigma))$, $A$ is forward complete for  
$\boldsymbol{Op}_{\ACTL}$ iff $A$ is forward complete for $\{\cup,
\pret_\sra\}$. Thus, by Lemma~\ref{shellequiv}, $\sS_{\{\cup ,
\pret_\sra\}} = \sS_{\boldsymbol{Op}_{\ACTL}}$, so that, by
Corollary~\ref{strai}, $\AD_{\fL_3} = \AD_\ACTL$. 
\end{proof}

Thus, by Proposition~\ref{onetwo}~(1), 
$P_\ACTL = \ok{\pr(\AD_{\ACTL})} =
\pr(\AD_{\fL_3}) = P_{\fL_3}$, so that $P_{\mathrm{simeq}} =
P_{\fL_3}$.  As a further consequence, by Corollary~\ref{spcb}, 
 any algorithm $\Algsimeq$ that computes simulation equivalence can be
viewed  as a partitioning abstraction of the
$\{\cup,\pret_\sra\}$-complete shell refinement: 

$$\Algsimeq(P) = \pr (\sS_{\{\cup , \pre_\sra\}}
(\cM(P))).$$

An instantiation of the generalized Paige-Tarjan-like procedure in \cite{rt05} for
the complete shell $\sS_{\{\cup , \pre_\sra\}}$ allows to design a new efficient
abstract intepretation-based 
algorithm for computing simulation equivalence \cite{rt06} whose space and
time complexity is comparable with that of state-of-the-art
algorithms  like \cite{bg03,gpp03}.

\subsubsection{Preorders as Abstract Domains}

Simulations give rise to preorders rather than equivalences like in
the case of bisimulations and dbs relations. Thus, in
order to characterize simulation for preorders as forward
completeness for abstract domains we need to view preorders as
abstract domains. This can be obtained by generalizing the abstraction
in Section~\ref{paa} from partitions to preorders. 

Let $R\in \PreOrd(\Sigma)$ and for any $x\in \Sigma$ let us define 
$R^{\pre} \ud \{ \pre_R(\{x\})
\subseteq \Sigma~|~ x\in \Sigma\}$. The preorder $R$ gives rise to
an abstract domain $\wp(R^{\pre})_\subseteq$ which is related to $\wp(\Sigma)_\subseteq$
through the following abstraction and concretization maps: 
$$\alpha_R(S)\ud\{\pre_R(\{x\}) \subseteq \Sigma~|~ x\in S\}
~~~~~~~
\gamma_R(\mathcal{X})\ud \cup_{X\in \mathcal{X}} X.$$
It is easy to check that from the hypothesis that $R$ is a
preorder it follows that $(\alpha_R,
\wp(\Sigma)_\subseteq, \wp(R^{\pre})_\subseteq,\gamma_R)$ is indeed a
GI. 
Hence, any $R\in \PreOrd(\Sigma)$ induces an abstract domain
denoted by
$\add(R) \in \Abs(\wp(\Sigma))$. Also, note that $\gamma_R\circ
\alpha_R = \pre_R$, namely $\pre_R$ is the closure associated to
$\add(R)$. 
The notation $\add$
comes from the fact that an abstract domain $A$ is equivalent to some
$\add(R)$ if and only if
$A$
is disjunctive. 
\begin{lemma}\label{lemmadisj}
$\{\add(R)\in \Abs(\wp(\Sigma))~|~ R\in
\PreOrd(\Sigma)\}=\{A\in \Abs(\wp(\Sigma))~|~ A~ \text{{\rm is disjunctive}}\}$. 
\end{lemma}
\begin{proof}
Observe that $\gamma_R$ is trivially additive, so that any $\add(R)$
is disjunctive. On the other hand, let $A\in \Abs(\wp(\Sigma))$ be
disjunctive and consider the relation $R^A = \{(x,y) ~|~ \alpha(\{x\})
\leq_A \alpha(\{y\})\}$ which is trivially a preorder. Thus,
$\add(R^A)$ is disjunctive so that in order to conclude that $\add(R^A)$
is equivalent to $A$ it is enough to observe that for any $y\in \Sigma$,
$\pre_{R^A}(\{y\}) = \gamma(\alpha(\{y\}))$: this is true because
$\gamma(\alpha(\{y\})) = \{x\in \Sigma~|~ \alpha(\{x\})\leq_A
\alpha(\{y\})\}=\pre_{R^A}(\{y\})$. 
\end{proof}

Let us observe that $\add$ indeed generalizes $\adp$ from partitions
to preorders because for any $P\in \Part(\Sigma)$, $\adp(P)=\add(R)$:
this is a simple consequence of the fact that for a partition $P$
viewed as an equivalence relation and
for $x\in \Sigma$,
$P_x$ is exactly a block of $P$ so that
$\alpha_P(S)=\{\pre_P(\{x\})~|~ x\in S\}$.  
On the other hand, an abstract domain $A\in \Abs(\wp(\Sigma))$ induces
a preorder relation $\preord(A)\in \PreOrd(\Sigma)$ as follows:
$$(x,y)\in \preord(A)  \text{~~~iff~~~} \alpha(\{x\}) \leq_A \alpha(\{y\}).$$
It turns out that the maps $\add$ and $\preord$ allows to view the lattice of
preorder relations as an abstraction of the lattice of abstract
domains. 
\begin{theorem}
$(\preord,\Abs(\wp(\Sigma))_\sqsupseteq,\PreOrd(\Sigma)_\supseteq,
\add)$. 
\end{theorem}
\begin{proof}
Let $A \in \Abs(\wp(\Sigma))$ and $R\in
\PreOrd(\Sigma)$. Let us prove that  
$R\subseteq \preord(A) \; \Lra\; \add(R) \sqsubseteq
\gamma\circ \alpha$. \\
$(\Rightarrow)$~Let $S\subseteq \Sigma$ and let us show that
$\add(R)(S)=\pre_R(S)\subseteq \gamma(\alpha(S))$. If $x\in
\pre_R(S)$ then $xRy$ for some $y\in S$, so that $(x,y) \in
\preord(A)$, i.e.\ $\alpha(\{x\}) \leq_A \alpha(\{y\})$. Thus, by
applying $\gamma$, $x\in \gamma(\alpha(\{x\})) \subseteq 
\gamma(\alpha (\{y\}))\subseteq \gamma(\alpha(S))$. \\
$(\Leftarrow)$~Let $(x,y) \in R$ and let us show that $\alpha(\{x\})
\leq \alpha(\{y\})$. Note that $x\in \pre_R(\{y\}) =\add(R)(\{y\})
\subseteq \gamma(\alpha(\{y\}))$, so that $\alpha(\{x\}) \leq_A
\alpha(\{y\})$, namely $(x,y)\in \preord(A)$. 
\end{proof}

Let us remark that $\mathbb{D}\ud \add \circ \preord$ is a lower closure operator on
$\tuple{\Abs(\wp(\Sigma)),\sqsubseteq}$ and that, by
Lemma~\ref{lemmadisj}, 
for any $A \in
\Abs(\wp(\Sigma))$, $A$ is disjunctive iff $\mathbb{D}(A)=A$. 
Hence, $\mathbb{D}$ coincides with the disjunctive-shell refinement,
also known as disjunctive completion \cite{CC79}, namely
$\mathbb{D}(A)$ is the most abstract disjunctive refinement of $A$.

We can now provide a characterization of simulation preorders in terms of forward
completeness. 
\begin{theorem}
Let $R\in \PreOrd(\Sigma)$. Then, $R$ is a simulation on $\cK$ iff
$\add(R)$ is forward complete for $\{\boldsymbol{p}~|~p\in \AP\}\cup \{\pret_\sra\}$.
\end{theorem}
\begin{proof}
Recall that $\pre_R$ is the closure associated to $\add(R)$. We first
observe that 
$(sRs' \: \Ra\: \ell(s') \subseteq \ell(s))$ iff $\pre_R$ 
is forward complete for  $\boldsymbol{\AP}$. On the one hand, if
$\boldsymbol{p}\in \boldsymbol{\AP}$ and $s\in \pre_R(\boldsymbol{p})$
then $sR s'$ for some $s' \in \boldsymbol{p}$, so that, from
$\ell(s')\subseteq \ell(s)$, we obtain $s\in \boldsymbol{p}$, and
therefore $\pre_R(\boldsymbol{p}) =\boldsymbol{p}$. On the other hand,
if $sRs'$ and $s'\in \boldsymbol{p}$, for some $\boldsymbol{p}\in
\boldsymbol{\AP}$, then $s'\in \boldsymbol{p}=\pre_R(\boldsymbol{p})$
so that $\pre_R(\{s'\})\subseteq \pre_R(\pre_R (\boldsymbol{p}) )
= \pre_R(\boldsymbol{p}) = \boldsymbol{p}$ and therefore from $s\in
\pre_R(\{s'\})$ we obtain $s\in \boldsymbol{p}$. \\
Thus, it remains to show that $R$ satisfies condition (2) of the
definition of simulation iff $\pre_R$ is forward complete for
$\pre_\sra$. \\
$(\Ra)$ We prove that for any $S$, $\pre_R (\pret_\sra
(\pre_R (S))) \subseteq \pret_\sra (\pre_R (S))$. Let $x\in  \pre_R (\pret_\sra
(\pre_R (S)))$ so that there exists some $y\in \pret_\sra (\pre_R
(S))$ such that $xRy$. If $x\sra x'$, for some $x'$, then, by
simulation, there exists some $y'$ such that $y\sra y'$ and $x'R
y'$. Hence, $y'\in \pre_R (S)$ and this together with $x'R y'$, as $R$
is transitive, gives $x'\in \pre_R(S)$. Therefore, $x\in \pret_\sra
(\pre_R(S))$. 
\\
$(\Leftarrow)$  Observe that in order to show that $R$ is a simulation it is
enough to show that if $xR y$ then $x\in \pret_\sra (\pre_R (\post_\sra
(\{y\})))$. The following implications hold, where
$\post_\sra (\{y\}) \subseteq \pre_R (\post_\sra (\{y\}))$ holds
because $\pre_R$ is a uco:
\begin{align*}
\post_\sra (\{y\}) \subseteq
\pre_R (\post_\sra (\{y\})) & ~~\Ra &
\text{~~~[as $\pret_\sra$ is monotone]}\\
 \pret_\sra(\post_\sra (\{y\})) \subseteq
\pret_\sra (\pre_R (\post_\sra (\{y\}))) & ~~\Ra &
\text{~~~[as $y\in \pret_\sra(\post_\sra(\{y\}))$]} \\
\{y\} \subseteq \pret_\sra (\pre_R (\post_\sra (\{y\}))) & ~~\Ra &
\text{~~~[as $\pre_R$ is monotone]} \\
\pre_R(\{y\}) \subseteq \pre_R(\pret_\sra (\pre_R (\post_\sra
(\{y\})))) & ~~\Ra &
\text{~~~[as $\pre_R$ is forward complete for $\pret_\sra$]} \\
\pre_R(\{y\}) \subseteq \pret_\sra (\pre_R (\post_\sra (\{y\}))) & ~~\Ra
&
\text{~~~[as $x\in \pre_R(\{y\})$]}\\
x \in \pret_\sra (\pre_R (\post_\sra (\{y\}))) & &
\end{align*}
and this closes the proof.
\end{proof}

\section{Related work} 
Loiseaux et al.~\cite{loi95} generalized the standard approach to
abstract model checking to more general abstract models where an
abstraction relation $\sigma
\subseteq \States \times A$  is used instead of a surjective
function $h:\States \ra A$. However,
the results of strong preservation given there (cf.~\cite[Theorems~3
and~4]{loi95}) require the hypothesis that the relation $\sigma$ is
difunctional, i.e.\ $\sigma = \sigma
\sigma^{-1}\sigma$. In this case the
abstraction relation $\sigma$ can indeed be derived from a function, so
that the class of strongly preserving abstract models in Loiseaux et al.'s 
framework is not really
larger than the class of standard partition-based abstract models (see the
detailed discussion by Dams et al.~\cite[Section~8.1]{dgg97}). \\
\indent
Giacobazzi and Quintarelli~\cite{GQ01} first noted that
strong preservation is related to
completeness in abstract interpretation  by 
studying the relationship between  complete abstract interpretations and
Clarke et al.'s \cite{cgjlv00,cgjlv03,cjlv02} spurious counterexamples. 
Given a formula $\varphi$ of $\ACTL$, a model checker running on a
standard abstract Kripke structure defined over a state partition $P$ may
provide a spurious counterexample $\ok{\pi^\sharp}$ for $\varphi$, 
namely a path of abstract
states, namely blocks of $P$, which does not correspond to a real concrete
counterexample.  
In this case, by exploiting the spurious
counterexample $\ok{\pi^\sharp}$, 
the partition $P$ is refined to $\ok{P'}$ by splitting a
single block of $P$.    As a result, 
this refined partition $\ok{P'}$ does not admit the spurious counterexample
$\ok{\pi^\sharp}$ for $\varphi$ 
  so that $\ok{P'}$ is given as a new refined 
abstract model for $\varphi$ to the model checker.  Giacobazzi and
Quintarelli \cite{GQ01} cast spurious counterexamples for a partition
$P$ as a lack of
(standard) completeness in the abstract interpretation sense for the
corresponding partitioning
abstract domain $\adp(P)$. Then, by applying the results in
\cite{jacm} they put forward a
method for systematically refining abstract domains in order to
eliminate spurious counterexamples. The relationship
between completeness and spurious counterexamples was further studied
in \cite{dp03}, where it is also shown that a 
block splitting operation in 
Paige and Tarjan~\cite{pt87} partition refinement algorithm  can be characterized
in terms of complete abstract interpretations.
More in general, the idea of
systematically enhancing the precision of abstract interpretations by
refining the underlying abstract domains dates back to the early works
by Cousot and Cousot~\cite{CC79}, and evolved to the systematic design
of abstract interpretations by abstract domain refinements
\cite{fgr96,GR97,jacm}.

\section{Conclusion}

This work shows how the abstract interpretation technique
allows to generalize the notion of strong preservation from standard
abstract models specified as abstract Kripke structures to generic
domains in abstract interpretation. For any inductively defined
language $\fL$, it turns out that strong preservation of $\fL$ in
a standard abstract model checking framework based on partitions of
the space state $\Sigma$ becomes a particular instance of the property
of forward completeness of abstract domains w.r.t.\ the semantic
operators of the language $\fL$. In particular, a
generalized abstract model can always be refined through a
fixpoint iteration to the most abstract domain that strongly
preserves $\fL$.  This generalizes in our framework the idea of
partition refinement algorithms that reduce the state space $\Sigma$
in order to obtain a minimal abstract Kripke structure that is
strongly preserving for some temporal language.

This work deals with generic temporal languages consisting of
state formulae only. As future work, it would be interesting to
study whether the ideas of our abstract interpretation-based approach
can be applied to linear languages like $\mathrm{LTL}$ consisting of 
formulae that are interpreted as
sets of paths of a Kripke structure. The idea here is to investigate
whether standard strong preservation of $\mathrm{LTL}$ can be
generalized to abstract interpretations of the powerset
of traces and to the corresponding completeness
properties.  Fairness can be also an interesting topic of
investigation, namely to study whether our abstract
interpretation-based framework allows to handle fair semantics and 
fairness constraints \cite{cgp99}. 

Finally, let us mention that the results
presented in this paper led to design a generalized Paige-Tarjan
refinement algorithm based on abstract interpretation for computing
most abstract strongly preserving domains~\cite{rt05}. 
As shown in Section~\ref{csp}, a most abstract strongly preserving
domain can be characterized as a greatest
fixpoint computation in $\Abs(\wp(\Sigma))$.
It is shown in \cite{rt05} that the Paige-Tarjan algorithm
\cite{pt87} can be viewed exactly 
as a corresponding abstract greatest fixpoint computation in
$\Part(\Sigma)$. This
leads to an   abstract
interpretation-based Paige-Tarjan-like refinement algorithm  that is parameteric on any
abstract interpretation of the lattice $\Abs(\wp(\Sigma))$ of abstract
domains of $\wp(\Sigma)$ and on any generic inductive language $\fL$.

\paragraph*{{\it Acknowledgements.}} We wish to thank Mila Dalla
Preda and Roberto Giacobazzi who contributed to the early stage of
this work. This paper is an extended and revised version of \cite{rt04a}.
This work was partially supported by the FIRB Project
``Abstract interpretation and model checking for the verification of
embedded systems'' and by the COFIN2004
Project ``AIDA: Abstract Interpretation Design and Applications''.

\end{document}